\documentclass[review]{elsarticle}
\usepackage[hmarginratio=1:1,top=32mm,left=20mm,columnsep=17pt]{geometry}

\usepackage{lineno,array}
\modulolinenumbers[5]
\usepackage{color}
\usepackage{tabularx}
\usepackage{graphicx,epstopdf}
\usepackage[dvipsnames]{xcolor}
\usepackage{amsmath}
\usepackage{amssymb}
\usepackage{amsfonts}	
\usepackage{moreverb}
\usepackage{dsfont}
\usepackage{tipa}
\usepackage{upgreek}
\usepackage{bm}
\usepackage{multirow}
\usepackage{soul}
\usepackage{ulem}
\usepackage{ textcomp }
\usepackage{relsize} 

\usepackage[textsize=footnotesize backgroundcolor=yellow!70, bordercolor=orange]{todonotes}

\newcommand\BibTeX{{\rmfamily B\kern-.05em \textsc{i\kern-.025em b}\kern-.08em
T\kern-.1667em\lower.7ex\hbox{E}\kern-.125emX}}

\usepackage{natbib}
\setcitestyle{square,numbers}

\usepackage[toc]{appendix}

\makeatletter
\usepackage[colorlinks]{hyperref}
\hypersetup{colorlinks=true}

\newcommand{\pd}{\mathcal{\partial}}

\renewcommand{\v}{\mathbf{v}}

\newcommand{\rmd}{{\rm d}}
\newcommand{\rmdt}{{\rm d}\, t}
\newcommand{\rmD}{{\rm D}}

\newcommand{\G}{\bm{G}}
\newcommand{\A}{\AAA}

\newcommand{\bdm}{\begin{displaymath}}
\newcommand{\edm}{\end{displaymath}}

\newcommand{\bea}{\begin{eqnarray} }
\newcommand{\eea}{\end{eqnarray} }

\newcommand{\dev}[1]{#1'} 

\newcommand{\AAA}{{\boldsymbol{A}}}

\newcommand{\DD}{{\mathbf{D}}}

\newcommand{\GG}{{\boldsymbol{G}}}

\newcommand{\vv}{{\boldsymbol{v}}}

\newcommand{\BB}{{\bm{B}}}

\newcommand{\FF}{{\bm{F}}}
\newcommand{\mm}{\bm{m}}
\newcommand{\LL}{{\bm{L}}}
\newcommand{\II}{{\bm{I}}}

\newcommand{\SSS}{{\bm{S}}}
\newcommand{\TT}{\bm{T}}
\newcommand{\asymvel}{\bm{\Omega}}	
\newcommand{\F}[2]{ F_{{#1}{#2}}\, }				
\newcommand{\Feff}[2]{ F^{\rm e}_{{#1}{#2}}\, }				
\newcommand{\Fplast}[2]{ F^{\rm p}_{{#1}{#2}}\, }				
\newcommand{\Fplastinv}[2]{ F^{\rm -p}_{{#1}{#2}}\, }				
\newcommand{\Feffvec}[1]{ \bm{F}^{\rm e}_{{#1}}\, }				
\newcommand{\dist}[2]{ A_{{#1}{#2}}\, }				
\newcommand{\ssigma}{\bm{\sigma}}
\newcommand{\GPR}{GPR }
\newcommand{\MG}{Mie-Gr{\"u}neisen\ }

\newcommand{\Id}{\bm{I}}

\newcommand{\tr}{\text{tr}}
\newcommand{\transpose}{{\rm {\mathsmaller T}}}
\newcommand{\eg}{e.g.\ }

\newcommand{\ie}{i.e.\ }
\newcommand{\css}{c_s^2}
\newcommand{\cs}{c_s}

\journal{Noname Journal}
\newfont{\numerikEleven}{ecrm1000}
\newfont{\numerikTen}{cmss10}
\newfont{\numerikNine}{cmss9}
\newfont{\numerikEight}{cmss8}

\begin{document}
\hypersetup{
    citecolor=RoyalBlue,
    linkcolor=WildStrawberry,
	urlcolor=NavyBlue
 }

\begin{frontmatter}
\title{Theoretical and numerical comparison of hyperelastic and hypoelastic formulations for 
Eulerian non-linear elastoplasticity} 
\author[IMT,NSC]{Ilya Peshkov}
\ead{peshenator@gmail.com}

\author[UniBZ]{Walter Boscheri$^{*}$}
\ead{walter.boscheri@unibz.it}
\cortext[cor1]{Corresponding author}

\author[IMB]{Rapha\"{e}l Loub\`{e}re}
\ead{raphael.loubere@u-bordeaux.fr}

\author[NSC,NSU]{Evgeniy Romenski}
\ead{evrom@math.nsc.ru}

\author[UniTN]{Michael Dumbser}
\ead{michael.dumbser@unitn.it}

\address[IMT]{{Institut de Math\'{e}matiques de Toulouse, Universit\'{e} Toulouse III, F-31062 Toulouse, France.}}
\address[UniBZ]{Faculty of Science and Technology, Free University of Bozen, Piazza Universit{\'a} 
1, 39100 Bolzano, Italy.} 
\address[IMB]{Institut de Math\'{e}matiques de Bordeaux (IMB), UMR 5219 Universit{\'e} de Bordeaux, 
F-33405 Talence, France}
\address[NSC]{{Sobolev Institute of Mathematics, 4 Acad. Koptyug Avenue, 630090 Novosibirsk, 
Russia}}
\address[NSU]{{Novosibirsk State University, 2 Pirogova Str., 630090 Novosibirsk, Russia}}
\address[UniTN]{Department of Civil, Environmental and Mechanical Engineering, University of 
Trento, Via Mesiano 77, 38123 Trento, Italy.} 

\begin{abstract}
The aim of this paper is to compare a hyperelastic with a hypoelastic model describing the Eulerian dynamics of solids in the context of non-linear elastoplastic deformations. Specifically, we 
consider the well-known hypoelastic Wilkins model, which is 
compared against a hyperelastic model based on the work of Godunov and Romenski.  
 First, we discuss some general conceptual 
differences of the two approaches. Second, a detailed study of both models is proposed, where 
differences are made evident at the aid of deriving a hypoelastic-type model corresponding to the 
hyperelastic model and a particular equation of state used in this paper. Third, 
using the same high order ADER Finite Volume and Discontinuous Galerkin  methods on fixed and moving unstructured meshes for both models, a wide range of numerical benchmark  test 
problems has been solved. 
The numerical solutions obtained for the two different models are directly compared with each other. For small elastic 
deformations, the two models produce very similar solutions that are close to each other. 
However, if large elastic or elastoplastic deformations occur, the solutions present 
larger differences. 
\end{abstract}

\vspace{5mm}
\end{frontmatter}




\section{Introduction} \label{sec:introduction}

%
%

Solid dynamics is naturally formulated in Lagrangian coordinates. However, 
the 
treatment of excessively large (finite) deformations in the Lagrangian frame is 
challenging from the computational viewpoint because of the large mesh 
distortion. Hydrodynamic-type effects like jets and  
vortexes may appear in many applications involving interactions of solids, 
e.g. high energetic interactions of metals involving fluidization, 
melting and solidification in metallurgy, complex flows 
such as granular flows and flows of viscoplastic fluids (yield stress fluids), which 
exhibit properties of both elastic solids and viscous 
fluids~\cite{Forterre2013,FrigaardReview2014}. Such hydrodynamic-type   
effects in solids are naturally treated in the 
Eulerian settings. It is therefore important to have also a proper Eulerian 
formulation of solid mechanics, since it is more suitable for treating large 
deformations.  
Moreover, an Eulerian formulation is also required for moving mesh  
techniques based on the Arbitrary Lagrangian Eulerian formulation, 
see ~\cite{BoscheriDumbser2016}, for example.  

In contrast to the Lagrangian frame, in which the description of the dynamics of elastic and plastic solids always relies on the use of strain measures (\eg deformation gradient) and therefore on a rigorous and objective geometric approach, there are historically 
two different possibilities to formulate the solid dynamics equations in the 
Eulerian frame: the \textit{hypoelastic} and the \textit{hyperelastic} one. In 
the hypoelastic-type models the stress tensor plays the role of the thermodynamic state variable 
along 
side with the mass, momentum and energy densities, and thus, an evolution equation for the stress 
tensor is employed (usually for its trace-less part). On the other hand, the hyperelastic Eulerian 
models, 
similar to their Lagrangian counterparts, rely on the use of a strain measure as the primary 
state variable. They are therefore also based on a rigorous geometric description of solid mechanics. The stress tensor in such models is considered merely as the constitutive momentum 
flux and should be computed from a stress-strain constitutive relation which is completely 
determined by defining the energy potential. The goal of this paper is to report about the results 
of a detailed analytical and numerical comparison of two Eulerian models for solid dynamics in the context of nonlinear elastoplastic deformations. In particular, we compare the hypoelastic Wilkins model \cite{Wilkins64,Maire2013,KulikPogorSemen,Gavrilyuk2008} against the 
 hyperelastic model proposed by Godunov and Romenski in~\cite{GodRom1972,God1978,GodRom2003}
 and by Peshkov and Romenski in \cite{HPR2016}. Throughout this paper, we will therefore refer to 
 the hyperelastic model as the \GPR model.  

In this study, we ignore the effects of strain hardening and heat 
conduction, in order to study only the principal parts of the models, \ie the evolution of elastic 
stress and strains and the impact of the terms modeling the inelastic deformations. We also note 
that the \GPR model belongs to the class of so-called rate dependent plasticity models,  
while the Wilkins model is formulated in the class of ideal (rate-independent) models. 
Nevertheless, we shall not modify these models in order to make them both either rate-dependent or 
rate-independent, but we will compare them in the forms as they are mainly used in the literature. 
We recall that both models, in the forms they are used in this paper, were designed for similar 
purposes, that is to describe the behavior of metals under high strain-rate loadings. 
To make such a comparison more informative, we shall run the \GPR model with  
parameters that make its rate-dependent properties less evident.

%
%
The structure of the paper is as follows. In Section~\ref{sec.preliminaries}, we discuss the 
conceptual differences between the hypoelastic and the hyperelastic approach. In 
Section~\ref{sec.PDEs}, we elaborate our comparative analysis of the two different models
by discussing in detail the differences between their governing partial differential 
equations. We also derive a hypoelastic-type model that corresponds to the hyperelastic 
\GPR model for a particular equation of state used in this paper. The high order 
ADER Discontinuous Galerkin (ADER-DG) and finite volume (ADER-FV) schemes employed in 
this paper are briefly presented in Section~\ref{sec:HOscheme}. In 
Section~\ref{sec:numerics}, we provide the results of the numerical comparison of the approaches on 
a large test case suite. Finally, we conclude the paper by summarizing the obtained 
results and discuss further perspectives in Section~\ref{sec:conclusion}.

\section{Conceptual differences between hyperelastic and hypoelastic models}\label{sec.preliminaries}

\subsection{Lagrangian viewpoint}

The Lagrangian viewpoint on the motion of a continuum implies the use of two coordinate systems. 
The one which labels the material elements, called the Lagrangian coordinate system, will be 
denoted by $ \xi_a $, $ a=1,2,3$. With respect to the Lagrangian system $ \xi_a $, the medium is 
always 
at rest. The second system of coordinates is a fixed (laboratory) coordinate system, with respect 
to which the basic characteristics of motion, such as the velocity, displacement, etc. are measured. This fixed coordinate system is called Eulerian coordinate system and will be denoted as $ x_i $, $ i=1,2,3 $. 
The 
Lagrangian and 
Eulerian coordinates relate to each other in a \textit{one-to-one manner} by 
the laws of motion, e.g.~\cite{Sedov1965},
\begin{equation}\label{laws.motion}
x_i = x_i(t,\xi_a), \qquad \xi_a = \xi_a(t,x_i),
\end{equation}
where $ t $ is the time.

In the Lagrangian description, the deformation gradient $ \F{i}{a} = \frac{\pd x_i}{\pd \xi_a} $, 
which contains the full information about the deformation and orientation of the material elements, 
is used as a primary state 
variable\footnote{We use on purpose different letters for indexes related to the 
Lagrangian reference frame ($1 \leq a,b,\ldots \leq 3 $) and to the current Eulerian frame ($ 1\leq  
i,j,\ldots \leq 3 $). Recall that the deformation gradient is not a second order tensor, but it  transforms as a vector under coordinate transformations. Such objects are also called two-point  
second order tensors. The summation over different indexes in $ \F{i}{a}\F{i}{b} $ and 
$ \F{i}{a}\F{j}{a} $ thus results in true tensors $ B_{ab} $ and $ B_{ij} $, respectively, which, 
in 
general, are defined on different spaces.}. The governing equations of motion of an arbitrary 
continuum (either fluid or solid) can be derived from the Hamilton principle of stationary 
action~\cite{DPRZ2017,SHTC-GENERIC-CMAT,GRGPR} and read as

\begin{equation}\label{lagr.PDE}
\frac{\rmd m_i}{\rmd \, t} - \frac{\pd U_{\F{i}{a}}}{\pd \xi_a} = 0, \qquad \frac{\rmd 
\F{i}{a}}{\rmd 
\, t} - \frac{\pd U_{m_i}}{\pd \xi_a} = 0,
\end{equation}
where $ \rmd/\rmd t = \pd/\pd t + v_k \pd/\pd x_k$ is the material time derivative, $ m_i = \rho_0 
v_i $ 
is the momentum density, $ \rho_0 $ is the reference mass density, $ v_i = \frac{\rmd x_i}{\rmdt}$ 
is the 
velocity field, $ 
\F{i}{a} = \frac{\pd x_i}{\pd \xi_a}$ is the deformation gradient, $ U = U(m_i,F_{ia}) $ is the 
total energy 
density (including the kinetic energy) of the system, and $ U_{m_i} = \pd U/\pd m_i $, $ U_{F_{ia}} 
= \pd U/\pd F_{ia} $. It is implied that $ U_{m_i} = v_i $. The energy conservation law 
\begin{equation}\label{lagr.energy.cons}
\frac{\rmd U}{\rmd\, t } - \frac{\pd }{\pd \xi_a} \left( U_{m_i} U_{\F{i}{a}}\right)  = 0
\end{equation}
is the consequence of equations~\eqref{lagr.PDE}, i.e. it can be obtained if \eqref{lagr.PDE}$ _1 $ 
is multiplied by $ U_{m_i} $ and added to \eqref{lagr.PDE}$ _2 $ multiplied by $ U_{\F{i}{a}} $.
Because of the isotropy assumption and the material frame indifference principle, the energy 
potential $ U $ may depend on the deformation gradient only via three invariants of a symmetric 
strain tensor obtained from $ \F{i}{a} $, e.g. $ B_{ab} = \F{a}{i}\F{i}{b} $, $ a,b=1,2,3 $ or $ 
\BB = 
\FF^\transpose\FF$. Thus, it is implied that 
\begin{equation}
U(\mm,\FF) = \tilde{U}(\mm,\BB).
\end{equation}

Note that from the definition of the deformation 
gradient, it follows that the identity 
\begin{equation}
\frac{\pd \F{i}{a}}{\pd \xi_b} - \frac{\pd \F{i}{b}}{\pd \xi_a} = 0,
\end{equation}
is fulfilled by the solutions of \eqref{lagr.PDE}. 

We note that in the momentum equation \eqref{lagr.PDE}$ _1 $, the non-symmetric stress tensor $ 
U_{F_{ia}} $ (the first Piola-Kirchoff stress tensor) is 
not a state variable but is completely determined by the strains $ F_{ia} $ and the specification 
of the energy potential $ U(m_i,F_{ia}) $. 
In such a formulation, the stress-free equilibrium configuration corresponds to
\begin{equation}\label{stress.free}
U_{\F{i}{a}} = 0
\end{equation}
and is achieved when $ \F{a}{i}\F{i}{b} = \delta_{ab} $, \ie when $ \F{i}{a} $ is an orthogonal 
matrix, or in other words when the length of the line elements $ \rmd x_i $ and $ 
\rmd \xi_a $ in the current and the reference configurations are equal.
This rigorous way of formulating the governing equations in terms of objective geometric quantities is 
referred to as  \textit{hyperelastic} in the solid dynamics community. 

Traditionally, instead of the first-order equations~\eqref{lagr.PDE} and \eqref{lagr.energy.cons}, the 
well-known second order formulation of solid dynamics with the displacement 
field as primary variable is typically discretized in computational codes based on the finite element method (FEM). However, in the last decades also the first-order formulation \eqref{lagr.PDE}, \eqref{lagr.energy.cons} gained some popularity in the context of transient dynamics of solids with the application of  
finite volume discretizations based on Godunov-type   
methods~\cite{Trangenstein1992,Peshkov2009,GodPesh2010,Kluth2010,Gil2013,Gil2017a}.

We now discuss how the Lagrangian equations 
\eqref{lagr.PDE} should be generalized in order to take into account also irreversible deformations.

\subsection{Inelastic media in the Lagrangian frame}\label{sec.plast.Lagr}

The key feature of inelastic media is their inability to recover the initial state. 
Such an irreversibility of deformations is due to the microscopic structural changes in the medium. 
The structural changes mean that the material elements 
(parcels of molecules) that were attached to each other in space may become 
disconnected after the irreversible process of material element rearrangements, which is, in fact,
the essence of any flow. An obvious consequence of the material element rearrangements is that the 
real stresses in the medium and the 
microscopic deformations of material elements always remain finite even though the observable 
macroscopic deformations 
encoded in the laws of motion \eqref{laws.motion} and hence, in the deformation gradient $ \F{i}{a} 
$, may grow unlimitedly (e.g. in simple shear flow). Another consequence is that the inelastic 
media might be highly inhomogeneous because at each time instant, in the zone of inelastic 
deformations, the neighboring material elements might be not connected in the previous time 
instants and hence may have a history of deformations that is completely independent from each 
other. The main question is therefore how to describe these facts on the mathematical level.  

These ideas can be mathematically expressed as follows. First of all, a \textit{local unstressed 
reference frame} is postulated to exist for each material element via a thought 
experiment~\cite{Eckart1948,God1978,GodRom2003,PlohrSharp1992plasticity}. Namely, if an 
infinitesimal 
volume is instantaneously cut out of the material and left to relax adiabatically, then it relaxes 
to a stress-free configuration. Because the material that suffered from inelastic deformations can 
be highly inhomogeneous, such an unstressed state can not be achieved simultaneously (globally) for 
all  material elements. In other words, the unstressed state is a purely local notion. 
Therefore, in the local unstressed state, the tangent space at every point of the continuum is 
assumed to be equipped with an orthonormal basis triad $ \bm{\delta}_{\mu} $, $ \mu=1,2,3, $
which, in the current deformed state, becomes the triad $ \Feffvec{\mu} $ (the superscript 'e' 
reference to 'elastic' and will be explained later). Mathematically speaking, 
the linear map $  \bm{\delta}_{\mu} \rightarrow \Feffvec{\mu}$, whose components are 
denoted by $ \Feff{i}{\mu} $, $ \det(\Feff{i}{\mu}) > 0$, defines a local frame in each material 
point. Eventually, it is assumed that the stresses in the medium are assigned to this local 
\textit{elastic} (recoverable) deformation of the material elements $ \Feff{i}{\mu} $. Hence, it is 
natural to assume that the energy $ U $ is not a function of $ \F{i}{a} $ but of $ 
\Feff{i}{\mu} $. In addition, because any two invertible matrices $ \F{i}{a} $ and $ \Feff{i}{\mu} 
$ can be related as
\begin{equation}\label{F.eq.EP}
\F{i}{a} = \Feff{i}{\mu} \Fplast{\mu}{a}
\end{equation}
for a certain matrix $ \Fplast{\mu}{a} $, $ \det(\Fplast{\mu}{a}) > 0$, we assume that 
\begin{equation}
\tilde{U}(\FF^{\rm e}) = \tilde{U}(\FF \FF^{\rm -p}) = U(\FF,\FF^{\rm p}),
\end{equation}
where $ \FF^{\rm - p} $ is the inverse to $ \FF^{\rm p} $, \ie $ \delta_{ab} = 
\Fplast{a}{\mu}\Fplastinv{\mu}{b} $, and hence, the Piola-Kirchhoff stress $ U_{\F{i}{a}} $ becomes
\begin{equation}\label{stress.plast}
U_{\F{i}{a}} = \tilde{U}_{\Feff{i}{\mu}} \Fplastinv{a}{\mu},\qquad \text{or} \qquad U_{\FF} = 
\tilde{U}_{\FF^{\rm e}}(\FF^{-\rm p})^\transpose.
\end{equation}
Lee and Liu~\cite{Lee1967} were among 
the first to introduce such a decomposition of the deformation gradient $ \F{i}{a} $ into 
a recoverable (elastic) $ \Feff{\mu}{a} $ and irrecoverable (plastic) $ \Fplast{\mu}{a} $ part.  
The formula above states a fundamental fact about the Lagrangian description of inelastic deformations, 
that is one has to know at least two matrices in \eqref{F.eq.EP} in order to compute 
the Lagrangian stress tensor \eqref{stress.plast}. This imposes severe limitations on using a pure 
Lagrangian approach for modeling very large plastic deformations or fluid flows because, even 
though the elastic part $ \Feff{i}{\mu} $ remains finite,  the total deformation gradient $ 
\F{i}{a} $ and its plastic part $ \Fplast{\mu}{a} $ 
potentially may grow unlimitedly and thus, it can be a source of numerical problems and errors. As we 
shall see later, the situation is quite different in the Eulerian frame. Namely, to compute the 
Eulerian stress (Cauchy stress), one needs to know only $ \FF^{\rm e} $, which is always finite.

The full Lagrangian system of governing equations for modeling inelastic deformations can be 
formulated as follows~\cite{GodPesh2010,Kondaurov1982}. It is convenient to chose the vector of 
state 
variables as
$ (\mm,\FF,\FF^{\rm p},s) $, where $ s $ is the specific entropy. The governing equations are

\begin{equation}\label{lagr.PDE.plast}
\frac{\rmd m_i}{\rmd \, t} - \frac{\pd U_{\F{i}{a}}}{\pd \xi_a} = 0, \quad \frac{\rmd 
\F{i}{a}}{\rmd 
\, t} - \frac{\pd U_{m_i}}{\pd \xi_a} = 0, \quad \frac{\rmd \Fplast{\mu}{a}}{\rmd \, t} = 
-\frac{1}{\theta} U_{\Fplast{\mu}{a}}, \quad \frac{\rmd s}{\rmd t} = \frac{1}{\theta 
U_s}U_{\Fplast{\mu}{a}}U_{\Fplast{\mu}{a}} \geq 0,
\end{equation}
where $ \theta = \theta(\FF,\FF^{\rm p},s) = \tilde{\theta}(\FF^{\rm e},s) > 0 $ is a parameter 
characterizing the rate of strain relaxation, which usually is taken as $ \theta \sim \rho_0 \tau 
c_s^2 $, where $ \rho_0 $ is the initial density, $ c_s $ is the shear sound speed, and $ \tau $ is 
the characteristic time of strain relaxation and will be discussed later. The energy potential 
should be defined as $ 
U(\mm,\FF,\FF^{\rm p},s) = \tilde{U}(\mm,\FF^{\rm 
e},s)$, where $ \tilde{U} $ may depend on $ \FF^{\rm e} $ only via its invariants. The energy 
conservation law has the same form as in \eqref{lagr.energy.cons} and still is the consequence of 
the governing equations~\eqref{lagr.PDE.plast}. It can be obtained as the linear combination of 
\eqref{lagr.PDE.plast} with the coefficients $ U_{m_i} $, $ U_{\F{i}{a}} $, $ U_{\Fplast{\mu}{a}} 
$, and $ U_s $ for \eqref{lagr.PDE.plast}$ _1 $, \eqref{lagr.PDE.plast}$ _2 $, 
\eqref{lagr.PDE.plast}$ _3 $, and \eqref{lagr.PDE.plast}$ _4 $ respectively. A Godunov-type 
numerical method for equations \eqref{lagr.PDE.plast} in the two-dimensional case was constructed 
in~\cite{GodPesh2010}. 

Given the above considerations, the Lagrangian description of solid mechanics is intrinsically of the 
\textit{hyperelastic} type. On the contrary, in the Eulerian framework, we do not 
have the information about the initial configuration of the continuum and, in particular, about the 
fields of labels $ \xi_a $. Nevertheless, the Eulerian evolution equations for the deformation 
gradient $ F_{ia} $ (or its inverse) and its parts $ \Feff{i}{\mu} $ and $ \Fplast{\mu}{a} $ can 
be easily obtained after the change of variables $ \xi_a \rightarrow x_i $, see e.g.  
 \cite{GodRom2003,SHTC-GENERIC-CMAT}. However, the history of Eulerian solid dynamics took a 
different route, in which the strain measures were disregarded and the stress tensor was promoted 
to an independent state variable. This route is the so-called \textit{hypoelastic} description 
and is discussed in the following section. 
Such a choice was motivated by the main interest of the solid dynamics community to formulate 
a solid dynamics theory capable of dealing with \textit{arbitrary} finite and even 
fluid-like irreversible deformations like jets and/or vortexes. Intuitively, one may think that 
in such cases, any strain-measure-based approach suffers from serious limitations. 
However, as recently shown in~\cite{HPR2016,DPRZ2016,DPRZ2017,HYP2016}, 
arbitrary motion of a continuum, either solid or fluid, can be successfully described also in 
the hyperelastic framework, i.e. in a strain-based theory, even for long times and for media 
undergoing large deformations. This will be discussed later in Section~\ref{sec.intro.hyper.euler}.

\subsection{Hypo-elastic solid dynamics}\label{sec.intro.hypo}

The hypoelastic approach to solid dynamics was the subject of active research in the 1950ies. Many 
authors contributed to its development, including Oldroyd~\cite{Oldroyd1947,Oldroyd1950}, Noll 
\cite{Noll1955}, Truesdell \cite{Truesdell1955} and Green \cite{Green1956}, to name just a few.
The theory of hypoelasticity aims to generalize the linear elasticity theory to general flows 
and large deformations by means of providing a general constitutive relation that is formulated 
as a time evolution equation relating the rate of stress to the rate of strains. A quite general 
evolution equation for the stress tensor $ \SSS $ can be 
found in~\cite{Oldroyd1950,Green1956,Noll1955}, which 
constitutes the basis for various hypoelastic models:
\begin{multline}\label{stress.rate.gen}
\frac{\rmD \SSS}{\rmD t} = \alpha_1 \tr (\asymvel) \Id + \alpha_2 \SSS + \alpha_3
\tr(\asymvel) \SSS + \alpha_4 \tr(\SSS \asymvel)\Id + \frac{1}{2}\alpha_5 (\SSS \asymvel + 
\asymvel \SSS) + \alpha_6 \tr(\asymvel) \SSS^2 + \alpha_7 \tr(\SSS\asymvel)\SSS+ \\
\alpha_8 \tr(\SSS^2\asymvel)\Id + \frac{1}{2}\alpha_9 (\SSS\asymvel + \asymvel \SSS^2) + 
\alpha_{10} \tr(\SSS \asymvel) \SSS^2 + \alpha_{11} \tr(\SSS^2 \asymvel)\SSS + 
\alpha_{12}\tr(\SSS^2 \asymvel)\SSS^2.
\end{multline}
Here, $\alpha_1, \, \alpha_2, \, \ldots, \, \alpha_{12} $ are regular functions of the 
three invariants of $ 
\SSS $, density, and temperature, $ \asymvel = (\LL - \LL^\transpose)/2$, $ \LL = \nabla \vv $, and 
$ \rmD/\rmD t $ is an 
objective time derivative. One may immediately notice that the PDE \eqref{stress.rate.gen} 
represents 
a rich source of uncertainties. First of all, the choice of the objective time 
derivative is fairly arbitrary and not unique, as discussed in \cite{Meyers2006}, for example. 
It is neither based on a firm physical ground nor on rigorous principles of differential 
geometry. Moreover, the discretization of these objective stress rates must be performed 
with great care~\cite{Maire2013} to ensure that the constitutive law at the
discrete level still satisfies the principle of material objectivity. This important property 
refers to as the incremental objectivity.   Let us also note that an objective stress rate such as 
the Jaumann rate leads to an evolution equation for the stress
which cannot be written under conservative form for multi-dimensional flows. This flaw renders the 
mathematical analysis
of discontinuous solutions questionable, as noticed in \cite{GavrFavr2008}.

The second source of uncertainties is the 
choice of constitutive functions $ \alpha_i $, $ i=1,2,\ldots,12 $ which also cannot be made based 
on 
firm physical principles. The functions $ \alpha_i $ should be specified by means of fitting 
experimental data. This fitting has to be done in a twelve-dimensional functional space which 
most likely makes the definition of $ \alpha_i $ not unique. In practice, as in particular in the Wilkins model considered here, only a few functions (one or two) are considered and usually they 
are assumed to be constant. However, for large elastic deformations the $ \alpha_i $ have to depend 
on the solution, as reported in \cite{Kim2016}.

We note that the hypoelastic-type models, which usually are obtained from \eqref{stress.rate.gen} by 
coupling with proper dissipative terms (usually of the relaxation type), have gained a lot of popularity in  
non-Newtonian fluid dynamics for modeling viscoelastic and viscoplastic fluids (yield stress 
fluids), see \eg \cite{Oldroyd1947,Oldroyd1950,Putz2009,Saramito2009,FrigaardReview2014}, in flows of 
granular media \cite{Forterre2013,Kamrin2015}, and in 
non-equilibrium relativistic~\cite{Muller1967,Israel1976,Stewart1977,MullerRuggeri1986} and 
non-relativistic \cite{MullerRuggeri1998,EIT2010,Torrilhon2016} gas dynamics, where the stress 
tensor evolution equation is obtained from the 
Boltzmann equation by means of the method of moments 
(such models generalize Maxwell's idea on modeling the viscoelasticity of 
gases~\cite{Maxwell1867,Noll1955}).

We discuss well-posedness and thermodynamics issues of the hypoelastic approach to continuum 
modeling of flowing media in Section~\ref{sec.thermo} and \ref{sec.well-posed}. 
However, we want to emphasize that, in spite of such drawbacks, hypoelastic 
models are able to reproduce many experimental observations where the data take the form of 
measurements of changes in stress with respect to changes in strain. That is why they are 
extensively used by many researchers and engineers and are currently used in many commercial 
computational codes such as LS-DYNA\footnote{\url{http://www.lstc.com/products/ls-dyna}}.


\subsection{Hyperelastic Eulerian solid and fluid dynamics}\label{sec.intro.hyper.euler}

In contrast to the aforementioned hypoelastic formulations, hyperelastic models rely on the direct   
evolution of a strain measure and thus on a geometric approach. Although the use of a 
strain-measure might seem to be counter-intuitive for the description of large deformation continuum mechanics (intense plastic deformations, viscous and inviscid fluid flows, etc.), as it was mentioned above, a 
hyperelastic GPR formulation of fluid and solid dynamics can nevertheless be also applied to 
model arbitrary fluid flows and solids undergoing large deformations. In what follows, we 
review the history of Eulerian hyperelastic-type models, not exhaustively though, with applications 
to inelastic deformations of solids and to fluid flows and we discuss the main features of such 
models. 

The main obstacle of the Lagrangian description of arbitrarily large inelastic deformations  
is the necessity to use two strain measures for computing the Lagrangian stress tensor $ 
U_{\F{i}{\mu}} = \tilde{U}_{\Feff{i}{\mu}} \Fplastinv{a}{\mu} $ (the first Piola-Kirchhoff stress), 
as discussed in Section~\ref{sec.plast.Lagr}. Thus, the key difference between the Eulerian and the 
Lagrangian formulations of continuum mechanics, which makes the practical use of Eulerian 
hyperelastic formulations possible for arbitrary flows, is that the Eulerian 
description, in principle, relies on the use of only one strain measure, namely the elastic strain $ 
\Feff{\mu}{i} $. This is sufficient to compute the Eulerian stress tensor (\ie the Cauchy 
stress), as will be clarified in what follows.

Perhaps, Eckart~\cite{Eckart1948} was the first to propose the idea of using an elastic 
(recoverable) strain measure in 
Eulerian inelasticity. In particular, he introduced the notion of the local relaxed state (exactly 
as discussed in Section~\ref{sec.plast.Lagr}) and suggested to characterize the deviation from this 
state by the non-Euclidean (\ie with non-vanishing curvature) metric tensor $ g_{ij} $ with the 
evolution equation 
\begin{equation}\label{Eckart.gij}
\frac{\rmd g_{ij}}{\rmd \, t} + g_{ik}\frac{\pd v_k}{\pd x_j} + g_{kj}\frac{\pd v_k}{\pd x_i} = - 
M_{ijkl}E_{g_{kl}},
\end{equation}
where $ E $ is the specific energy of the system, $ E_{g_{kl}} = \pd E/\pd g_{kl} $, and $ 
M_{ijkl}E_{g_{ij}}E_{g_{kl}} $ is a positive definite quadratic form which was not specified in 
\cite{Eckart1948}. The right-hand side represents the rate of change of the metric tensor due to the
microscopic process of structural relaxation and not due to the macroscopic motion of the 
continuum. Equation  \eqref{Eckart.gij} was derived based on the first and second laws of 
thermodynamics. In the same thermodynamics spirit, the concept of the local relaxed state was later 
discussed by Sedov in \cite{Sedov1965}, who also suggested to use the metric tensor as a 
thermodynamic state variable. The next important contribution to the Eulerian description of 
nonlinear inelastic deformation was made by Besseling \cite{Besseling1968}, see also the 
book~\cite{Besseling1994}. In contrast to Eckart 
and Sedov, Besseling suggested to use not the metric tensor $ g_{ij} $ but a non symmetric 
strain $ b_{\mu i} $ (in notations of \cite{Besseling1968}) that is related to the metric tensor 
as $ g_{ij} = b_{\mu i} b_{\mu j} $ and is defined as a transformation of the line elements $ \rmd 
x_i $ in the current deformed state and $ \rmd a_{\mu} $ in the local relaxed 
reference state, \ie $ \rmd a_{\mu} = b_{\mu i } \rmd x_i $. As it is clear now, Besseling's $ 
b_{\mu i} $ is exactly the elastic distortion field in the \GPR model proposed by Godunov and 
Romenski 
\cite{GodRom1972,God1978,Romenski1979,Rom1984,GodRom2003} and Peshkov and Romenski~\cite{HPR2016} 
and further 
denoted by $ \dist {\mu}{i} $, as in our previous papers~\cite{HPR2016,DPRZ2016,DPRZ2017,BoscheriDumbser2016},  
and which is governed by the evolution equation
\begin{equation}\label{intro.dist.PDE}
\frac{\rmd \dist{\mu}{i} }{\rmd \, t} + \dist{\mu}{j} \frac{\pd v_j}{\pd x_i} = 
-\frac{1}{\theta(\tau)}E_{\dist{\mu}{i}},
\end{equation}
which can be recast into the form
\begin{equation}\label{intro.dist.PDE2}
\frac{\pd \dist{\mu}{i}}{\pd t} + \frac{\pd (\dist{\mu}{j}	v_j)}{\pd x_i} + v_j\left(\frac{\pd 
	\dist{\mu}{i}}{\pd 	x_j} - \frac{\pd \dist{\mu}{j}}{\pd x_i}\right)
	=-\dfrac{ E_{\dist{\mu}{i}} }{\theta(\tau)},
\end{equation}
where $ E_{\dist{\mu}{i}} = \pd E/\pd \dist{\mu}{i} $, and $ \theta(\tau) > 0$ is a positive 
function of the strain relaxation time $ \tau $ and $ E $ 
is the total energy potential.  
Originally, Godunov and Romenski presented their 
model~\cite{GodRom1972} also in terms of the metric tensor $ g_{ij} $ (effective 
metric tensor), while later \cite{God1978,Romenski1979}, they started to use the elastic distortion 
$ \dist{\mu}{i} $
as the primary state variable. One of the main motivations for such a model was the deformation of 
metals at high strain rates, which may exhibit hydrodynamic effects in the case of welding, see e.g.~\cite{Godunov1970welding}. Another  
important motivation was to obtain a mathematically well-posed system of equations that can be 
solved numerically. In particular, they require that the system of governing equations is
hyperbolic, and if possible even \textit{symmetric hyperbolic}. 
As it has become clear soon, the use of the metric tensor $ g_{ij} $ 
as the elastic strain measure of deviation from the local relaxed state does not lead to a  
symmetric hyperbolic model. On the other hand, the use of the elastic distortion does allow to 
symmetrize the model, see \cite{Rom1984,GodRom1995,GodRom1996,GodRom2003}. We also note the works by 
Leonov~\cite{Leonov1976,Leonov1987}, who, similar to Eckart \cite{Eckart1948}, Sedov 
\cite{Sedov1965} and Godunov and Romenski \cite{GodRom1972}, proposed a relaxation model in the 
context of 
non-Newtonian polymeric fluids that also employs only the metric tensor $ g_{ij} $  
as elastic strain measure and as primary thermodynamic state variable, and does not use 
any other additional total or plastic strain measures.  

The dissipative effect due to the inelastic deformations in the \GPR formulation  was 
introduced as a relaxation source term (right hand side in \eqref{intro.dist.PDE}) in the evolution 
equation for the elastic distortion (see further
details in Section~\ref{sec.hyper.present}). Such a 
treatment of inelastic deformations attributes the \GPR model to rate-dependent 
plasticity models with the relaxation parameter $ \tau $ being the characteristic time of 
relaxation of tangential strains. Interpolation formulas for $ \tau $ for metals at high 
strain-rates were studied in \cite{Godunov1976a,Godunov1976b}. Those take into account also the 
temperature effect and melting. In this paper, we use a simplified version of the dependence of $ 
\tau $ on the state variables, which does not take into account the temperature effect. The shock 
structure in a relaxed medium modeled with the \GPR model was studied in 
\cite{Godunov1976,God1978,GodRom2003}. In the 1980ies, the 
\GPR 
model was intensively studied numerically by means of a Godunov-type method in works of Merzhievsky 
and Resnyansky 
~\cite{MerzhResnyan1985,MerzhResnyan1985a,MerzhResnyan1987,MerzhResnyan1992,MerzhResnyan1995}, 
including one-dimensional and two-dimensional simulations of high-velocity impacts of metals with a 
moving mesh technique. The linearized version of the \GPR model was also extended to 
modeling of anisotropic composite viscoelastic media in papers by Resnyanski, Romenski and 
co-authors, \eg ~\cite{ResnyanRom1992,RomResnyan1993}. The question of hyperbolicity of the 
\GPR formulation for Eulerian non-linear elasticity was studied 
in~\cite{Romenskii1984,GodRom1995,GodRom1996,GodRom1996a,GodRom2003,GodRom2003}. Relations between 
the hyperelastic and hypoelastic formulations for Eulerian non-linear elasticity was 
investigated by Romenski in~\cite{Rom1973}.

Independently of the aforementioned references, an Eulerian approach to finite-strain inelasticity 
was also developed by Plohr and Sharp in \cite{PlohrSharp1988,PlohrSharp1992plasticity} in 
the 1990ies and was implemented in computational codes based on Godunov-type finite volume 
methods in~\cite{Trangenstein1991,Plohr1993,MillerColella2001}. In contrast to the \GPR 
formulation of Eulerian inelasticity, which is based on the use of only one strain 
measure, namely the elastic distortion field, the Plohr-Sharp formulation employs two strains, 
the total deformation gradient $ \F{i}{a} $ and the plastic strain $ \Fplast{\mu}{a} $, 
which measures the change of the 
line elements with respect to the original undeformed state. Such an approach however eliminates the 
advantage of the Eulerian framework of describing very large inelastic deformations and fluid-like 
effects, because it essentially represents the Lagrangian equations~\eqref{lagr.PDE.plast} directly 
written in Eulerian coordinates. 
Moreover, although plastic strain $ \Fplast{\mu}{a} $ is a legitimate mathematical quantity, it 
bears no physical relevance. Thus, a plastically processed material should not remember its initial 
shape, e.g. see the discussion on p. 249 in \cite{Besseling1994}. Also, such an approach 
cannot be applied to plastic flows of solid-like materials, \eg viscoplastic fluids, granular 
media, while the \GPR formulations has no such restrictions and can be applied to 
arbitrary inelastic deformations, including flows of viscous fluids, and in particular Newtonian 
flows~\cite{HPR2016,DPRZ2016,HYP2016}.

Because of the unified character of the \GPR model to describe solids and fluids, it 
attracts certain  attention in the last decade for the use in Eulerian interface tracking 
computational techniques in compressible multi-material simulations. 
Thus, the model 
was incorporated into the 
diffuse interface approach by Gavrilyuk, Favrie et al 
\cite{Gavrilyuk2008,FavrGavr2012,Ndanou2015,Hank2017}, a level set 
method by Barton et al~\cite{Barton2013a,Barton2018} as well as by Gorsse, Iollo at 
al~\cite{Iollo2014} (in 
the elastic limit), an Arbitrary-Lagrangian-Eulerian (ALE)
technique by Boscheri et al in~\cite{BoscheriDumbser2016} and in material Riemann solvers by Michael and 
Nikiforakis \cite{Michael2018a}. 

The work hardening effect is not considered in this paper, but it can 
be incorporated in a standard phenomenological isotropic hardening manner via the evolution of a  
hardening scalar, or in a more sophisticated manner according to \cite{Barton2012}. Also, 
the impact of the dislocation dynamics can be taken by means of a direct evolution of the 
dislocation density tensor 
(Burgers tensor)~\cite{SHTC-GENERIC-CMAT}, which is the subject of future research.

An important extension of the applicability of the \GPR model was recently proposed 
in~\cite{HPR2016,DPRZ2016}. It was realized, that in fact this model can deal not only with 
inelastic deformations in solids, but it can be also applied to arbitrary flows of solids and fluids, 
including  
Newtonian fluids, provided that the dissipative terms (the right-hand side 
in~\eqref{intro.dist.PDE}) are properly defined. In this regard, it is important to emphasize an 
intrinsic rate dependent 
character of the model, which is represented by the relaxation source terms in the elastic 
distortion evolution equation. For example, the ideal plasticity law employed in other formulations
\cite{Trangenstein1991,Plohr1993,MillerColella2001} cannot be generalized to viscous fluid flows. 

One may also note that, in principle, the evolution equation
\begin{equation}\label{intro.PDE.Fe}
\frac{\pd \left (\rho \Feff{i}{\mu}\right )}{\pd t} + \frac{\pd}{\pd x_j}\left (\rho \Feff{i}{\mu} 
v_j - 
\rho 
\Feff{j}{\mu} v_i\right ) + v_i \frac{\pd \left (\rho \Feff{j}{\mu}\right )}{\pd x_j} = 
-\frac{E_{\Feff{i}{\mu}}}{\theta(\tau)},
\end{equation}
for the inverse elastic distortion, which is 
in fact the elastic strain $ \Feff{i}{\mu} = (\dist{\mu}{i})^{-1}$ introduced earlier in 
Section~\ref{sec.plast.Lagr}, can be used instead of the evolution equation \eqref{intro.dist.PDE2} 
for $ \dist{\mu}{i} $. This equation has been considered in 
\cite{God1978,GodRom2003,Kondaurov1981,Kondaurov1982,Barton2009,BartonRom2010}. One may expect that 
for smooth 
solutions and in the absence of inelastic deformations, equations \eqref{intro.dist.PDE2} and 
\eqref{intro.PDE.Fe} are equivalent. Their 
equivalence for weak solution is an open question. Also, in the presence of inelastic deformations 
the 
equivalence of these equations has not been established yet. Most likely, they are not equivalent 
because the measure of the deformation compatibility for the elastic distortion $ 
\dist{\mu}{i} $ is the Burgers tensor $ B_{\mu ij} = \frac{\pd \dist{\mu}{j}}{\pd x_i} 
- \frac{\pd 
\dist{\mu}{i}}{\pd x_j}$, which explicitly emerges in the time evolution~\eqref{intro.dist.PDE2},  
while for $ \Feff{i}{\mu} $ it is the vector 
\begin{equation}\label{burg.vector}
b_{\mu} = \frac{\pd \left (\rho \Feff{i}{\mu}\right )}{\pd x_i} = \rho 
\Feff{j}{\mu}\Feff{k}{\nu}B_{\nu jk}, 
\end{equation}
which means that it may happen that for a non-zero Burgers tensor $ B_{\mu ij} \neq 0  $ (\ie the 
deformation is inelastic in terms of $ \dist{\mu}{i} $), vector $ b_\mu $ may vanish (the 
deformation is elastic in terms of $ \Feff{i}{\mu} $). 

An important remark should be made here about the form of the time 
evolution equations \eqref{intro.dist.PDE2} and \eqref{intro.PDE.Fe} in the elastic limit, 
where one has $ \F{i}{a} = \Feff{i}{a} $. As a consequence, the distortion field $ \dist{\mu}{i} $ 
and the elastic strain $ \Feff{i}{\mu} $ represent compatible deformations, meaning that 
$ \frac{\pd \dist{\mu}{j}}{\pd 
x_i} 
- \frac{\pd 
\dist{\mu}{i}}{\pd x_j} \equiv 0$ and $ \frac{\pd \rho \Feff{i}{\mu}}{\pd x_i} \equiv 0 $, and 
hence one may think that these terms, which explicitly emerge in equations~\eqref{intro.dist.PDE2} 
and \eqref{intro.PDE.Fe} respectively, can be dropped out. This, however, is not recommended, or at 
least should be done with a great care, because these terms are parts of the structure of the 
equations, \eg see the discussion in~\cite{SHTC-GENERIC-CMAT}. In particular, by ignoring these 
terms, equations \eqref{intro.dist.PDE2} 
and \eqref{intro.PDE.Fe} are not Galilean invariant anymore and have non-physical characteristics 
that are not co-moving with the media~\cite{Trangenstein1991,MillerColella2001,Barton2009}. The 
situation is  identical to the 
magnetohydrodynamics (MHD) equations, as discussed in~\cite{Powell1999} and which was already 
recognized by Godunov in \cite{God1972MHD}.

Eventually, we emphasize an important theoretical property of the \GPR formulation. 
As was recently demonstrated~\cite{SHTC-GENERIC-CMAT}, the non-dissipative part of the time 
evolution (i.e. all the differential terms) of such this formulation admit a fully Hamiltonian 
formulation of continuum mechanics. Namely, it can be derived from the Hamilton principle of 
stationary action and it further admits a representation via Poisson brackets. The latter means 
that this model is fully compatible with the GENERIC (General Equation for Non-Equilibrium 
Reversible-Irreversible Coupling) formulation of non-equilibrium 
thermodynamics~\cite{GrmelaOttingerI,GrmelaOttingerII} and hence, potentially a link between the 
\GPR unified 
formulation of continuum mechanics \cite{HPR2016,DPRZ2016,DPRZ2017,HYP2016,GRGPR} and the 
fundamental equation of statistical physics, the Liouville equation, can be 
established. One may notice that both theories, the microscopic one represented by the Liouville 
equation and the macroscopic one represented by the \GPR equations 
\cite{HPR2016,DPRZ2016,DPRZ2017,HYP2016,GRGPR}, are applicable to all three states 
of matter, gaseous, liquid and solid. Also, note that the Hamiltonian formulation allows a  
genuinely nonlinear coupling between different physical processes, \eg transfer processes, multi-phase 
formulations, coupling with electromagnetic fields~\cite{DPRZ2017}, and even allows an extension to 
the general relativistic case \cite{GRGPR}.

\section{Hypoelastic and hyperelastic models in the Eulerian frame}\label{sec.PDEs}

In this section we discuss the mathematical features of the hypoelastic-type model of Wilkins and the 
hyperelastic-type \GPR model. Next, we provide a direct comparison of the two models. This comparison  may infer some situations from which those two models differ.

%
%
\subsection{Hypoelastic model of Wilkins}

The hypoelastic Wilkins model is formulated in terms of the state variables $ (\rho, \varepsilon, 
\rho \vv, \SSS)$, \eg see \cite{Maire2013,KulikPogorSemen,Gavrilyuk2008} for a modern description, 
where $ \rho $ is the mass density, $ \varepsilon = E - \frac{1}{2}\vv^2$ is the 
specific internal 
energy, $ E $ is the specific total energy of the system, $ \vv $ is the velocity field, $ \SSS = 
[S_{ij}]$ is 
the deviatoric or trace-less part of the symmetric Cauchy stress tensor $ \bm{T} $
\begin{equation} \label{eq:T}
\bm{T} = -p \bm{I} + \SSS, \qquad \tr(\SSS) = 0,
\end{equation}
with $p = p(\rho,\varepsilon)$ being the hydrodynamic pressure and $\bm{I}$ is the identity tensor. 

The system of governing equations for the Wilkins model is the conventional mass, momentum, and 
energy conservation, supplemented with the time evolution equation for the deviatoric stress $ 
S_{ij} $
\begin{subequations}\label{eqn.Wilkins}
  \begin{align}
    &\frac{\pd \rho}{\pd t}+\frac{\pd (\rho v_k)}{\pd x_k}=0,   
    \label{eqn.W.conti}\\[2mm]
    &\frac{\pd (\rho v_i)}{\pd t}+\frac{\pd \left(\rho v_i v_k + p 
    \delta_{ik} - S_{ik} \right)}{\pd x_k}=0,  
      \label{eqn.W.momentum}\\[2mm]
    &\frac{\pd( \rho E)}{\pd t} + \frac{\pd}{\pd x_k}\left( v_k\rho E +  v_i(p\delta_{ik}  - 
    S_{ik})\right) = 0,  \label{eqn.W.energy}\\[2mm]
    & \frac{\rmD S_{ij} }{\rmD t} + \mu \left( \frac{\pd v_i}{\pd x_j} + \frac{\pd v_j}{\pd x_i} - 
    \frac{2}{3}\frac{\pd v_k}{\pd x_k}\delta_{ij}\right)   = 2\mu D^{\text{p}}_{ij}, \label{W.S}
  \end{align}
\end{subequations}
where we have employed Einstein's summation convention over repeated indices. The pressure $ p 
$ should be determined from a hydrodynamic equation of state $ p = \mathcal{P}(\rho,\varepsilon) $. 
The \textit{stiffened gas} or \textit{Mie-Gr{\"u}neisen} equations of state are used in 
this paper, see Section~\ref{sec.EOS.hydro}. The constitutive law for 
plasticity is formulated as a partial differential equation for the deviatoric stress $ S_{ij} $ 
and is discussed in what follows.


\paragraph{Constitutive law for plasticity}
The evolution equation for the deviatoric stress $\SSS$ is given by an incremental constitutive 
law which applies to elastic-perfectly-plastic materials \cite{Gurtin09}. In such a theory,
the strain rate tensor  $\bm{D}$, i.e. the symmetric part of the velocity gradient $\LL = \nabla 
\vv$,
\bea
\bm{D} = \frac12 ( \LL + \LL^{\transpose}).
\eea
admits the additive decomposition $\bm{D}=\bm{D}^{\text{e}}+\bm{D}^{\text{p}}$ between elastic and 
plastic strain rates.
For the plastic strain rate, it is assumed that
$\text{tr}(\bm{D}^{\text{p}}) = 0$ and also the dissipation inequality $\SSS:\bm{D}^{\text{p}} \geq 
0$ holds, 
where $:$ denotes the inner product of tensors, i.e $\bm{R}:\bm{Q}=\text{tr}( 
\bm{R}^{\transpose}\bm{Q} ) = R_{ij}Q_{ij}$ for any two arbitrary tensors
$\bm{R}$ and $\bm{Q}$.
Equipped with these notions, the incremental constitutive law for the deviatoric stress writes as
\begin{equation}\label{eqn.W.eqns.constitutive}
\frac{\rmD \SSS }{\rmD t} + 2\mu( \dev{\DD}-\DD^{\text{p}}) = 0, 
\end{equation}
where $\mu$ is the Lam{\'e} material dependent coefficient 
also called shear elastic module, and $\dev{\DD}$ denotes the deviatoric part of the strain
rate tensor
\begin{equation}\label{eqn.W.D0}
\dev{\DD} = \DD - \frac13 \text{tr}\,(\DD)\bm{I},
\end{equation}
and $\DD^{\text{p}}$ is given by
\begin{equation}\label{eqn.W.Dp}
\DD^{\text{p}} = \chi(f,\SSS) \left( \frac{\SSS}{||\SSS||}:\DD\right) 
\frac{\SSS}{||\SSS||}\,, \qquad ||\SSS|| = \sqrt{\tr \SSS^2}.
\end{equation}
The function $\chi$ is a switch function such that for a symmetric tensor $ \bm{\sigma} $
\begin{equation}\label{eqn.W.chi}
\chi(f,\bm{\sigma}) = \left\{
\begin{array}{ll}
 0,& f < 0, \\ [2mm]
 0,& f = 0\ {\rm and}\ \frac{\bar{\sigma}}{\sqrt{\text{tr}(\bm{\sigma}^2)}}\leq 0,\\ [2mm]
 1,& f = 0\ {\rm and}\ \frac{\bar{\sigma}}{\sqrt{\text{tr}(\bm{\sigma}^2)}} > 0,
\end{array} 
\right.
\end{equation}
where $ f = \sqrt{\dfrac{3}{2} \text{tr}(\bm{\sigma}^2)} - \sigma_Y $, $\sigma_Y$ is the yield 
strength which is a constant in the case of elastic perfectly plastic materials and 
$\bar{\sigma}=\text{tr} (\bm{\sigma}^\transpose \bm{D} )$. \\
In \eqref{eqn.W.eqns.constitutive} and \eqref{W.S}, $ \rmD/\rmD t $ is a so-called 
\textit{objective} time 
derivative. 
Though, many objective derivatives are known and a specific choice cannot be justified based on a 
physical ground. Typically, the Jaumann derivative 
\begin{equation}\label{eqn.W.Jaumann}
\frac{\rmD \SSS}{\rmD t} = \frac{\pd \SSS}{\pd t} + \vv\cdot\nabla\SSS + 
\SSS\asymvel - \asymvel\SSS,
\end{equation}
is widely employed in the Wilkins model, where the anti-symmetric part of the velocity gradient is
\begin{equation}\label{eqn.W.spin}
\asymvel = \frac12 \left( \nabla \vv - \nabla \vv^{\transpose} \right) .
\end{equation}
Let us notice that the case $f < 0$ corresponds to a behavior in the elastic range, whereas the case 
$f\geq 0$ corresponds to the
different behaviors depending on the sign of $\left( \frac{\SSS}{||\SSS||}:\DD\right)$.
In the case of negative $\left( \frac{\SSS}{||\SSS||}:\DD\right)$, the material deforms 
elastically, while 
for a strictly positive value, material deforms plastically. 


%
%
\subsection{\GPR hyperelastic-type model}\label{sec.hyper.present}

We now describe the hyperelastic-type \GPR model in which inelastic deformations are 
modeled via relaxation terms. This model is formulated in terms of mass density $ \rho $, momentum 
density $ \rho \vv $, total energy density $ \rho E $, and the elastic distortion field $ \AAA $. 
The governing equations are the mass, momentum, and energy conservation laws, which are 
supplemented with the evolution equation 
for the distortion field
\begin{subequations}\label{GPR.eqns}
	\begin{align}
	& \frac{\pd \rho}{\pd t}+\frac{\pd (\rho v_k)}{\pd 
	x_k}=0,\label{GPR.conti}\\[2mm]
	&\frac{\pd \rho v_i}{\pd t}+\frac{\pd 
	\left(\rho v_i v_k + p \delta_{ik} - \sigma_{ik} \right)}{\pd x_k}=0, 
	\label{GPR.momentum}\\[2mm]
	&\frac{\pd \rho  E}{\pd t}+\frac{\pd \left(v_k \rho  E + v_i 
	(p 
	\delta_{ik} - \sigma_{ik}) \right)}{\pd x_k}=0.
	\label{GPR.EnergyCons} \\[2mm]
	&\frac{\pd \dist{\mu}{k}}{\pd t} + \frac{\pd (\dist{\mu}{j}	v_j)}{\pd x_k} + v_j\left(\frac{\pd 
	\dist{\mu}{k}}{\pd 	x_j} - \frac{\pd \dist{\mu}{j}}{\pd x_k}\right)
	=-\dfrac{ \psi_{\mu k} }{\theta(\tau)},\label{GPR.deformation}
	\end{align}
where $\dist{\mu}{i}$ are the components of the distortion field and $\bm{\psi} = [\psi_{\mu i}]$ 
is the dissipative term due to material element rearrangements and will be specified below as 
well as the denominator $ \theta $, which is assumed to be proportional to the characteristic strain 
dissipation time $ \tau $. The definitions of the pressure $ p = \rho^2 E_\rho $ and the elastic 
stress $ 
\bm{\sigma} = -\rho \A^\transpose E_{\A} $ are conditioned by the requirements of the 
thermodynamical 
compatibility~\cite{SHTC-GENERIC-CMAT} and hence depend on the specification of the energy 
potential $ E(\rho,s,\vv,\A) $. 
An entropy inequality can also be derived and reads as
\bea
\frac{\pd \rho s}{\pd t}  
     + \frac{\pd (\rho s v_k )}{\pd x_k}
	=\frac{1 }{ \theta T } \psi_{\mu i}\psi_{\mu i} \geq 0 ,\label{GPR.entropy}
\eea
\end{subequations}
where $T = E_{s}$ denotes the temperature.
The total energy $ E $ is assumed to consist of three parts, each of 
which represents an energy distributed on one of the three different 
scales~\cite{DPRZ2016,HYP2016}: the 
molecular scale 
(microscale), the mesoscale of the material elements, and the macroscale:
\begin{equation}\label{eqn.GPR.TotalEnergy}
E(\rho,s,\vv,\A)= \underbrace{E^1(\rho,s)}_{\text{microscale}} + 
\underbrace{E^2(\rho,s,\A)}_{\text{mesoscale}} 
+ \underbrace{E^3(\vv)}_{\text{macroscale}} .
\end{equation}  
The microscale energy $ E^1(\rho,s) $ is given by the stiffened gas or Mie-Gr{\"u}neisen equations 
of state, the macroscopic energy $ E^3 $ is simply the kinetic energy $ E^3 = \frac{1}{2}\vv^2 $. 
From 
the requirement of the thermodynamic compatibility~\cite{SHTC-GENERIC-CMAT}, the pressure is given 
by 
\bea\label{eqn.GPR.Pressure}
   p = \rho^2 E_\rho \equiv \rho^2 \left( E^{1}_\rho + E^2_\rho \right) ,
\eea
and the temperature by
\bea\label{eqn.GPR.Entropy}
   T =  E_s \equiv E^{1}_s + E^2_s  ,
\eea
where, we recall, $E^{j}_\rho$ denotes the partial derivative  $\pd E^{j} /\pd \rho$
and $E^{j}_s = \pd E^j/\pd s$. However, in this paper we shall consider $ E^2 $ which depends only 
on $ \AAA $ and does not depend on $ \rho $ and $ s $, \ie $ E^2 = E^2(\AAA) $.

For the purpose of this paper, it is sufficient to use the following simplified mesoscopic energy $ 
E^2 $ 
\begin{equation}\label{eqn.GPR.E_2}
E^2(\A)=\dfrac{\css}{4}G'_{ij}G'_{ij}\equiv \dfrac{\css}{4} \left 
(I_2-\frac13 I_1^2\right ), \qquad c_s = const > 0,
\end{equation}
with 
\begin{equation}
\GG'= [G_{ij}'] = \GG - \frac{1}{3} {\rm tr}(\GG) \bm{I},  \qquad 
 \textnormal{ and } \qquad \GG=\A^\transpose\A. 
\end{equation} 
Here, $\GG'$ is the deviatoric, or \textit{trace-less}, part of the metric
tensor $\GG=\A^\transpose\A$, 
and $ c_s > 0 $ is the characteristic velocity of propagation of transverse perturbations. In the 
following we shall refer to it as the \textit{shear sound 
velocity}. In general, $ c_s$
may depend on the density and the entropy~
 but we do not consider this possibility here. 
The principle of material frame indifference implies that the total energy can only depend on 
vectors 
and tensors
by means of their invariants. Thus, we note that
\bea
G'_{ij}G'_{ij} \equiv I_2 - \frac13 I_1^2 ,
\eea 
where $ I_1={\rm tr}(\GG) $ and $ I_2={\rm tr}(\GG^2)$.
As such, $E^2$ and the total energy $E$ are functions of the invariants 
of $\A$.
The algebraic dissipative source term, $-\frac{\bm{\psi}}{\theta}$ 
on the right-hand side of \eqref{GPR.deformation}
describes the shear strain dissipation due to material element rearrangements. As discussed in 
\cite{HPR2016,DPRZ2016,SHTC-GENERIC-CMAT}, this term should be proportional to $ E_{\A} = \pd 
E/\pd\A $ and hence, 
we 
define $\bm{\psi} =  E_{\A}$, which thus has the meaning of a stress (similar to the Lagrangian 
first 
Piola-Kirchhoff stress).
Once the total energy potential is specified, all fluxes
and source terms have an explicit form. 
Thus, for the energy $E^2$  given by (\ref{eqn.GPR.E_2}), 
the elastic stress reads as
\begin{equation}\label{eqn.GPR.ElasticStress} 
  \ssigma = -\rho\A^\transpose \boldsymbol{\psi} 
  = -\rho\A^\transpose E_{\A} = -\rho\, \css \GG\, \dev{\GG}.
\end{equation}
For the further analysis, another expression of the stress tensor $ \ssigma $ will be useful
\begin{equation}\label{GPR.ElasticStress2}
\ssigma = -\rho \css \left( \kappa \dev{\GG} + \dev{\GG}^2 \right), \qquad \kappa = \frac{1}{3} 
\tr(\GG).
\end{equation}
Notice that $ \tr(\bm{\sigma}) = -\rho \css\tr(\dev{\GG}^2) \neq 0$, therefore the overall 
pressure is not $ p $, but $ p + \frac{1}{3}\tr(\bm{\sigma}) $ instead. 
The dissipation term is expressed as
\begin{equation}\label{eqn.GPR.psi} 
-\dfrac{\boldsymbol{\psi}}{\theta} = 
-\dfrac{E_{\A}}{\theta}=-\dfrac{3}{\tau } \left| \A 
\right|^{\frac{5}{3}} \A \dev{\GG}, \qquad \quad
\theta = \frac{1}{3}\tau \, \css \, |\A|^{-\frac{5}{3}} ,
\end{equation}
where $|\A|=\det(\A) > 
0$ being the determinant of $\A$. The strain relaxation time $\tau=\tau(\rho,s,\AAA)$ is the 
continuum interpretation of the Frenkel time~\cite{Frenkel1955}, which can be interpreted as a 
characteristic time of material element rearrangements, see~\cite{HPR2016,HYP2016}, and thus, in 
metals, the model for $ \tau $ can be obtained based on the physics of dislocations which are the 
source of the structural rearrangements in crystalline solids. For 
example,
the dislocation velocity is known to be well approximated by~\cite{Greenman1967,Johnston1959}
\begin{equation}\label{dislocation.vel1}
v = v_0 \exp(-\sigma_0/\sigma)
\end{equation}
under a wide range of conditions, where $ v_0 $ and $ \sigma_0 $ are the material constants, 
while $ 
\sigma $ is the resolved shear stress. For moderate strain rates, the experimental data is also 
adequately approximated by the expression~\cite{Vitman1949,Greenman1967}
\begin{equation}\label{dislocation.vel2}
v = v_0 (\sigma/\sigma_0)^n.
\end{equation}
Thus, as shown in~\cite{GodDenisenko1975,Godunov1976a} for several metals and for the deformation 
rate ranging in $ 0 \leq \dot{\varepsilon} \leq 10^7 $ s$ ^{-1} $ the following interpolated 
formula for $ 
\tau $ gives a good agreement with the experimental data
\begin{equation}\label{eqn.GPR.tau}
\tau = \tau_0\left(\frac{\sigma_Y}{\sigma}\right)^n, \qquad \sigma = \sqrt{\dfrac{3}{2} 
\text{tr}(\dev{\bm{\sigma}}^2)}, \qquad \dev{\bm{\sigma}} = \bm{\sigma} - \frac{1}{3}\tr 
(\bm{\sigma}) \II,
\end{equation}
where the temperature dependent terms from \cite{GodDenisenko1975,Godunov1976a} are ignored in this 
paper. Here, $ \tau_0={\rm const}>0 $ is a material 
dependent parameter with the physical dimension of time, typically a small
time-scale ($ < 10^{-6}$s) for metals, $ n> 0 $ can be a function of the state parameters, in 
general, but for simplicity is assumed to be constant in this study. It indicates the degree of 
rate-dependency of the elastic-to-plastic transition. 
The parameter $\sigma_Y > 0 $ is the static
yield strength. As in real media~\cite{Greenman1967}, the yield strength in the \GPR model depends 
on the rate of 
deformations and is determined by both $ \sigma_Y $ and $ n $, see the discussion in 
Section~\ref{closure.plastic}. 
For large $ n $, the dynamic yield strength approaches $ \sigma_Y $, 
that is, the elastic-to-plastic transition approaches the ideal plasticity law with the 
Von Mises yield criterion, while for small $ n $, the dynamic yield strength 
becomes rate dependent, as will be shown in the numerical examples. 

We now proceed with a more detailed comparison of the models in the next section.

%
%

%
%
\subsection{Comparison with the Wilkins hypoelastic model}\label{sec.Wilkins}

Although both systems of PDEs (\ref{eqn.Wilkins}) and (\ref{GPR.eqns}) were originally designed 
for modeling of the same physical processes, behavior of metals under high strain-rate, they 
obviously differ. In this section,
we ought to compare these models and discuss their differences.
For the sake of clarity let us first rewrite both of them side-by-side \footnote{here we use the compact notation  
$\nabla (\A \vv) + \vv \cdot (\nabla  \A - \nabla \A^T) = \frac{\pd (\dist{\mu}{j}	v_j)}{\pd x_k} + v_j\left(\frac{\pd 
	\dist{\mu}{k}}{\pd 	x_j} - \frac{\pd \dist{\mu}{j}}{\pd x_k}\right)$ in the evolution equation for 
$\A$}:
\begin{subequations}\label{GPR.vs.Wilk}
  \begin{align}
    & \frac{\pd \rho}{\pd t} + \nabla \cdot (\rho \vv ) =0,  &  
    & \frac{\pd \rho}{\pd t} + \nabla \cdot (\rho \vv ) = 0, 
      \label{vs.conti2} \\[2mm]
    & \frac{\pd \rho \vv}{\pd t}+ \nabla \cdot ( \rho \vv\otimes \vv) - \nabla \cdot 
    \bm{T}   =0,  &
    & \frac{\pd \rho \vv}{\pd t}+ \nabla \cdot ( \rho \vv\otimes \vv) - \nabla \cdot 
    \bm{T}   =0, 
      \label{vs.momentum2}\\[2mm]
    & \frac{\pd \rho E}{\pd t} + \nabla \cdot ( \rho E \vv ) - \nabla \cdot (\bm{T}\vv) = 0,  &
    & \frac{\pd \rho E}{\pd t} + \nabla \cdot ( \rho E \vv ) - \nabla \cdot (\bm{T}\vv) = 0, 
      \label{vs.energy2}\\[2mm]
    & \frac{\pd \A}{\pd t} + \nabla (\A \vv) + \vv \cdot (\nabla  \A - \nabla \A^T)    
      =-\dfrac{ \bm{\psi} }{\theta(\tau)}, &
    & \frac{\pd \SSS}{\pd t} + \vv\cdot\nabla\SSS +  \SSS\asymvel - \asymvel\SSS + 2\mu        
    \dev{\DD} = 2\mu\DD^{\text{p}},
      \label{vs.elast2} 
  \end{align}
  \begin{align}
    & \text{Closure relations}  & & \text{Closure relations} \nonumber\\
    & E = E^1(\rho,s) + \frac{c_s^2}{4}\dev{G}_{ij}\dev{G}_{ij} + \frac12 \vv^2  && E = E^1(\rho,p) 
    + \frac12 \vv^2   \label{vs.energy}\\
    & \bm{T} = -p \bm{I} + \bm{\sigma}, \quad \tr (\bm{\sigma}) \neq 0  && \bm{T} = -p \bm{I} + 
    \SSS,\quad     \tr (\SSS) = 0  \\
    & \bm{\sigma} = -\rho \css \G\dev{\G}, \quad \G = \A^{\transpose}\A,  &
    &  \DD=\frac12\left ( \nabla\vv + \nabla\vv^{\transpose}\right),\quad 
       \asymvel=\frac12\left ( \nabla\vv - \nabla\vv^{\transpose}\right)
    \\
    &  \bm{\psi} = \css  \A \dev{\G}, \quad  \theta = \tau \frac{\css}{3} \, 
    |\A|^{-\frac{5}{3}}&
    &   \DD^{\text{p}} = \chi(f,\SSS) \left( \frac{\SSS}{||\SSS||}:\DD\right) 
    \frac{\SSS}{||\SSS||}      \\
    & \tau=\tau_0\left(\frac{\sigma_Y}{\sigma }\right)^n, \quad \sigma = \sqrt{ \frac32 
    \,\tr\left (\dev{\bm{\sigma}}^2\right )}   
    &
    &  \chi(f,\bm{\SSS}) = \left\{
      \begin{array}{l} 0 ,\\ 1 , \end{array}  \right. \quad 
    f=\sqrt{\frac32 \tr\left (\bm{\SSS}^2\right )} - \sigma_Y \\
    & \text{Material dependent parameters:}  & & \text{Material dependent parameters:} \nonumber\\
    & c_s > 0, \quad \sigma_Y>0, \quad n>0, \quad \tau_0>0,   &
    & \mu \geq 0, \quad \sigma_Y>0,\\
    & \text{Hydrodynamic EOS:}  & & \text{Hydrodynamic EOS:} \nonumber\\
    & \text{Stiffened gas/Mie-Gr{\"u}neisen}   &
    & \text{Stiffened gas/Mie-Gr{\"u}neisen} .
  \end{align}
\end{subequations}

\subsubsection{Closure relations for the hydrodynamic part}

In the absence of elastic and elastoplastic effects, both models reduce to the conventional Euler 
equations of ideal fluid and the stress tensor reduces to $ \TT =-p\II $. Thus, the hydrodynamic 
parts of both models are equivalent, in which the pressure is defined from a hydrodynamic equation 
of state (EOS). Two equations of state, stiffened gas and Mie-Gr{\"u}neisen, are used in this paper 
and are summarized in Section~\ref{sec.EOS.hydro}.

Note that in the \GPR model, the mass density $ \rho $ and the distortion field $ \A $ 
are not genuinely independent as the mass density and the stress deviator $ \SSS $ are independent 
in the Wilkins model. Thus, the continuity equation \eqref{vs.conti2} is, in fact, the consequence 
of the equation \eqref{vs.elast2} \cite{GodRom2003}. In particular, it is implied that 
\begin{equation}\label{rho.detA}
\rho = \rho_0 \det(\A),
\end{equation}
where $ \rho_0 $ is the mass density in the reference configuration.

\subsubsection{Closure relations for the elastic part}\label{sec.closure.elast}

While the fundamental continuity equation \eqref{vs.conti2} is identical in both models, the next 
fundamental conservation law, the linear momentum conservation \eqref{vs.momentum2} requires a 
closure relation for the stress tensor. Thus, the stress tensor in the \GPR model is 
defined from the requirement of the thermodynamical consistency with the first law of 
thermodynamics, \eg see~\cite{SHTC-GENERIC-CMAT}, the energy conservation, which is equivalent to 
the entropy conservation in the 
absence of dissipative processes. The general form of the stress tensor is 
\begin{equation}\label{vs.stress.GPR}
T_{ij} =-\rho \left(\rho E_\rho \delta_{ij} + \dist{\mu}{i} E_{\dist{\mu}{j}} \right)
\end{equation}
and thus, it is completely defined by the specification of the energy potential $ 
E(\rho,s,\vv,\AAA) $. The 
form~\eqref{vs.stress.GPR} of the stress tensor is also conditioned by the Hamiltonian nature of 
the non-dissipative part of the \GPR model~\cite{SHTC-GENERIC-CMAT,GRGPR} and is 
invariant for any types of materials, gases, liquids, or solids. Therefore, the specification of  
the energy potential $ E(\rho,s,\vv,\AAA) $ is the critical step in the \GPR model specification. 
However, in many cases, 
and in particular in this study, the elastic strains are small and it is sufficient to use a 
simplified formula for the elastic part $ E^2 $ 
\eqref{eqn.GPR.E_2}. 
Another requirement to the energy $ E(\rho,s,\vv,\AAA) $ is that it should be a convex function of 
all the state variables in order to guaranty local well-posedness of the initial value 
problem~\cite{SHTC-GENERIC-CMAT}. However, the problem of non-convexity of the energy with respect 
to the two-point second order tensors such as the distortion $ \dist{\mu}{i} $ field is a 
well-known issue~\cite{Ciarlet1988,Dafermos2005} in non-linear 
elasticity, see the discussion in Section~\ref{sec.well-posed}. 

In contrast to the \GPR model, the closure for the stress tensor of the Wilkins model 
is not a scalar 
potential, but is given by an evolution equation for the deviatoric part $ \SSS $ of the stress 
tensor $ \bm{T} $, which is a particular case of the general evolution equation 
\eqref{stress.rate.gen}. Moreover, as 
already mentioned by several authors \cite{Rom1973,Gavrilyuk2008,Maire2013},
there is a lack of 
connection of hypoelastic formulations of continuum mechanics with a thermodynamic potential. Thus,  
the deviatoric stress $ \SSS $ and the energy $ E $ are usually assumed to be \textit{independent} 
quantities and, for 
example, the total energy of the Wilkins model \eqref{vs.energy} does not have an elastic part $ 
E^2 $ and is set to 
$ 
E(\rho,p,\vv) = E^1(\rho,p) + 
E^3(\vv) $. Because the total energy is conserved in both models, the lack of elastic energy 
in the total energy of the Wilkins model results in the fact that the amount of energy stored 
in $ E^1 $ and $ E^3 $ might be different in the two models,  
which of course then results in differences in the density, temperature and velocity fields. 
However, for the case of small elastic deformations, such a discrepancy might be not noticeable. 
Furthermore, the absence of the elastic energy in the Wilkins model results in the violation of 
the entropy conservation for elastic deformations, as discussed in~\cite{Rom1973,Gavrilyuk2008,Maire2013} which may lead to 
incomplete recovery of the undeformed state~\cite{Kojic1987}. Thus, one may conclude that it is 
necessary to avoid the application of the Wilkins model to finite-strain elastic deformations 
which, in fact, was not designed to be used in such a regime. Extension of the Wilkins model to 
finite-strains is not a trivial task and requires the use of a more complex evolution equation for 
$ \SSS $ with a solution dependent elastic modulus $ \alpha_1,\alpha_2,\ldots,\alpha_{12} $ in 
\eqref{stress.rate.gen}, see e.g. \cite{Kim2016} for more details.  

Since both models are assumed to be consistent with the linear elasticity limit, 
the elastic modulus $ \mu $ in the 
Wilkins model and the shear sound velocity $ c_s $ in the \GPR model are related by  
\begin{equation}\label{cs.mu}
\mu = \rho_0 c_s^2,
\end{equation}
where $ \rho_0 $ is the mass density in the undeformed state.

We also note that our choice for the elastic energy $ E^2 $ gives the non-vanishing trace of $ 
\sigma = -\rho 
\AAA^\transpose E_{\AAA} $ and therefore the total pressure of the \GPR model is $ P = 
\frac13 \tr (\TT) = p + \frac13 \tr (\ssigma) = p -\frac13 \rho \css\tr(\dev{\GG}^2)$, while we 
have $ P = \frac{1}{3} \tr (\TT) = p $ for the 
Wilkins model. In fact, in the case of small elastic deformations (small deviator $ \dev{\G} $), 
i.e. in the 
region of 
applicability of the Wilkins model, the contribution $ \frac13 \tr (\ssigma) = -\frac13 \rho 
\css\tr(\dev{\GG}^2)$ is quadratic in $ \dev{\G} $ and two or three orders 
of magnitude smaller than the hydrodynamic pressure $ p $, as will be shown in the numerical 
examples. 
We note that in general, and especially in shock physics, the elastic energy $ E^2(\A) $ 
should depend not only on shear deformations 
but also on compression and temperature~\cite{Steinberg1974,Steinberg1980,Steinberg1989,Plohr1995}. 
The toy equation of state for the elastic part $ E^2 $ employed in this model only weakly accounts 
for the coupling of the compression and shear effects, but more sophisticated equations of 
state can be designed, \eg 
see~\cite{BartonRom2010}. However, if the flow is weakly compressible, energy 
potentials in the so-called separable form~\cite{Ndanou2014} can be used, which eliminate the 
non-linear coupling of the compression and shear deformations and may help to simplify the 
theoretical analysis and the numerical implementation of the model. 

We finally note that the lack of a connection between the hypoelastic-type models (Wilkins model 
in particular) and a thermodynamic potential makes it also unclear whether there exists 
the possibility of deriving such models from the fundamental Hamilton principle of stationary 
action, or not.

\subsubsection{Closure relations for the inelastic part}\label{closure.plastic}

In the \GPR model, inelastic deformations are modeled by the source term in the 
distortion evolution equation \eqref{GPR.deformation} or \eqref{vs.elast2}$ _1 $. Because the 
inelastic deformations are due to the irreversibility of the micro-structural rearrangements 
(dynamics of dislocations), which 
is a thermodynamically irreversible process, the source term should not violate the second law of 
thermodynamics, which states that physical entropy should not decrease. Furthermore, because of the 
exceptional role of 
the energy potential $ E $ in the formulation of the \GPR 
model, such a source term should have a certain structure, see 
\cite{SHTC-GENERIC-CMAT}. Namely, it has to be proportional to $ E_{\dist{\mu}{i}} $, which 
automatically guarantees that the entropy is not decreasing for inelastic deformations.

Although all thermodynamically irreversible processes (including inelastic deformations) are due to a 
certain \textit{dynamics} happening at the microscales and hence are fundamentally rate-dependent, the 
avoidance of the  
rate-dependency in the mathematical model might be a reasonable approximation in many situations. 
Nevertheless, the \GPR model is intrinsically rate-dependent, because otherwise neither 
well-posedness nor thermodynamical consistency can be guaranteed. 

Yet the model \eqref{eqn.GPR.tau} for $ \tau(\rho,s,\AAA) $ represents a simple empirical 
model~\cite{GodDenisenko1975,Godunov1976a} and 
admits strain-rate dependency of the effective yield strength, and the ideal plasticity is 
recovered in the limit of $ n \rightarrow \infty $. Indeed, the material parameter $ \sigma_Y $ 
should be considered as the yield strength in the limit of vanishing strain rate (creep flows) 
while, as discussed in details in \cite{GodRom2003}, the effective yield strength is the 
result of a combination of the model parameters $ n $, $ \sigma_Y $ and the flow parameter $ \LL = 
\nabla\vv $, and can be obtained as the steady-state solution of the equation for the distortion
\begin{equation}\label{yield.strength}
\frac{\rmd \A}{\rmd \, t} + \A \LL = -\frac{\bm{\psi}}{\theta(\tau)},
\end{equation}
in which the rate of strain tensor $ \LL $ is considered as a parameter (constant),  see \eg 
Fig.1 in~\cite{BartonRom2010}. Such a rate-dependency property of the \GPR model in 
conjunction with the smallness of the elastic strains $ \dev{\GG } $ is the key 
feature that allows treating viscous Newtonian fluids as materials with zero yield 
strength $ \sigma_Y = 0 $, see Fig.1 in \cite{HPR2016} and \cite{DPRZ2016}.

 
On the other hand, the Wilkins model employs the ideal plasticity law with the von Mises yield 
criterion. Thus, the intensity of tangential stresses $ \sigma = \sqrt{\frac32 \tr\left 
(\bm{\SSS}^2\right )} $ cannot exceed the static yield strength $ \sigma_Y $, while as we shall see 
in the numerical examples,
it is usually the case that $ \sigma $ can be larger than $ \sigma_Y $ in the intrinsically 
rate-dependent \GPR model. Finally, let us note that such a notion as the yield 
strength is a purely static notion and can not be determined for genuinely transient phenomena.

We also note that the time $ \tau $ represents a mesoscopic time scale in the \GPR 
model and can be related to the mesoscopic length scale $ \ell $ of material elements as $ \tau 
\sim \ell $ as discussed in \cite{HYP2016}. This feature of the model might be useful for the 
modeling of inelastic strongly heterogeneous deformations, \eg in crack propagation.

\subsubsection{Objectivity and a hypoelastic form of the \GPR model}

In contrast to the stress tensor, the distortion field $ \A $ is not a characteristic of the 
material response, but is a geometrical object. It is therefore not surprising that its time 
evolution  automatically fulfills the principle of material frame indifference. In fact, the 
non-dissipative part of the distortion time evolution (all the differential terms) can be obtained 
as the integrability condition for the equations of motion derived from the Hamilton principle of 
stationary action~\cite{GRGPR}. Moreover, the non-dissipative part of the distortion time evolution 
is the Lie derivative along the four-velocity and hence, is invariant under arbitrary 
transformation of the time and spatial coordinates~\cite{GRGPR}. It therefore represents an 
objective time rate by construction.

On the other hand, in order to obtain a hypoelastic model satisfying the principle of material 
frame invariance, an objective stress rate $ \rmD\SSS/\rmD t $ has to be used. As already 
mentioned before, its selection is difficult and cannot be made on a physical basis. 
Thus, the choice of the objective time derivative is the source of uncertainty in 
hypoelastic-type models. This choice,  
however, is a very important step in the model formulation. Furthermore, it is well-known that 
a wrong choice may result in the loss of hyperbolicity, which implies that the solution to the equations can exhibit catastrophic, high-frequency  Hadamard instabilities, \eg see 
\cite{Rutkevich1970,Rutkevich1972,DupretMarchal1986,Joseph1986a} and \cite{Beris1994}, \S8.1.6.

Being formulated in different terms, the \GPR and Wilkins model are not directly 
available for a comparison. It is clear that the Wilkins model as it is used in this paper cannot 
be reformulated as a hyperelastic model, because the energy potential does not contain the elastic 
energy. However, any hyperelastic model can be rewritten as a hypoelastic one, if only the 
stress-strain relations are invertible~\cite{Rom1973,Favrie2015Hypo}. Thus, in what follows, we 
obtain a hypoelastic version of 
the \GPR model under the assumption of small elastic deformation, \ie the deviator $ 
\dev{\GG} $ is small, which is the region of applicability of the Wilkins model, and we compare the 
obtained evolution equation for the stress deviator with its Wilkins' counterpart 
\eqref{vs.elast2}$_2 $. We note that the obtained PDE will be the result of the energy 
specification. Each time, the form of the energy (its elastic part) is changed, the resulting 
equation for the 
stress deviator changes as well.

Let us consider the stress-strain relation \eqref{GPR.ElasticStress2} defined by our choice 
of the elastic part of the energy potential \eqref{eqn.GPR.E_2}. If $ \dev{\G} $ is small, 
relation 
\eqref{GPR.ElasticStress2} can be approximated as (we ignore second order terms in $ \dev{\G} $)
\begin{equation}\label{stress.strain.small}
\ssigma = -\rho \, \css \kappa \, \dev{\GG}, \qquad \kappa = \frac{1}{3} 
\tr(\GG),
\end{equation}
which in this approximation results in $ \tr(\ssigma) = 0$. Hence, the time evolution for $ 
\ssigma $ can be obtained by differentiating 
\eqref{stress.strain.small}$ _1 $:
\begin{equation}\label{stress.evol}
\dot{\ssigma} = -c_s^2(\dot{\rho} \, \kappa \,  \dev{\GG} + \rho \,  \dot{\kappa} \,  \dev{\GG} + 
\rho \,  \kappa \, \dev{\dot{\GG}}),
\end{equation}
where the dot denotes the material time derivative $ \rmd/\rmd t $. In order to find the rates $ 
\dot{\rho} $, $ \dot{\kappa} $, and $ \dev{\dot{\G}} $ we may use the continuity equation and the 
evolution equation for the metric tensor $ \G = \A^\transpose \A $
\begin{equation}\label{G.PDE}
\dot{\G} = -\G \LL - \LL^\transpose \G - \bm{\mathcal{S}}, \qquad \bm{\mathcal{S}} = 
-\frac{1}{\tau} g^{5/6} \G \dev{\G} =  
-\frac{1}{\tau} g^{5/6} (\kappa \dev{\G} + \dev{\G}^2), \qquad g = \det(\G),
\end{equation}
which is the direct consequence of the distortion time evolution \eqref{GPR.deformation}. Recalling 
that we ignore second order terms in $ \dev{\G} $, the relaxation source term in 
\eqref{G.PDE} can be approximated as $ \bm{\mathcal{S}} = -\frac{1}{\tau} g^{5/6} \kappa \, 
\dev{\G} $. Therefore, one may obtain
\begin{equation}\label{rates}
\dot{\rho} = -\rho\, \tr(\LL), \qquad \dot{\kappa} = - \frac23 \left (\kappa \, \tr(\LL) + 
\tr(\dev{\GG}\LL)\right ), \quad 
\dev{\dot{\G}}=-\left(\dev{\G} \LL^\transpose + \LL \dev{\G}  -\frac23 \tr \left(\LL 
\dev{\G} \right) + 2 \kappa \dev{\DD}\right) - \dev{\bm{\mathcal{S}}},
\end{equation}
where $ \LL = \nabla\vv $, and $ \dev{\bm{\mathcal{S}}} $ is the deviatoric part of $ 
\bm{\mathcal{S}} = -\frac{1}{\tau} 
g^{5/6} \kappa \, 
\dev{\G} $ and, in fact, $ 
\dev{\bm{\mathcal{S}}} = \bm{\mathcal{S}}  $.

Hence, plugging \eqref{rates} into \eqref{stress.evol}, the evolution equation for $ \ssigma $, in 
the limit of small elastic deformation, becomes
\begin{equation}\label{GPR.stress.rate}
\underbrace{ \frac{\pd \bm{\sigma} }{\pd t} + \vv\cdot\nabla\bm{\sigma}  + 
\bm{\sigma}  (\nabla \vv) + (\nabla \vv)^\transpose\bm{\sigma} + \frac{5}{3}\tr(\nabla 
\vv)\bm{\sigma} - 
\frac{2}{3}\tr(\bm{\sigma} \nabla \vv)\bm{I}}_{\text{Objective 
derivative}} \, - \, 2 \, \mu \, \dev{\DD} = -\frac{1}{\tau}\kappa g^{5/6}\ssigma, \qquad \mu 
= \rho c_s^2 \kappa^2.
\end{equation}
Note that in the passage from \eqref{stress.evol} to \eqref{GPR.stress.rate}, we also ignore the 
term $ \tr(\dev{\G}\LL)\dev{\G} $ which is second order in $ \dev{\G} $. The underlined 
differential terms in \eqref{GPR.stress.rate} form an objective time derivative because the first 
four terms on the left constitute the Lie derivative of a two times covariant tensor (which 
transforms as 
a tensor), while the remaining two underlying terms also transform as a tensor.

The widely used variant of the Wilkins model (\ref{vs.elast2})$ _2 $ with the Jaumann objective 
derivative, on the other hand, if expressed in terms of $\nabla \vv$, yields
\begin{equation}\label{Wilkins.stress.rate}
\underbrace{ \frac{\pd \SSS}{\pd t} + \vv\cdot\nabla\SSS + 
\frac12 \left( 
\SSS \nabla\vv  + (\nabla \vv)^\transpose  \SSS - \SSS (\nabla \vv)^\transpose  -   \nabla\vv \ \SSS
\right) }_{\text{Jaumann objective derivative}} \, - \, 2 \, \mu \, \dev{\DD} = \DD^{\rm p}.
\end{equation}
The two objective stress rates \eqref{GPR.stress.rate} and \eqref{Wilkins.stress.rate} are apparently 
very different. However, as one may expect and as will be shown later via numerical results, 
these differences in the governing PDE do not result in remarkable differences in the solutions 
in the linear elasticity limit. On the other hand, in the case of large elastic deformations, which, however, is not the range of applicability of the Wilkins model and will be shown only for  
comparison purposes, the solutions of both models differ significantly. In the finite-strain 
elastoplastic range, such a difference in the definition of the objective stress rates will be less 
pronounced, see Section~\ref{sec:numerics}.  


\subsubsection{Thermodynamics}\label{sec.thermo}

The \GPR model was deliberately developed within the class of the so-called SHTC 
equations (Symmetric Hyperbolic and 
Thermodynamically Compatible equations)~\cite{God1961,God1972MHD,GodRom1996a,Rom1998,Rom2001}, or 
see the recent review of the SHTC formulation of continuum mechanics~\cite{SHTC-GENERIC-CMAT}. As 
it follows from the name 
of 
the SHTC equations, a model belonging to such a class is thermodynamically compatible, which 
means that the first and second laws of thermodynamics are fulfilled for such system of equations 
by definition. In particular, the \GPR model is characterized by the quantities
\begin{equation}\label{var.GPR}
\rho, \ s,\ \vv, \ \A,\ E
\end{equation}
but in fact the total energy $ E $ is not an independent state 
variable but is a potential which depends on $ (\rho,s,\vv,\A) $, that is $ E = E(\rho,s,\vv,\A) $, 
and hence the energy conservation 
law \eqref{GPR.EnergyCons} is not an  
independent equation but, in fact, it is the consequence of all the other equations, including the 
entropy law \eqref{GPR.entropy}, (it can be 
obtained as the sum of equations \eqref{GPR.eqns} multiplied the derivatives of $E$ with respect to the state variables, see \eg \cite{SHTC-GENERIC-CMAT}). 

On the other hand, the mathematical structure of the Wilkins model is less constrained. Thus, the system is characterized by the quantities 
\begin{equation}\label{var.Wilk}
\rho, \ p,\ \vv, \ \SSS,\ E
\end{equation}
but contrarily to the \GPR model, the energy $ E $ depends only on $ (\rho,p,\vv) $ and 
does not depend on $ \SSS $. In other words, the hydrodynamic part of the Wilkins model is 
thermodynamically consistent, while its elastic part is fully decoupled from the thermodynamic 
potential, and the evolution equation for $ \SSS $ is 
rather arbitrary and may take various forms~\eqref{stress.rate.gen}, which includes the uncertainty 
in the choice of the objective time derivative $ \rmD/\rmD t $. Such an extra degree of freedom of 
the Wilkins model may also lead to difficulties in selecting a physically admissible solution. For example, hypoelastic-type models may exhibit an unphysical dissipative behavior within the purely elastic regime, see \eg \cite{Kojic1987}. In other words, reversible transformations of an elastic material are characterized by a non-zero rate of entropy~\cite{Rom1973,Gavrilyuk2008}. As a consequence, certain 
processes can cause hypoelastic laws to give pathological results, such as a non-zero stress at 
zero deformation after cyclic loading, e.g. see \cite{Kojic1987}. Let us point out that these 
difficulties did not prevent from developing numerical methods for simulating \textit{transient} 
elastoplastic deformations with the Wilkins model over the years 
\cite{Wilkins64,Howell2002,KulikPogorSemen,Maire2013,Fridrich2017}.

\subsubsection{Extensions}

Because the non-dissipative part of the \GPR model can be derived from the Hamilton 
principle~\cite{SHTC-GENERIC-CMAT}, it can be coupled in a consistent way with various physical 
phenomena such as interaction with the electromagnetic fields~\cite{DPRZ2016}, non-equilibrium heat 
conduction~\cite{SHTC-GENERIC-CMAT}, mass transfer and 
poroelasticity~\cite{Romenski2011porous,Romenski2014,PeshGrmRom2015}, general relativistic 
flows~\cite{GRGPR}, etc.
Moreover, since plastic deformations are due to the nucleation, motion and interaction of the  
dislocations, more accurate continuous plasticity models can be designed by taking into account 
such a microscopic dislocation dynamics, \ie by building a continuum finite strain theory of 
dislocations. Thus, the use of the distortion field as thermodynamic state variable in 
principle allows to build such a theory, because the Burgers tensor $ B_{\mu ij} = \frac{\pd 
\dist{\mu}{j}}{\pd x_i} 
- \frac{\pd 
\dist{\mu}{i}}{\pd x_j}$ discussed in Section~\ref{sec.intro.hyper.euler} has the meaning of 
the density of continuously distributed 
dislocations~\cite{GodRom2003,SHTC-GENERIC-CMAT} and is directly available 
in the theory. Also, such a dislocation based theory of plasticity should be of great importance 
for developing of accurate and physically consistent models for damage and fracture of solids. 
The lack of a connection of the Wilkins model with the Hamilton principle of stationary action 
makes the above mentioned extensions in the framework of the Wilkins model rather questionable.

\subsubsection{Well-posedness and hyperbolicity}\label{sec.well-posed}

A system of time-dependent PDEs representing a continuum mechanics model has to have a well-posed 
initial value problem (IVP), that is, for sufficiently regular initial data, the solution 
exists, it is unique, and depends continuously on input data. It is known that hyperbolic 
systems of PDEs have well-posed IVP~\cite{Dafermos2005,Serre2007}. The well-posedness of the IVP is 
also a fundamental property of a time-dependent system in order to be solved numerically.

The Wilkins model, if used with the Jaumann objective derivative and in the limit of small elastic 
deformations (small $ \SSS $), has well-posed IVP because it is proven to be hyperbolic in that 
case, see~\cite{Maire2013}. Nevertheless, it cannot be guaranteed that a change of the objective 
stress rate will not affect the well-posedness of the Wilkins model, \eg see~\cite{Joseph1986a}. In 
general, 
the question of hyperbolicity 
of a hypoelastic model~\eqref{stress.rate.gen} with solution-dependent elastic moduli $ 
\alpha_i $, $ i=1,2,\ldots,12 
$ should be considered separately.

In contrast to the hypoelastic-type models, and the Wilkins model in particular, the form of the 
equations in 
the hyperelastic-type \GPR model is fixed and the hyperbolicity of the model entirely depends on 
the specification of the energy potential $ E $. An important feature of the \GPR model is its 
Galilean invariance and hence, invariance with respect to   coordinate rotations. For such 
PDEs, the question of hyperbolicity of 3D equations is equivalent to the hyperbolicity of 1D 
equations~\cite{Ndanou2014}. Thus, it is known that the 1D (say for the $ x_1 $-direction) 
hyperelastic Eulerian equations written 
in terms of  either the distortion 
field $ \dist{\mu}{i} $~\eqref{GPR.eqns} or elastic strain $ \Feff{i}{\mu} $ (inverse distortion) 
are known to be hyperbolic if the so-called \textit{acoustic tensor} $ \Omega_{ij} = 
-\rho^{-1}\frac{\pd T_{1i}}{\pd \dist{\mu}{1}}\dist{\mu}{j}$ or $ \Omega_{ij} = 
\rho^{-1}\frac{\pd T_{1i}}{\pd \Feff{1}{\mu}}\Feff{j}{\mu}$, respectively, is positive definite 
\cite{MillerColella2001,Barton2009,BartonRom2010}. 
Here, $ T_{ij} = -\rho \dist{\mu}{j} E_{\dist{\mu}{i}} = \rho \Feff{i}{\mu} E_{\Feff{j}{\mu}}$ is 
the total Cauchy stress and besides, it is implied that the density is not treated as an 
independent state variable but as $ \rho = \rho_0 \det(\A) = \rho_0/\det(\bm{F}^{\rm e}) $. 
However, 
the conditions on the energy potential which may guarantee the positive definiteness of the acoustic 
tensor are not known in general. Our working hypothesis is that the energy should be a convex 
function not of all the nine components of $ \dist{\mu}{i} $ but be a convex function of each 
column of 
$ \A $ separately. At least this guarantees that the one-dimensional Riemann problem is 
solvable~\cite{GodPesh2010}. Also, by Ndanou et al~\cite{Ndanou2014}, it has been shown that for a 
certain class of 
equations of state (when the volumetric and shear parts of energy are separated), the positive 
definiteness of the acoustic tensor is equivalent to the convexity of the volume shear energy 
$ \det(\dist{\mu}{i}) E^2 $ with respect to the first column of $ \dist{\mu}{i} $ (second or third 
if 
the 1D 
problem is considered in the other directions). It is interesting to extend the results of 
\cite{Ndanou2014} to more general equations of state in which the volumetric and shear effects can 
be coupled as discussed in Section~\ref{sec.closure.elast}.

\section{Hydrodynamic equation of state for both models}\label{sec.EOS.hydro}

In this section, we summarize the hydrodynamic equation of state that will be used for both 
models in the numerical examples in Section~\ref{sec:numerics}.

The internal energy $E^1(\rho,s)$ is related to the kinetic energy of the molecular motion.
In this paper, for $ E^1 $ we will 
use the \textit{stiffened gas equation of state} for solids
\begin{equation}\label{eqn.GPR.stiff_gas_eos}
E^1(\rho,s)=\dfrac{c^2_0}{\gamma(\gamma-1)}\left ( \dfrac{\rho}{\rho_0}\right 
)^{\gamma-1}e^{s/c_V}+\dfrac{\rho_0 c_0^2-\gamma p_0}{\gamma\rho}, \qquad c_0 = 
const
\end{equation}
where $c_0$ is an adiabatic sound speed, $p_0$ is the reference (atmospheric) pressure,
 $\rho_0$ the reference mass density and $c_v, c_p$ are 
the specific heat capacities at constant volume and pressure respectively, which are
related by their ration $\gamma=c_p/c_v$. 
The last EOS we consider in this work is the \textit{Mie-Gr{\"u}neisen equation of state}
\begin{equation}\label{eq:mie_gruneisen_eos}
E^1(\rho,p)= \frac{p-\rho_0 c_0^2 \ f(\nu)}{\rho_0 \Gamma_0}, \quad f(\nu) = 
\frac{(\nu-1)(\nu-\frac12\Gamma_0(\nu-1))}{(\nu - s(\nu-1))^2}, \quad \nu=\frac{\rho}{\rho_0}.
\end{equation}
The pressure is always given by 
\begin{equation}\label{eqn.pressure}
  p = \rho ^2 E^1_{\rho } ,
  \end{equation}
which in the case of perfect gas equation of state leads to 
\begin{equation}\label{eqn.ideal.pressure}
  p = \mathcal{P}_{ID} =  \rho^\gamma e^{\frac{s}{\text{cv}}}
\end{equation}
while for the stiffened gas equation of state (\ref{eqn.stiffgas})
\begin{equation}\label{eqn.stiffgas.pressure}
  p = \mathcal{P}_{SG} 
    = \frac{\rho_0 \, c_0^2e^{\frac{s}{\text{cv}}} \left(
           \frac{\rho }{\rho_0}\right)^{\gamma}}{\gamma}
      -\left(\frac{\rho_0\, c_0^2}{\gamma } -p_0 \right),
\end{equation}
with $\pi_\infty=\frac{\rho_0\,c_0^2 }{\gamma } -p_0 $, 
and for the Mie-Gr{\"u}neisen equation of state
\begin{equation}\label{eqn.mg.pressure}
  p = \mathcal{P}_{MG} = \rho_0 \Gamma_0 E^1(\rho,s) + \rho_0 c_0^2 f(\nu).
\end{equation}

The hydrodynamic equation of state can also be written in the form 
\bea
p = \mathcal{P}( \rho, \varepsilon),
\eea
which in the case of the \textit{stiffened gas} equation of state becomes 
\bea \label{eqn.stiffgas}
p = \mathcal{P}_{\text{SG}}( \rho, \varepsilon) = (\gamma - 1) \rho \varepsilon 
- \gamma \pi_\infty,
\eea
where $\gamma, \pi_\infty$ are two material dependent constants.

\section{High order accurate numerical methods} \label{sec:HOscheme}

In this section we present a brief summary of the numerical methods employed to solve the 
Wilkins model and the \GPR model:  
we employ a high order accurate direct Arbitrary-Lagrangian-Eulerian (ALE) ADER-WENO scheme 
on moving unstructured meshes, see  
\cite{Lagrange3D,LagrangeNC,LagrangeQF,WB_PhD,ALEMOOD1,ALEMOOD2} and Section~\ref{sec:ALEscheme},  
as well as a high order ADER Discontinuous Galerkin (DG) scheme supplemented with 
\textit{a posteriori} subcell finite volume (FV) limiter operating on fixed meshes, 
see Section~\ref{sec:DGscheme} and 
\cite{Dumbser2008,GassnerDumbserMunz,DumbserNSE,DGLimiter1,DGLimiter2}. 
Both schemes are based on the ADER approach \cite{schwartzkopff,toro3,titarevtoro,Toro:2006a,Dumbser2009} and solve
general hyperbolic systems of PDEs, possibly with stiff source terms and non-conservative products 
\cite{DumbserEnauxToro,ADERNC,HidalgoDumbser,LagrangeNC,ALEMOOD2}.
The limiter in the ALE scheme is based on nonlinear WENO reconstruction  
\cite{shu1,shu_efficient_weno,balsara,DumbserKaeser07,Lagrange2D,Semplice2016,ADER_CWENO}, 
while a novel \textit{a posteriori} subcell FV limiter based on the MOOD approach \cite{DGLimiter1,DGLimiter3,ALEDG} is employed for our high order ADER-DG schemes. 

Both numerical schemes are in principle of arbitrary order of accuracy in space and time.
For the Wilkins model, all terms in the constitutive equation (\ref{vs.elast2})$ _2 $ are non-conservative. Also the term involving the curl of $\A$  
in Eqn. (\ref{vs.elast2})$ _1 $ of the \GPR model $ \vv \cdot (\nabla \A - \nabla \A^T)$ is non-conservative. All these non-conservative terms are consequently 
treated with the path-conservative approach of Castro and Par\'es, see  
\cite{Castro2006,Pares2006,Castro2007,Castro2008}.  
An important difference between the two models is the way the plasticity is dealt with.
At the end of one timestep the plastic threshold and radial return is computed for the Wilkins model.
Contrarily, for the \GPR model, the solution at $t^{n+1}$ is not further modified by the 
scheme, because the plastic threshold is already embedded in the source term.

Although all parts of those schemes have already been described thoroughly in the aforementioned 
references, the next subsections recall the main ingredients of their design.

\subsection{Direct Arbitrary-Lagrangian-Eulerian ADER-WENO finite volume method} \label{sec:ALEscheme}

This numerical method is a Finite Volume scheme on moving meshes working on triangles or 
tetrahedra. 
As such data are piecewise constants, that are the mean values of the conserved 
 variables within a control volume.
The mesh, constituted of moving cells, is described by the position of the vertices and their 
connectivity.
We assume that the displacement of the mesh is always linear and continuous. 
Therefore a simplex cell remains simplex during its motion even if compression, dilatation, 
rotation and translation may occur.

The numerical scheme is built following the strategy described in 
\cite{Lagrange3D,LagrangeNC,LagrangeQF,WB_PhD} which can be summarized as follows.
The system of PDEs is integrated over a space-time control volume 
between time $t^n$ and $t^{n+1}$, knowing the cell averages at $t^n$ in the neighbor cells.
The ADER method requires the knowledge of a high-order piecewise polynomial reconstruction 
of the data at time $t^n$, which is obtained by a high order nonlinear WENO reconstruction. 
This reconstruction is further evolved in time to construct the so-called space-time predictor polynomial in each cell \cite{DumbserEnauxToro,Dumbser2008,Dumbser2009,HidalgoDumbser}. 
In order to deal with algebraic source terms we use a local space-time discontinuous 
Galerkin predictor, which allows to take into account also possible stiffness. 
As already mentioned before, non-conservative products are dealt with a path 
conservative approach, see \cite{Castro2006,Pares2006,Castro2008,DumbserEnauxToro,ADERNC,HidalgoDumbser,ALEMOOD2} 
for details.

Then the space-time predictor polynomials are employed to feed a suitable numerical flux 
function based on an exact or approximate Riemann solver at the element boundaries within 
each finite volume step. The numerical flux takes into account the interactions between 
neighbor space-time control volumes, see for instance \cite{WB_PhD,ALEDG}.   
The corrector step of the ADER approach is obtained by directly integrating a weak  
form of the governing PDE in space and time at the aid of the predictor. 

Moreover, in the current context of this direct ALE scheme the displacement velocity is enforced to 
be close to the computed material  velocity. Indeed a vertex is displaced by a weighted average of 
the material velocities of the neighbor cells. 
We are referring to this direct ALE displacement field as being 'quasi-Lagrangian'. 
The time step is restricted by a classical CFL-type stability condition for explicit schemes,  
which is common for the two models \cite{Lagrange3D,LagrangeNC}. 

\subsection{ADER-DG scheme with \textit{a posteriori} subcell FV limiter} \label{sec:DGscheme}
The ADER-DG scheme is implemented on quadrilateral meshes following \cite{Dumbser2014}. 
More precisely it is an unlimited one-step ADER-DG scheme. As previously described, 
within the ADER approach a local space-time predictor is computed, starting from the 
 known piecewise polynomial $\mathbb{P}_k$ ($k>0$) data representation of the 
 underlying state variables.
As before, a local space-time discontinuous Galerkin method is used for the construction 
of an element-local predictor solution of the PDE in the small, hence neglecting
the influence of neighbor elements. 
This predictor solution is subsequently inserted into the corrector
step, which then provides the appropriate coupling between neighbor elements
via a numerical flux function, which in this paper we choose to be of the Rusanov type. 
Likewise for the ALE scheme, a path-conservative jump term for the discretization of the 
non-conservative product is used, see \cite{DumbserEnauxToro,ADERNC,HidalgoDumbser}.
The obtained ADER-DG scheme is of $(k+1)$th order of accuracy in space and time, and, so far has no
embedded limiting mechanism to damp spurious oscillations.
Recently a novel family of \textit{a posteriori} subcell FV based limiters have been designed in  
\cite{DGLimiter1,DGLimiter2,Zanotti2015b,DGLimiter3,ALEDG}, based on the 
MOOD framework put forward in \cite{CDL1,CDL2,CDL3,ADER_MOOD_14}.  
This limiter uses the correspondence between a $\mathbb{P}_k$ DG polynomial and 
its projection onto an appropriate number of subcells. Those projections are exactly the mean value 
subcell-based data used by a FV scheme which would act on subcells. 
The troubled cells in the DG candidate solution, that is computed by the ADER-DG unlimited scheme at 
$t^{n+1}$, are flagged as not acceptable according to
some user-defined or developer-given physical and numerical detection criteria. 
This step is similar to other troubled cell detectors for DG limiters 
\cite{cbs3,cbs4,Qiu_2005,balsara2007,Zhu_hadap_2013}.
The candidate DG solution in these troubled cells is then discarded and recomputed 
starting again from the previous time level $t^n$, but this time using a more robust Finite Volume 
(FV)
scheme operating on a sufficiently large number of sub-cells, as to conserve the intrinsic subcell
resolution capability of a DG scheme. 
The fact that the solution is locally recomputed
by re-starting from a valid discrete solution at $t^n$ is radically different from
existing classical DG limiters.


\section{Numerical experiments} \label{sec:numerics}

In this section we gather the numerical experiments carried out in order to:
\begin{enumerate}
\item 
	Show that the hypo-elastic model of Wilkins can be simulated within our high order 
	Eulerian and direct ALE framework by means	of some sanity test problems;
\item
	Demonstrate that the hyperelastic \GPR model can be used to solve elasto-plastic situations 
	usually dealt with
	the hypoelastic model of Wilkins in a pure Lagrangian or indirect ALE framework;
\item
	Compare the two models under the same numerical framework to analyze the possible differences 
	for simplified
	elasto-plastic simulations. These differences must result from the models themselves, hence 
	directly assess their
	prediction capabilities.
\end{enumerate}
Recall that the hypo-elastic model of Wilkins as well as the hyper-elastic \GPR model 
are solved by the very same numerical schemes, under the same platform, leading thus to a 
relatively fair comparison. 
Here, we systematically employ either an ADER-WENO finite volume scheme on moving meshes, 
or a high order Eulerian ADER-DG scheme with finite volume subcell limiter. Both schemes 
are nominally third or fourth order accurate.
The numerical experiments are made as simple as possible because the goal is to illustrate the 
differences, and not to verify the models against laboratory experiments. This important point 
is however planed as a future work. In table~\ref{tab:test}, we recall the test suite employed 
in this paper.
%
\begin{table}[ht!]
  \begin{center}
    \begin{tabular}{|c|c|c|c|c|c|}
    	\hline
      \multirow{2}{*}{\textbf{\#} }   & \multirow{2}{*}{\textbf{Problems} }      & 
      \multirow{2}{*}{\textbf{Dimensions}} & \multirow{2}{*}{\textbf{Regime}} & 
      \multirow{2}{*}{\textbf{Purpose/goal}} & \multirow{2}{*}{\textbf{Diagnostics}} \\
      & & & & & \\
    	\hline
      	\ref{ssec:Piston}	&Elasto-plastic piston 		& 1D 	&Lagrangian		& Sanity 		&  
      	Convergence/accuracy \\
      	\hline
      	\ref{ssec:Be_plate}&Beryllium bending plate 	& 2D	&quasi-Lagrangian& Simulation & 
      	Convergence/accuracy \\
        \ref{ssec:Shell}	&Elasto-plastic shell		& 2D 	&quasi-Lagrangian& Validation	& 
        Convergence/accuracy \\
      	\ref{ssec:Taylor}	&Taylor rod 				& 2D	&quasi-Lagrangian& Verification & 
      	Model comparison \\
    	\hline
        \ref{ssec:shear} 	& Shear layer		   		& 1D 	& Eulerian & Elasticity			& 
        Model comparison\\
        \ref{ssec:rotor} 	& Solid Rotor 		   		& 2D 	& Eulerian & Elasticity			& 
        Model comparison\\
        \hline
      \end{tabular}
    \caption{ Test suite employed to compare Wilkins' and \GPR models.}
    \label{tab:test}
  \end{center}
\end{table}
Physical units are based on the $[m,kg,s]$ unit system and the \MG EOS 
\eqref{eq:mie_gruneisen_eos} is used for solids as usually done 
\cite{Kluth2010,Burton2013,Maire2013}.  
For the last two tests we consider a simplified material under the stiffened gas EOS.
In Table \ref{tab:dataEOS}, we report some mechanical constants as well as the parameters needed in 
the \MG EOS for the materials considered in the test cases for solid mechanics 
presented in this paper.  
\begin{table}
\caption{Material parameters: reference density $\rho_0$, reference (atmospheric) pressure $p_0$, 
adiabatic sound speed $c_0$, shear wave speed $c_s$, Yield stress $\sigma_0$ and the coefficients 
$\Gamma_0$ and $s$ appearing in the Mie-Gr{\"u}neisen equation of state 
\eqref{eq:mie_gruneisen_eos}.}
\centering
\begin{tabular}{|c||ccccccc|cc|}
\hline
			& $\rho_0$	& $p_0$ 	& $c_0$ 	& $c_s$ 	& $\sigma_0$ 	& $\Gamma_0$ 	& s & 
			$\mu$ (Pa) & $\sigma_Y$ (Pa)\\ 
\hline
Copper     	& $8.930$ 	& $0.0$ 	& $0.394$ & $0.225$ & $0.004$  		& 
$2.00$ 	& $1.480$ & $45 \times 10^9$ & $90\times 10^6$\\
Beryllium  	& $1.845$ 	& $0.0$ 	& $1.287$ & $0.905$ & $1$			& $1.11$ 	& $1.124$ & 
$151.11\times 10^9$ & $330 \times 10^6$ \\
Aluminum 	& $2.785$ 	& $0.0$ 	& $0.533$ & $0.305$ & $0.003$		& $2.00$ 	& $1.338$ &  
$27.6\times 10^9$ & $300\times 10^9$ \\
\hline
\end{tabular}
\label{tab:dataEOS}
\end{table}
%
The models are run with structured  quadrangular mesh or unstructured meshes made of simplices.
However, the very same mesh is systematically employed for a given simulation when comparing the 
two 
models. 

 \textbf{Initialization of both models.}
The initialization for Wilkins model consists in providing at $t=0$, a computational domain 
$\Omega$ paved with 
a mesh $\mathcal{M}$, the material parameters in each cell
$(\rho_0, p_0, c_0, c_s, \sigma_0, \Gamma_0, s, \mu)$, the conservative state vector in each 
cell,
$\bm{Q}=(\rho, \rho \vv, \rho E )$ and the deviatoric part of the stress tensor 
$\bm{\sigma}$,
generally set to $\bm{0}$.
By means of the EOS we can compute the pressure $p$ and deduce the Cauchy stress tensor 
$\bm{T}$.
Boundary conditions  and a final time $t_{\text{final}}$ must  be prescribed.

Starting from the same data, we can initialize the hyper-elastic \GPR model as follows.
Material parameters and state variables are kept alike, $\A$ is initialized with $\II$.
The shear sound speed $c_s$ is computed thanks to 
$
\css = \frac{\mu}{\rho} | \A^\transpose  \A |^{-2/3},
$
while the relaxation time $\tau$ is given by
$\tau = \tau_0 \left(  \frac{\sigma_Y}{\sigma}  \right)^n$, 
with $n=10$, and
$ \tau_0=10^{-5} t_{\text{final}}$,
where $\sigma=\sqrt{\frac32 \tr(\dev{\bm{\sigma}}^2)}$, and $ \dev{\ssigma} = \ssigma - 
\frac13\tr(\ssigma)\II$.
Otherwise noticed, this is the way we have systematically initialized the \GPR model.

 \textbf{Visualization.} 
Because we can express $\bm{\sigma}$ as a function of $\A$, $\rho$ and $\cs$ then we use the 
components of $\bm{\sigma}$ and its norm for visualization purposes. Primitive variables may also 
be displayed
(density, pressure, velocity component/norm).
Moreover, plastic regions for which $\sigma>\sigma_Y$ may be emphasized.
When appropriate the final mesh may also be plotted.

\subsection{Elasto-plastic piston} \label{ssec:Piston}
This test is characterized by a homogeneous stress-free material at rest compressed by a piston or 
explosively (instantaneously) generated wave at a material boundary, 
leading to a two-wave structure for moderate stresses (exceeding the Hugoniot elastic limit) with 
the first wave being an elastic precursor followed by a 
plastic wave \cite{zeldovich67,Clayton2018}. It is experimentally observed (\eg 
see \cite{Johnson1970a,Gurrutxaga-Lerma2015}) that the amplitude of 
the elastic precursor decreases from the initial (impact) stress to an often steady minimum value 
which can be thought as a decrease in the transient Hugoniot elastic limit of the material. In the 
plastic wave, relaxation of the tangential stress occurs due to the structural rearrangements, and 
as a result the uniaxial deformed 
state is transformed into a triaxial stress state corresponding to the yield surface. Because 
the process of structural rearrangements has a finite characteristic time scale, the width of the 
plastic wave is non-zero but has a finite thickness.

The analytical solution to the Wilkins model can be derived \cite{Maire2013,Wayne2016}. This 
solution has the expected two-wave elasto-plastic structure presented by two discontinuous waves 
which, however, do not possess the attenuating behavior of the elastic precursor and the finite 
width of the plastic wavefront due to the rate-independent character of the model. Yet the Wilkins 
model 
provides 
a reasonable approximation in many practical situations. The solution to the \GPR model, on the 
other hand, 
is genuinely time-dependent and its analytical expression is unknown. However, an 
analytical expression of the asymptotic solution can be obtained when the waves are infinitely 
far from the 
boundary and the solution reaches a self-similar steady 
structure~\cite{Godunov1976,GodRom2003,Romenskii2007} with discontinuous precursor and smooth 
plastic wave. 
In the transient near-boundary zone, the solution to the \GPR model has been studied 
numerically~\cite{MerzhResnyan1985,Barton2012} and was shown to have the 
experimentally observed features of the elastic precursor and the plastic wave due to the model
intrinsic rate-dependent character.

\begin{figure}[!htbp]
	\begin{center}
		\begin{tabular}{cc} 
			\includegraphics[draft=false,width=0.47\textwidth]{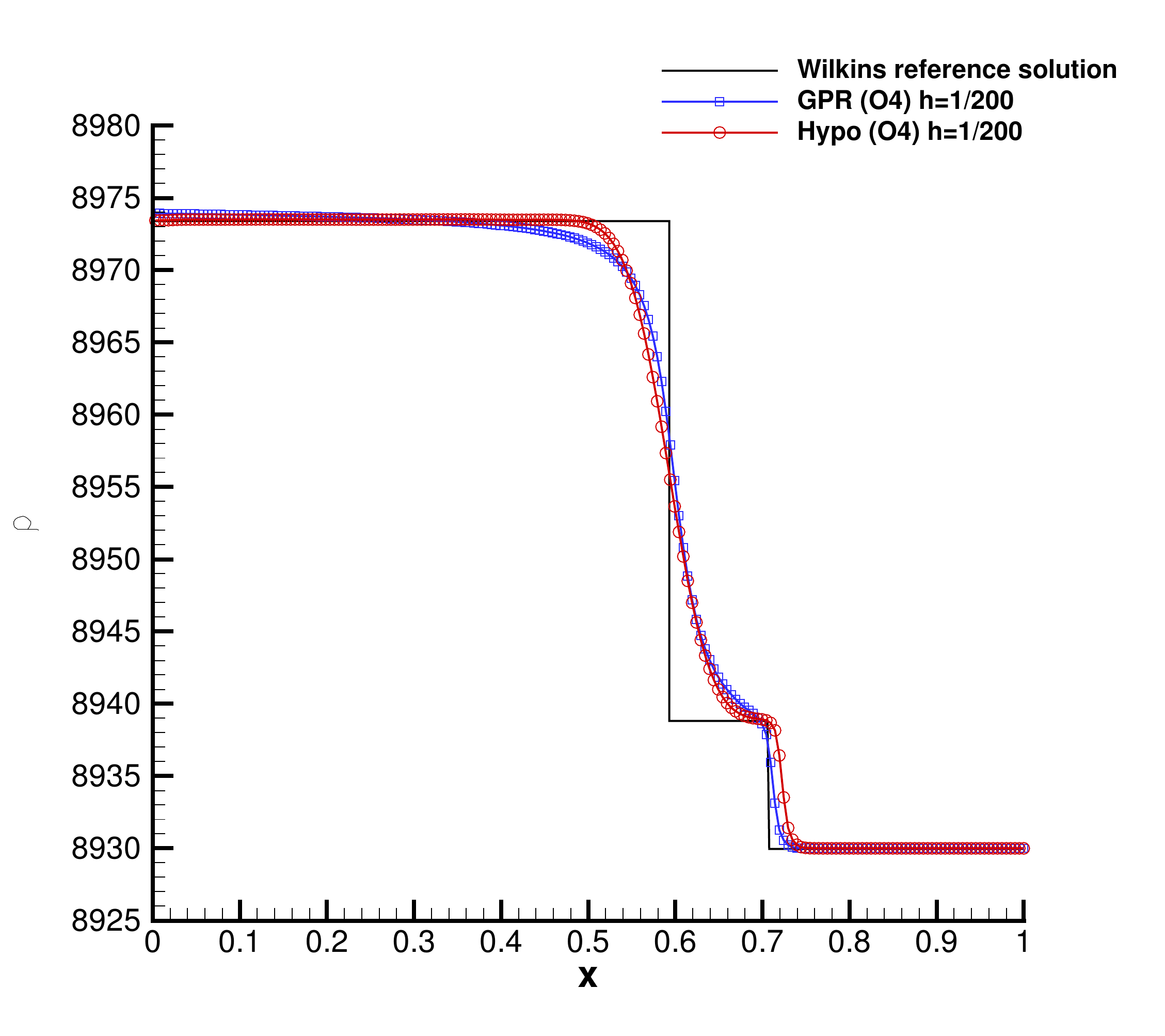}
			  &           
			\includegraphics[draft=false,width=0.47\textwidth]{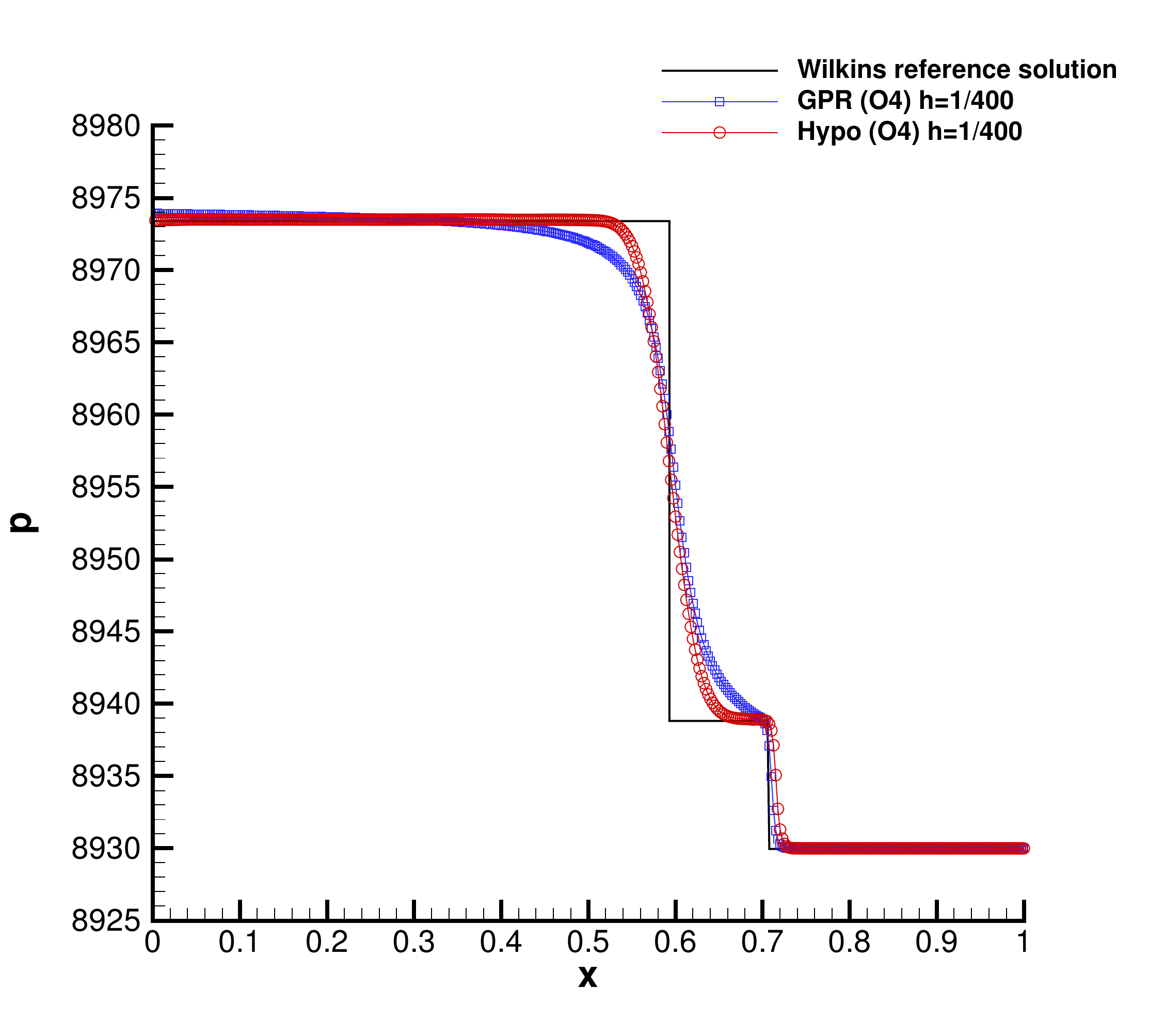}
			 \\
			\includegraphics[draft=false,width=0.47\textwidth]{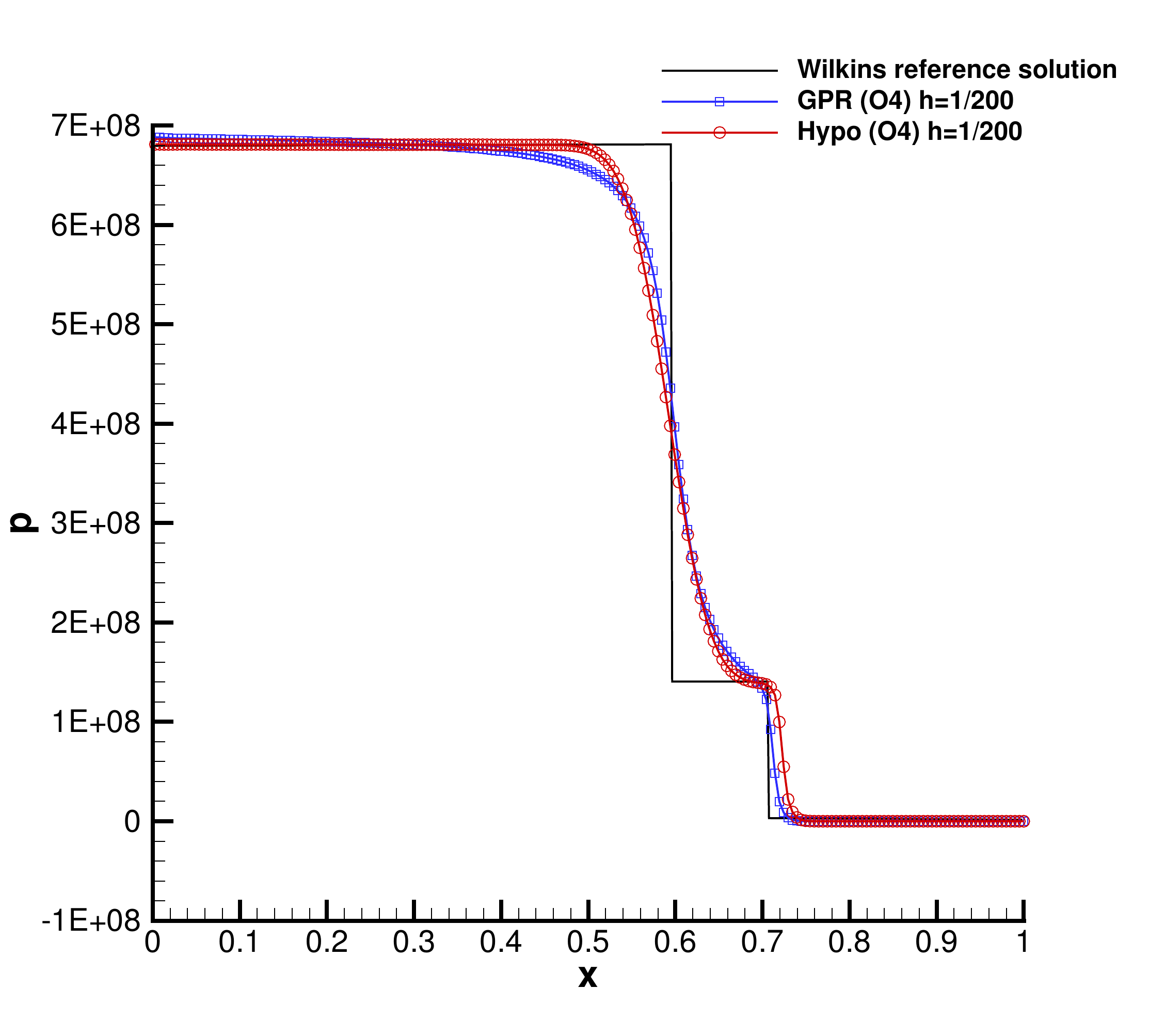}
			  &           
			\includegraphics[draft=false,width=0.47\textwidth]{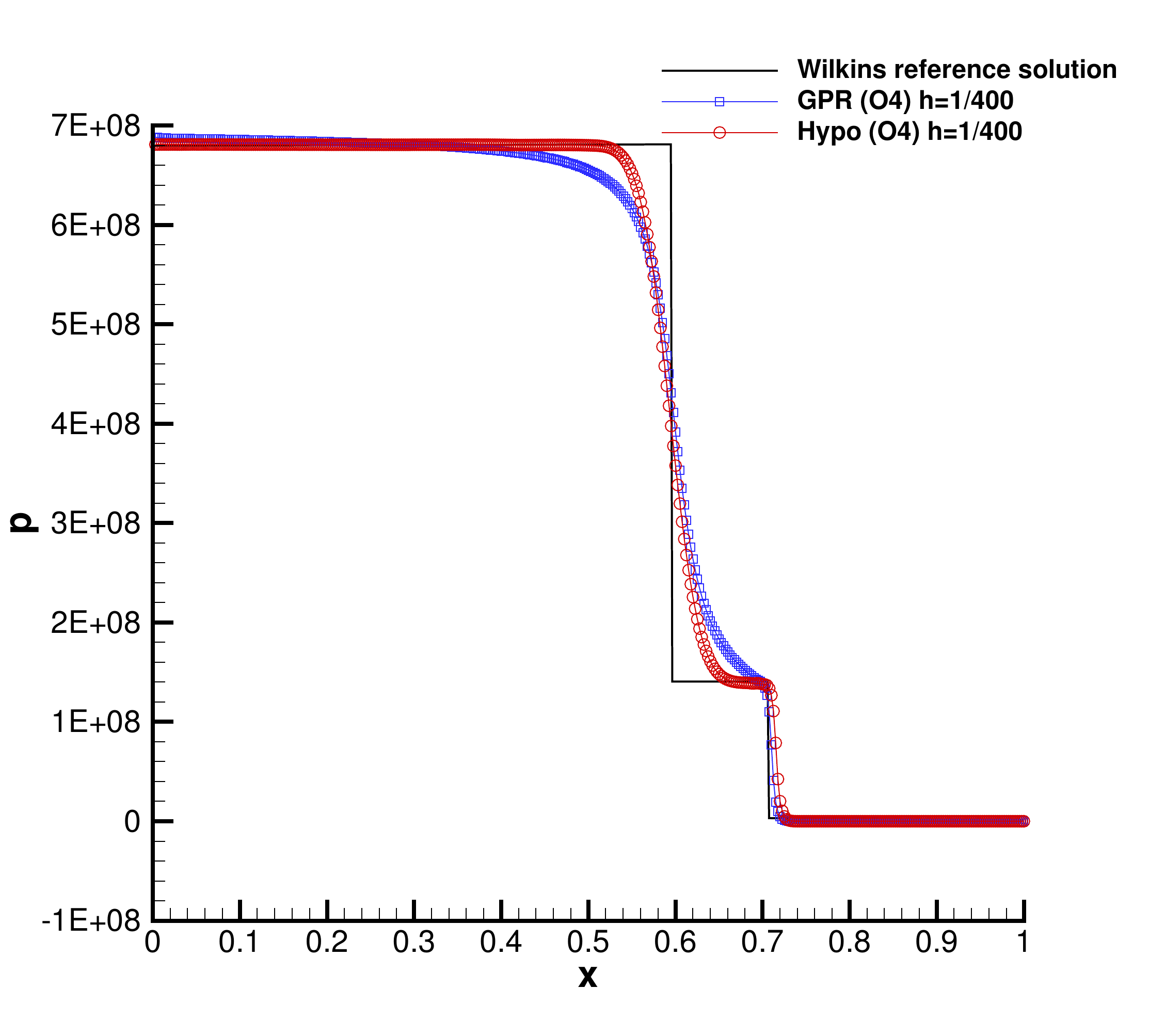}
			 \\
		\end{tabular} 
		\caption{Comparison of the numerical solutions to the elastoplastic piston problem for the 
		Wilkins and \GPR model obtained with ALE ADER-WENO fourth order schemes. The density 
		distribution (top row) and pressure distribution (bottom row) at final time 
		$t_{\text{final}}=150\cdot 10^{-6}$ are shown along side with the exact solution (solid 
		lines) of the Wilkins model. Each plot 
		shows the numerical results computed with the \GPR model (squares) and the Wilkins model 
		(circles) for two different computational grids with characteristic mesh size of $h=1/200$ 
		and $h=1/400$.} 
		\label{fig.EPP2D}
	\end{center}
\end{figure}

\begin{figure}[!htbp]
	\begin{center}
		\begin{tabular}{cc} 
			\includegraphics[draft=false,width=0.47\textwidth]{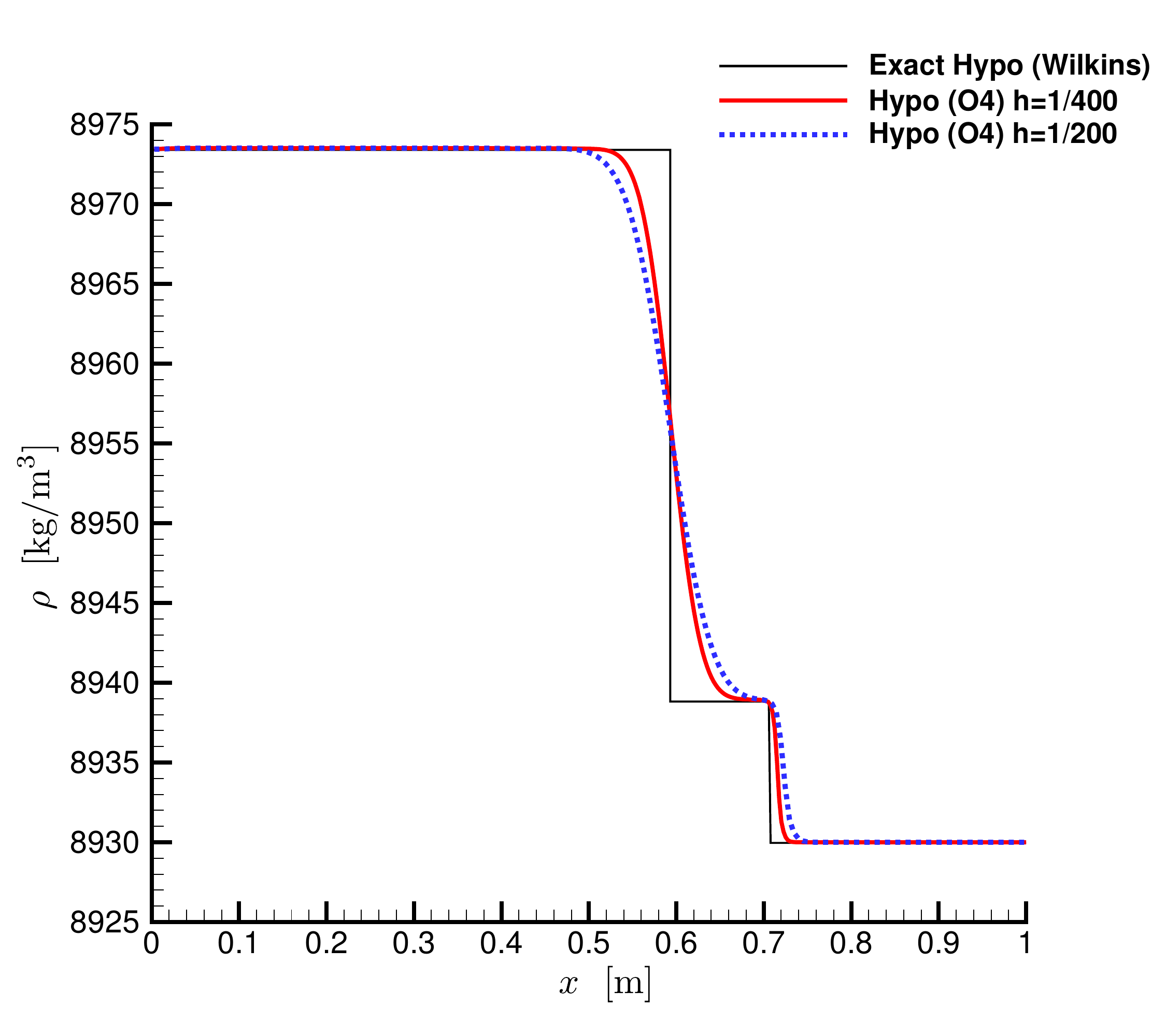}		
			  &           
	\includegraphics[draft=false,width=0.47\textwidth]{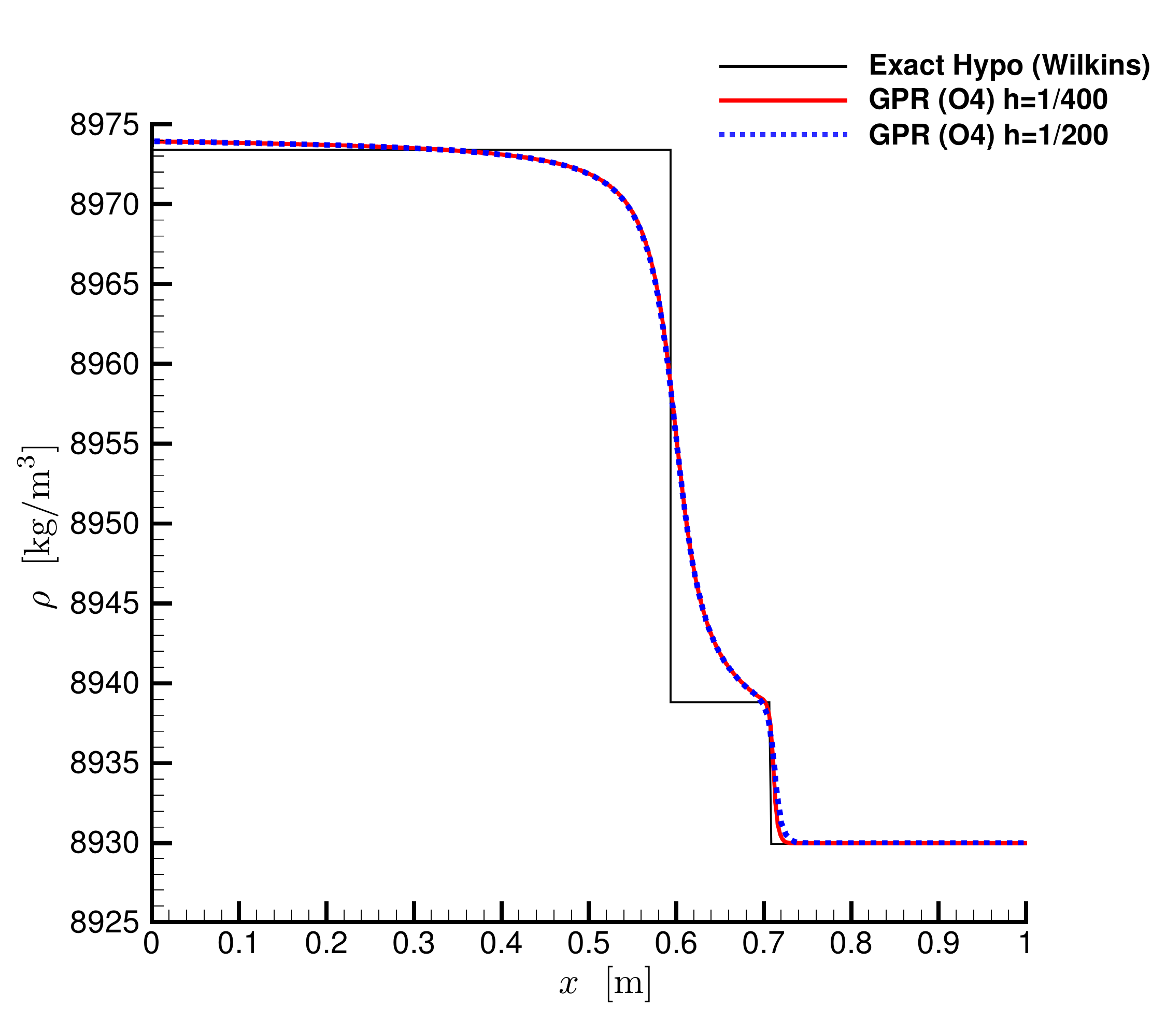}
		\end{tabular} 
		\caption{Comparison of the solution convergence rate for elastoplastic piston problem of  
		the hypoelastic-type Wilkins model
		(left) and the hyperelastic-type \GPR model (right). Because the plastic wave is 
		continuous in the relaxation \GPR model, the solutions to the \GPR model obtained on a 
		coarse and fine meshes with 4-th order method are almost indistinguishable in the plastic 
		wave.} 
		\label{fig.EPP2D.400vs200}
	\end{center}
\end{figure}

\begin{figure}[!htbp]
	\begin{center}
			\includegraphics[draft=false,width=0.47\textwidth]{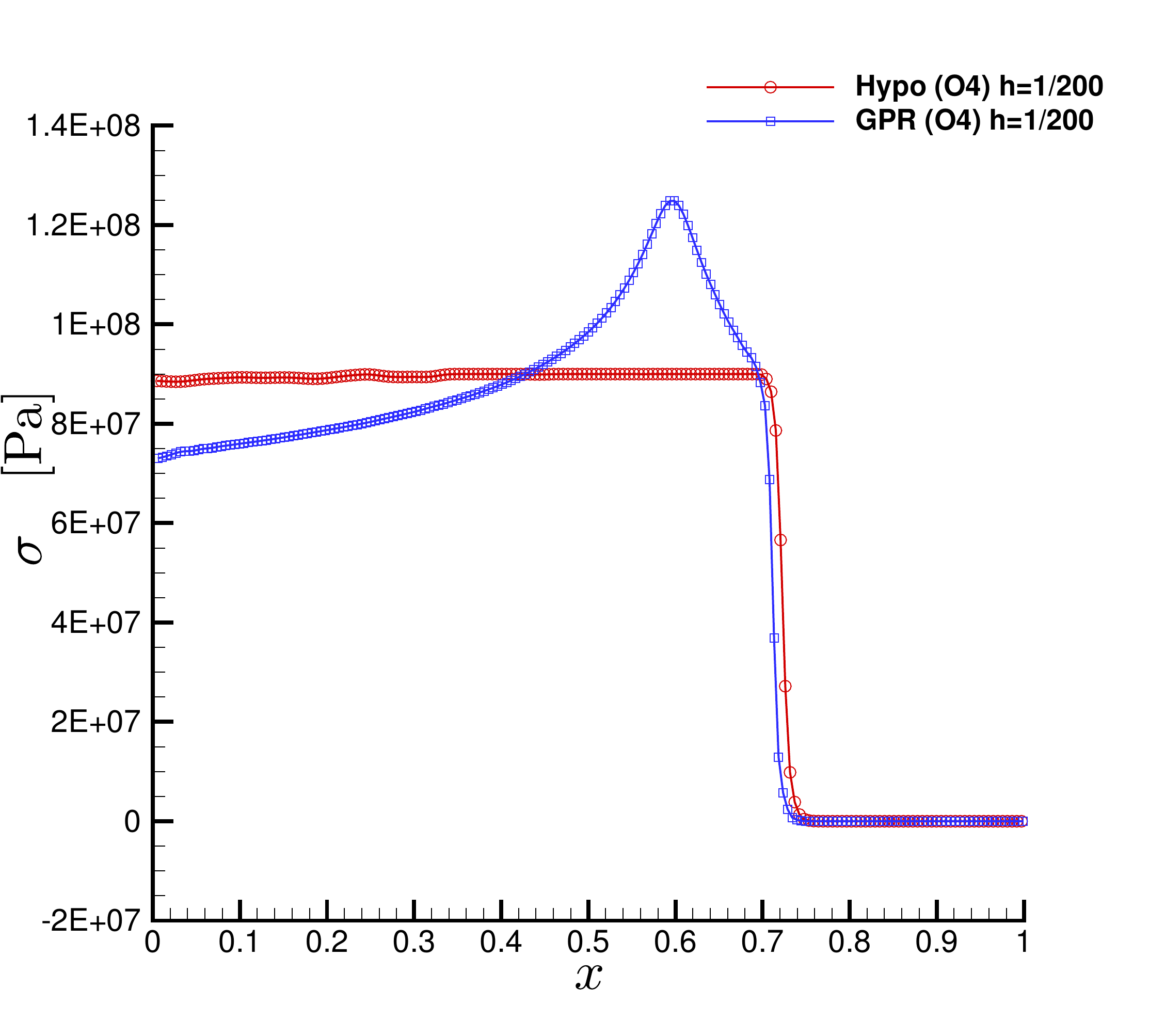}
		\caption{The intensity of tangential stresses $\sigma=\sqrt{3 \tr(\dev{\bm{\sigma}}^2)/2}$  
		for the elastic-plastic piston problem at final time 
		$t_{\text{final}}=150\cdot 10^{-6}$ for Wilkins model and \GPR model.} 
		\label{fig.EPP2Dstress}
	\end{center}
\end{figure}

\begin{figure}[!htbp]
	\begin{center}
		\begin{tabular}{ccc} 
			\includegraphics[draft=false,width=0.37\textwidth]{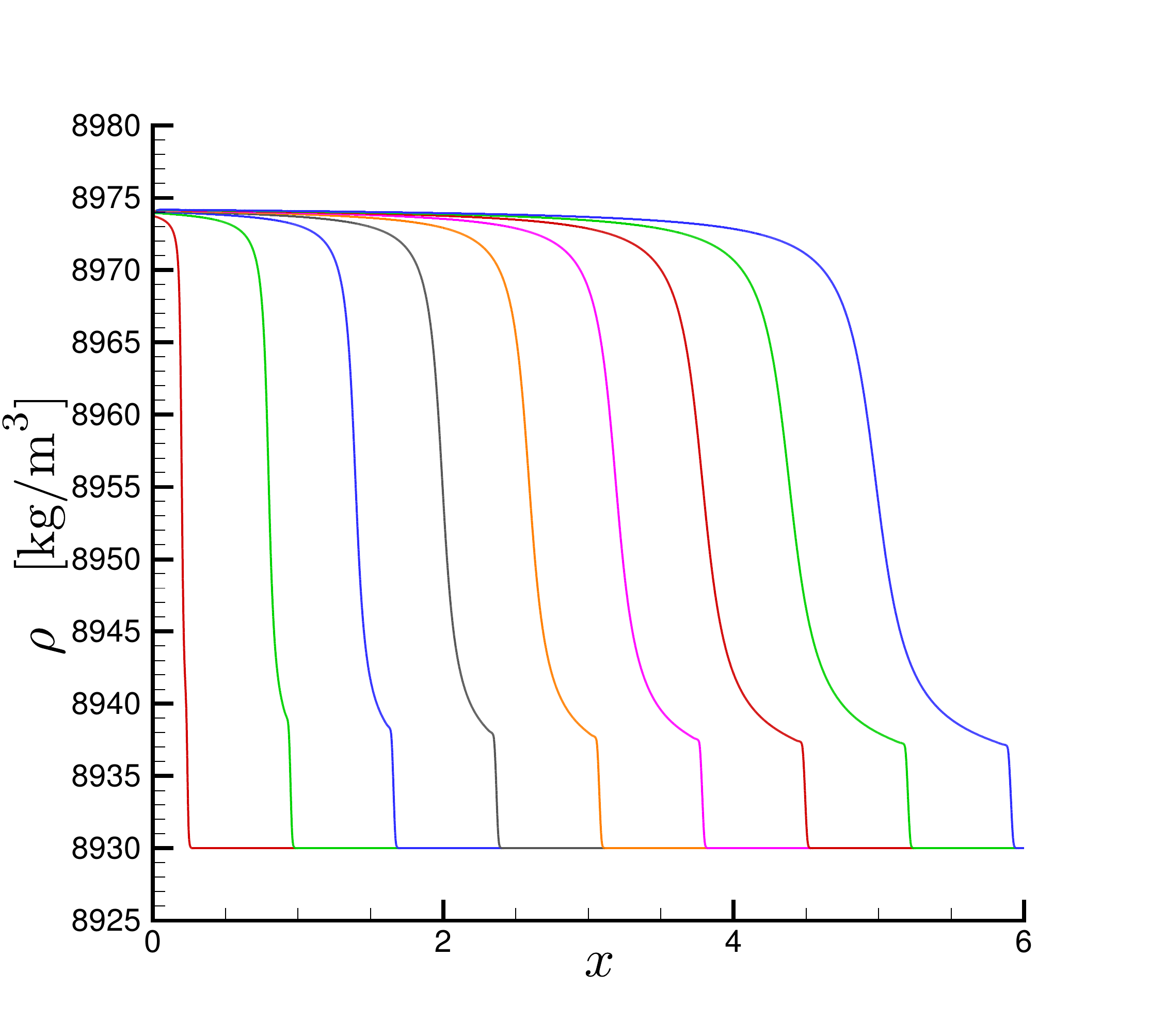} 
			&
			\includegraphics[draft=false,width=0.37\textwidth]{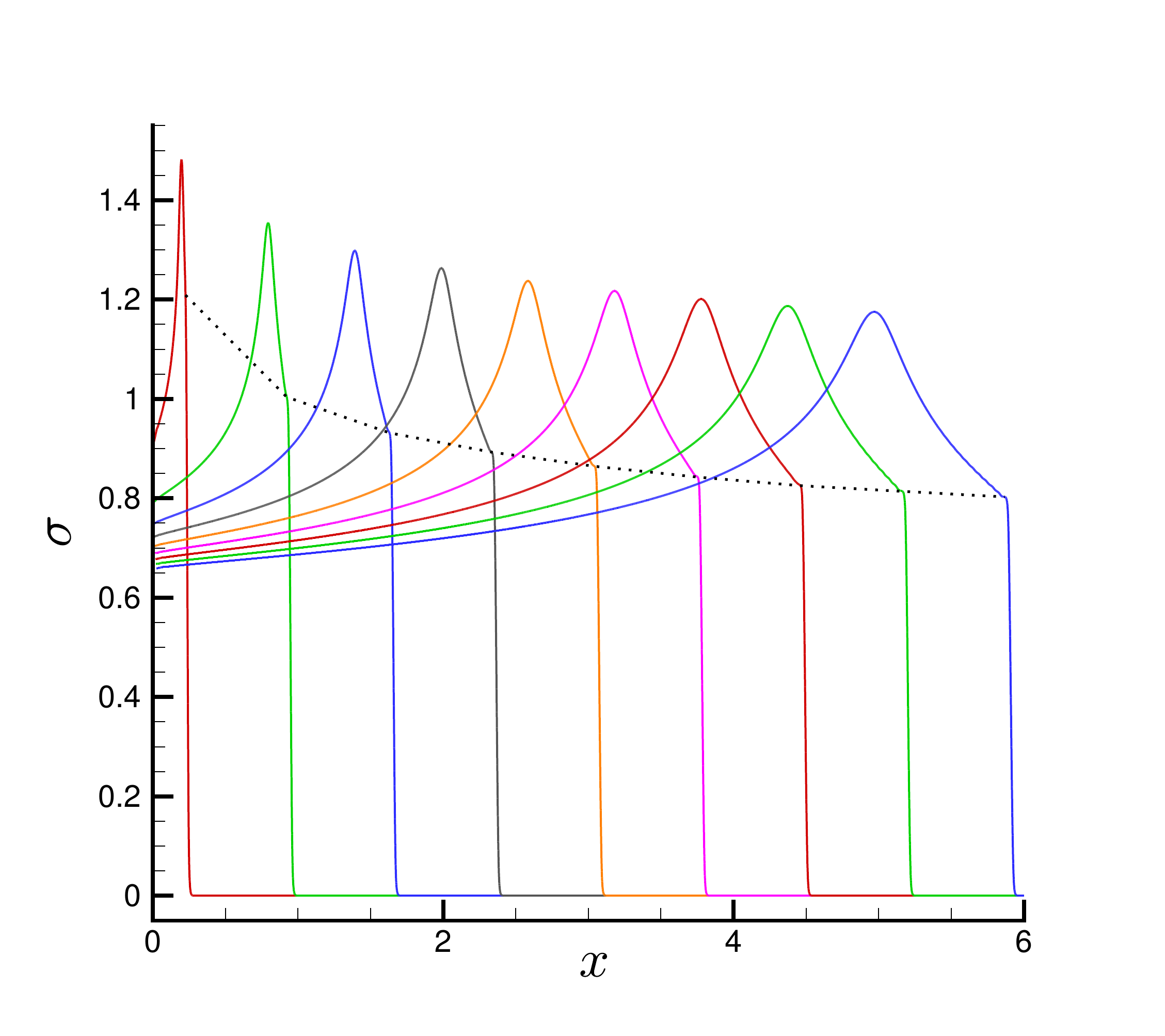} 
		\end{tabular} 
		\caption{Long-time integration of the elastic-plastic piston problem for the \GPR model. An 
		unsteady evolution of the elastic precursor magnitude (dotted line on the right plot) is 
		observed.  The third order accurate scheme on a mesh with 
		the 
		characteristic size of $h=1/200$ was used.} 
		\label{fig.EPP2D-long}
	\end{center}
\end{figure}

The material under consideration in this test case is copper modeled by the \MG EOS with the static 
yield strength set to $\sigma_Y=9\times 10^{-4}$.
The computational domain is $\Omega=[0:1.0]\times [0:0.1]$ and the mesh is made using a 
characteristic 
length $h=1/200$
or $h=1/400$. 
The initial density and pressure correspond to the reference values, see Table~\ref{tab:dataEOS}, 
and the initial velocity field is zero, 
while the distortion is simply set to $\AAA=\II$, and $\tau_0=0.001$ and $n=10$ for the 
hyperelastic \GPR model.
The  piston on the left boundary moves with a horizontal velocity $\mathbf{v}_p=(20,0,0)$.
Fig.~\ref{fig.EPP2D} shows a comparison of the two models on a coarse and a fine mesh as well as
the exact solution to the Wilkins model. 
The elastic precursors are discontinuous in both models. 
Contrarily, the plastic wave is continuous in the \GPR model and 
discontinuous in the Wilkins model. The continuity of the plastic wave in the \GPR model can be also 
seen from the fact that the numerical solution does not show remarkable variation in the plastic 
wave when the mesh is refined, see Fig.~\ref{fig.EPP2D.400vs200}, while the mesh refinement 
steepens 
the plastic wave for Wilkins model, which tends to the discontinuous profile of the exact solution. 
The behavior of the norm of the 
deviatoric stress $\sigma=\sqrt{3 \tr(\dev{\bm{\sigma}}^2)/2}$ is depicted in 
Fig.~\ref{fig.EPP2Dstress}. A very different behavior can be observed. Thus, in the \GPR model it 
is 
allowed that in the plastic wave $ \sigma > \sigma_Y $ while it is forbidden in the Wilkins model. 
From the dislocation dynamics standpoint, the peak in stress in the plastic wave 
is a result of a delay in the dislocation densities and velocities reaching the maximum values, and 
hence the rate of increase of tangential stress momentarily exceeds the decay.
The elastic precursor attenuation is shown in Fig.~\ref{fig.EPP2D-long} which qualitatively 
coincides with the experimentally observed behavior, while a quantitative comparison may require a 
refinement of the model for $ \tau $ and, in particular, of taking into account of the full model 
for $ \tau $~\cite{GodDenisenko1975,Godunov1976a}, including the thermal terms, but more
importantly it may require an accounting for the dislocation dynamics in a more accurate way, \eg 
see \cite{Clayton2018,Barton2012}. A discrepancy in the precursor location given by two models can 
be 
observed, \eg see Fig.~\ref{fig.EPP2Dstress} which can be explained by the fact that the precursor 
velocity is constant in the Wilkins model, while it is not constant in the \GPR model due to the 
precursor attenuation.

\begin{figure}[!htbp]
	\begin{center}
		\begin{tabular}{cc} 
			\includegraphics[draft=false,width=0.4\textwidth]{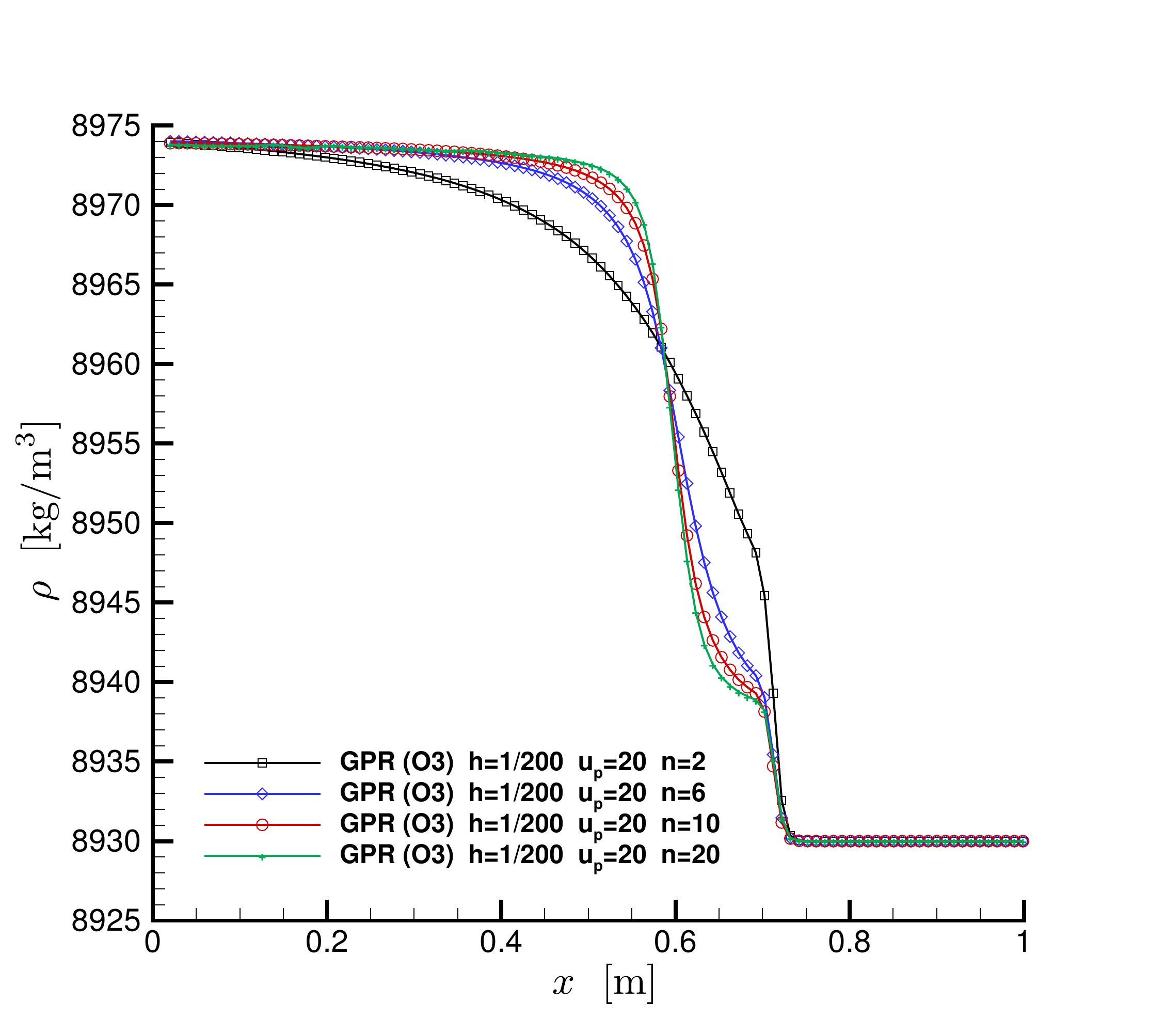}
			&
			
\includegraphics[draft=false,width=0.4\textwidth]{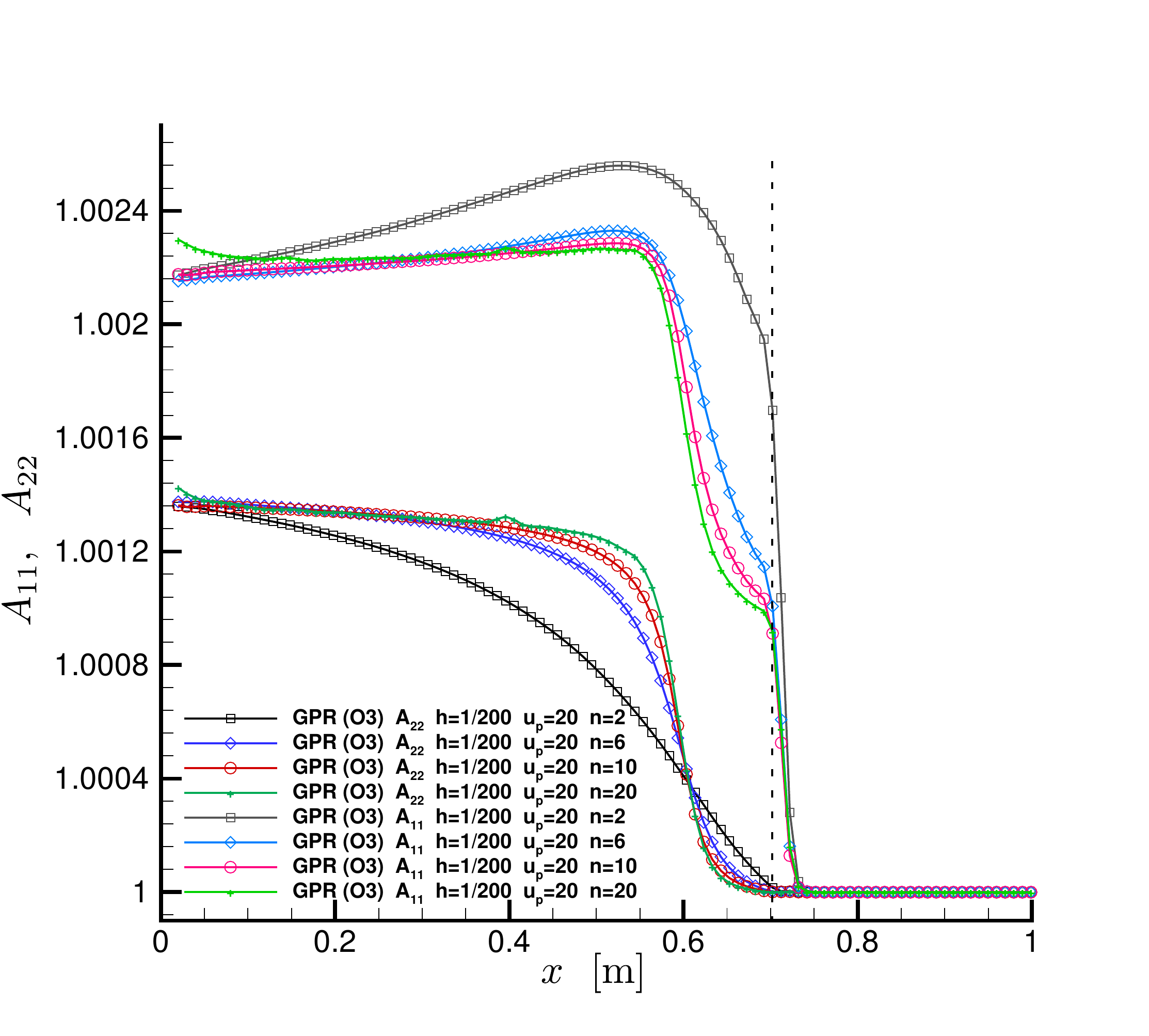}\\
			
\includegraphics[draft=false,width=0.4\textwidth]{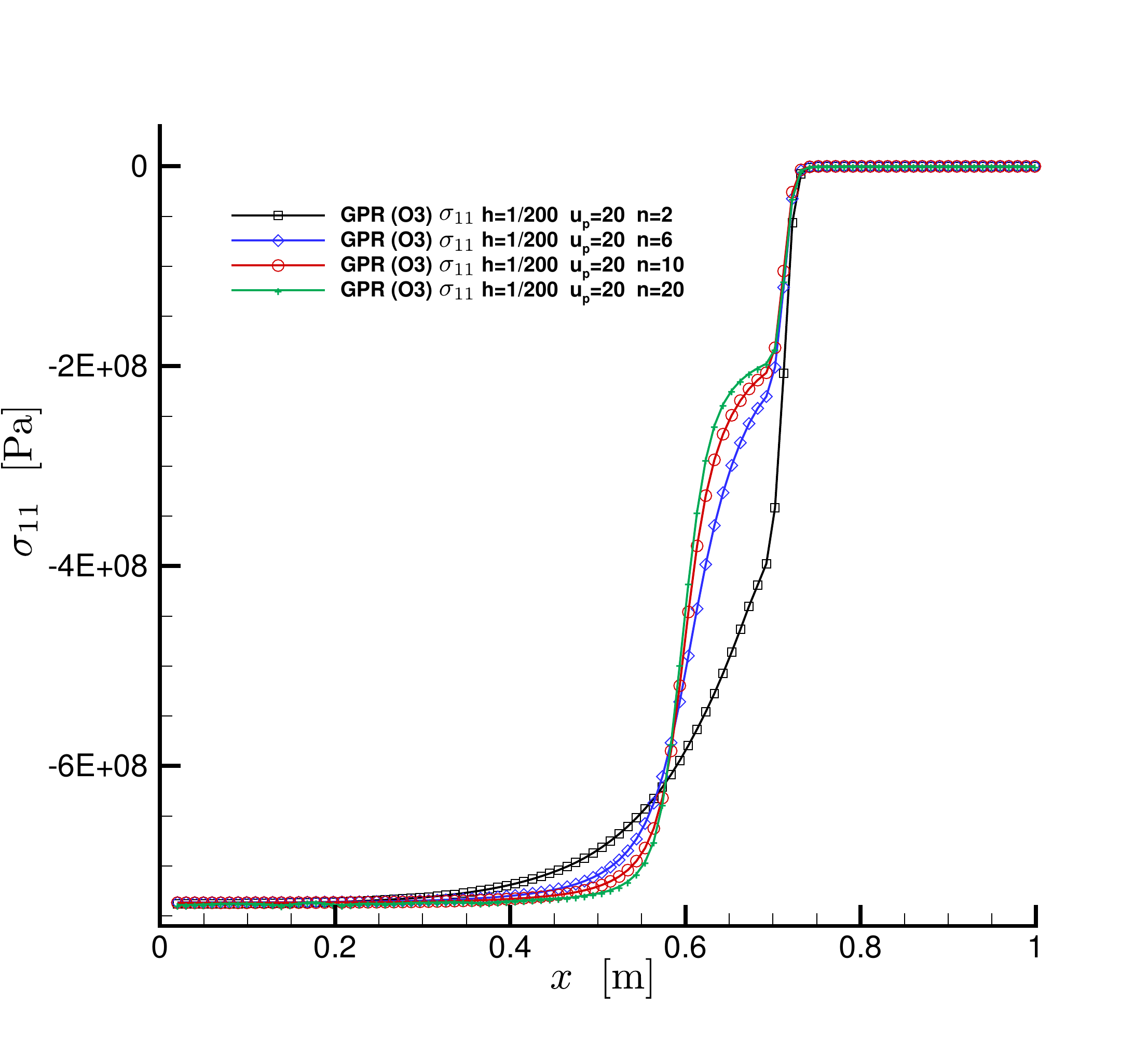}
			&
			\includegraphics[draft=false,width=0.4\textwidth]{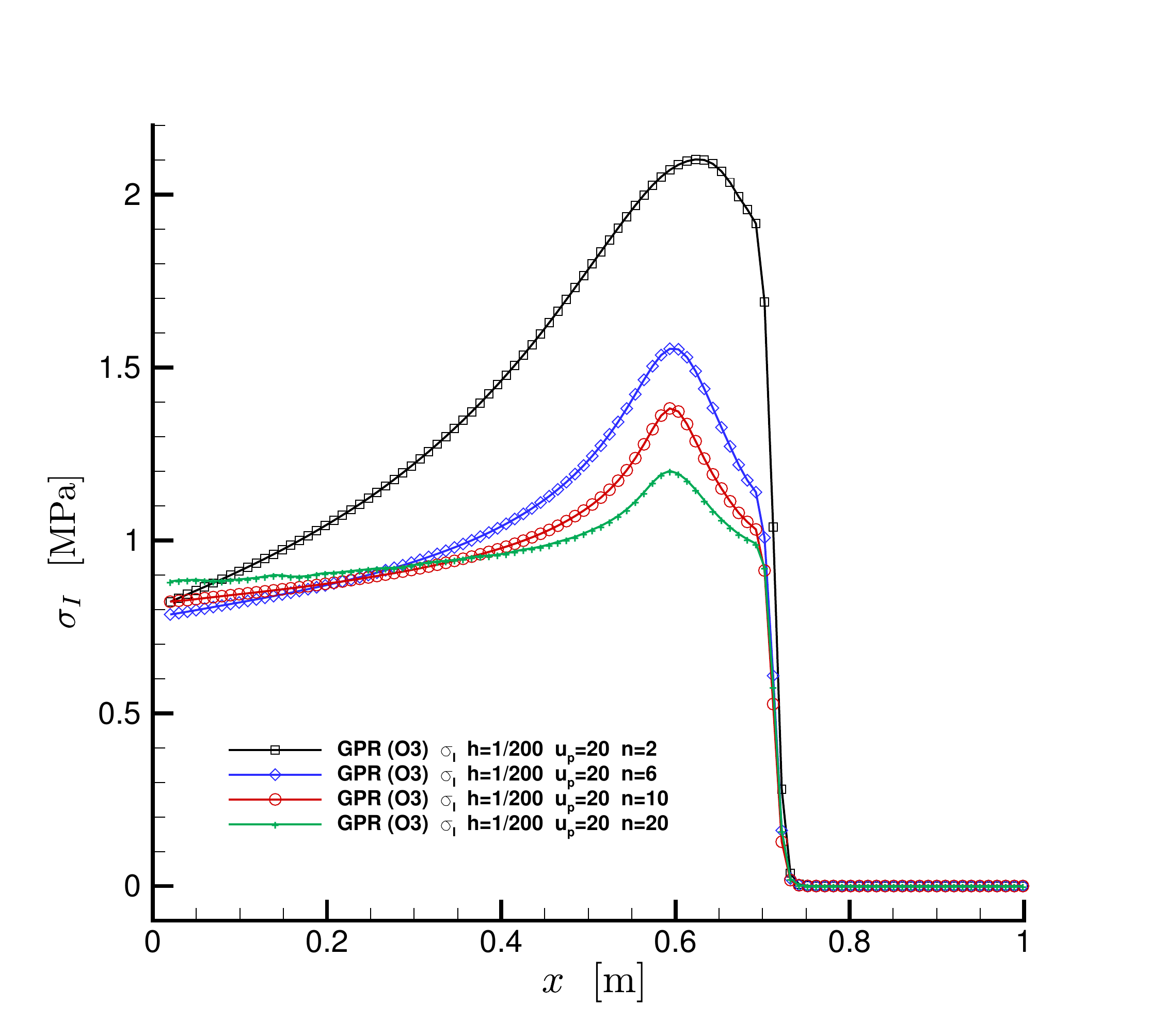}
		\end{tabular} 
		\caption{Elastic-plastic piston problem for the \GPR model run with the same piston 
		velocity 
$u_p=20$ for different values of the parameter $n$. The third order accurate scheme on a mesh with 
the 
characteristic size of $h=1/200$ was used.} 
		\label{fig.EPP2D-comparison-n}
	\end{center}
\end{figure}

From Fig.~\ref{fig.EPP2D-comparison-n}, one can judge about the sensitivity of the dynamic yield 
strength of the \GPR model with respect to different values of the power law index $ n $  from the 
precursor amplitude. 
Such a behavior 
cannot be observed in the rate-independent Wilkins model. Finally, it is important to remark about 
the behavior of the 
$ A_{22} $ component of the distortion field $ \dist{\mu}{i} $. Because the velocity $ v_2 = 0 $, 
the component $ A_{22} $ does not change in the elastic precursor, see 
Fig.~\ref{fig.EPP2D-comparison-n} (top right). However, it changes in the plastic wave due to the 
work of the relaxation source terms which tends to reduce the difference $ A_{11} - A_{22} > 0 $

%
%
%
%

%
%


\subsection{Elastic vibrations of a Beryllium plate} \label{ssec:Be_plate}
\begin{figure}[!htbp]
	\begin{center}
		\begin{tabular}{cc} 
			\includegraphics[draft=false,width=0.47\textwidth]{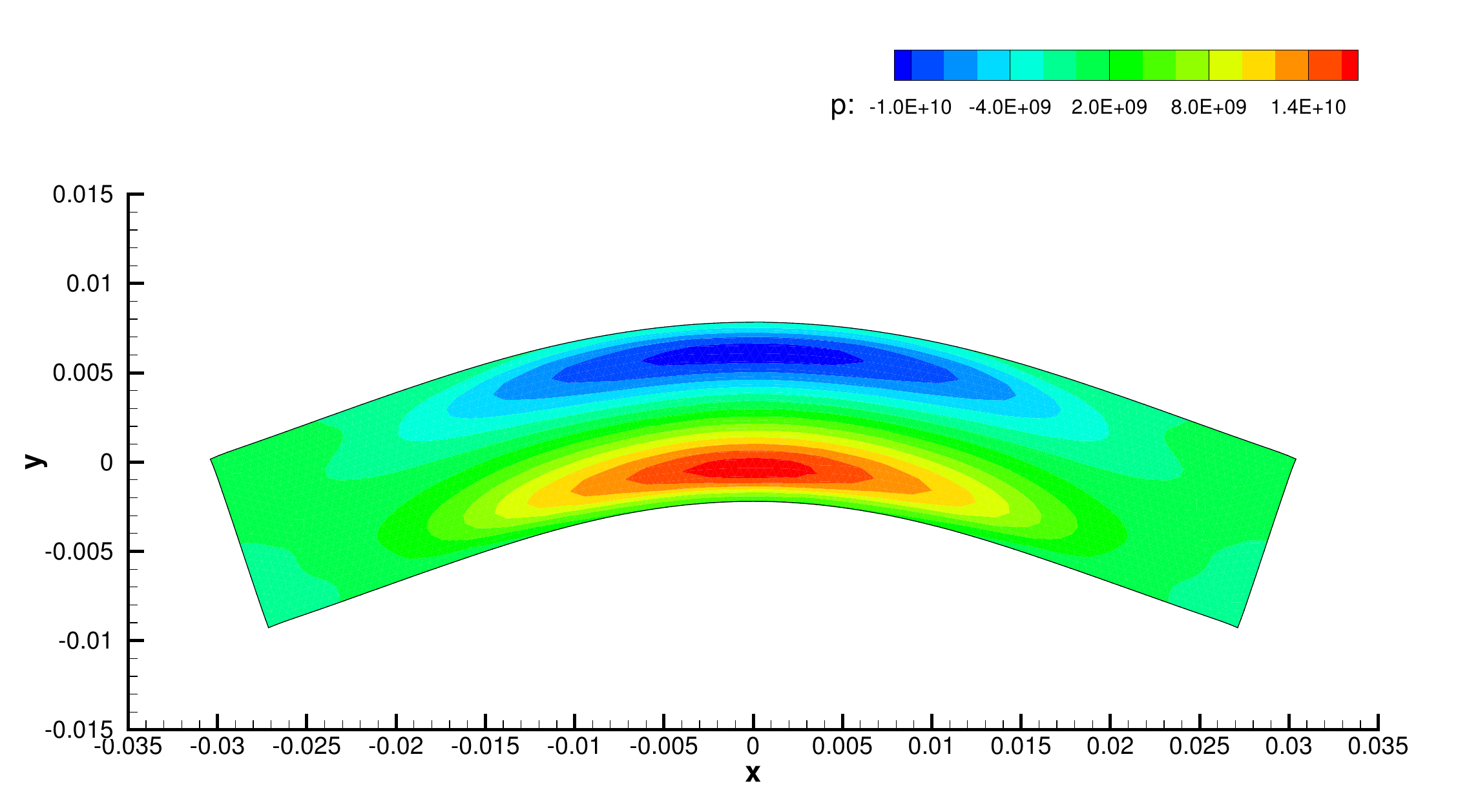}
			  &           
			\includegraphics[draft=false,width=0.47\textwidth]{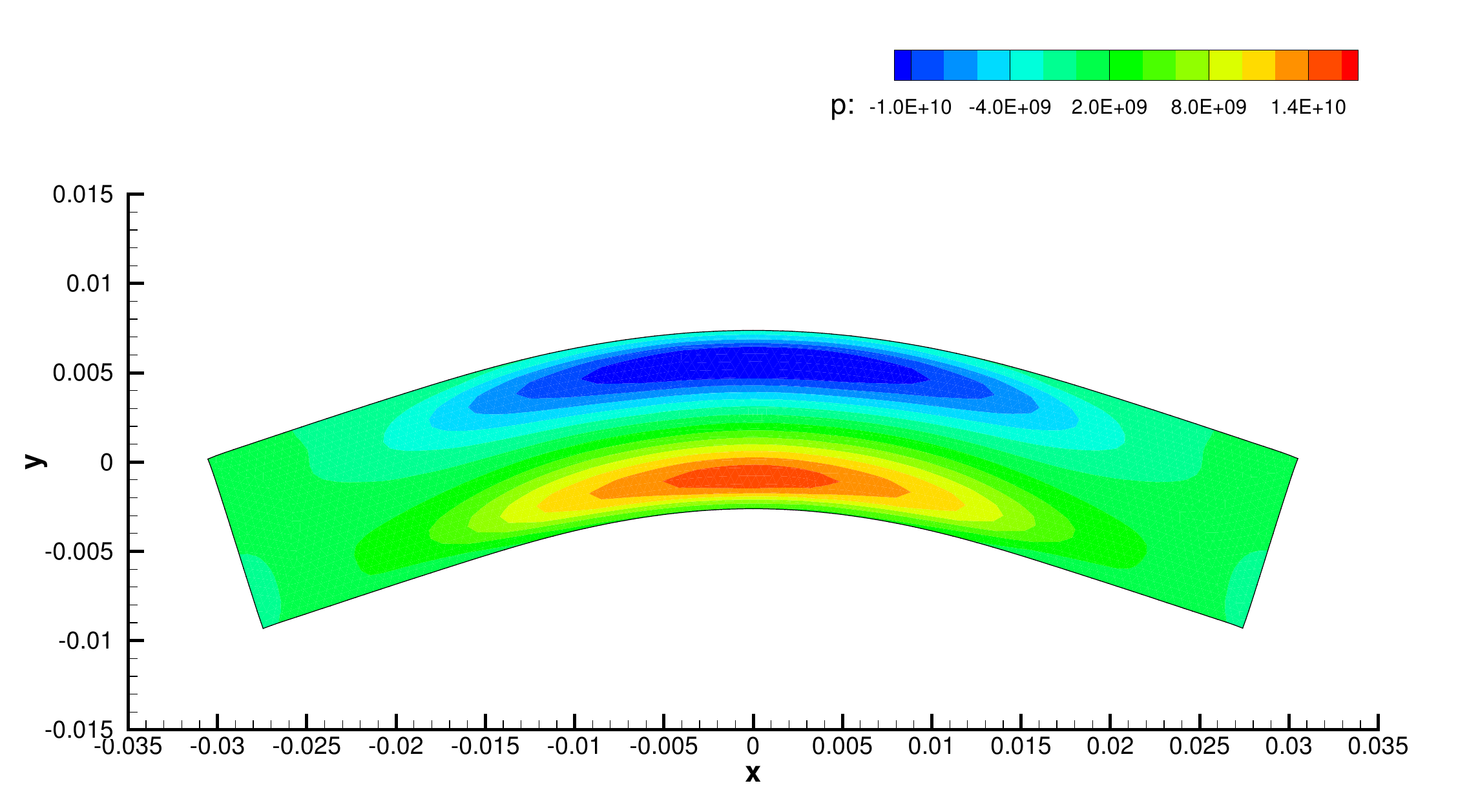}
			 \\
			\includegraphics[draft=false,width=0.47\textwidth]{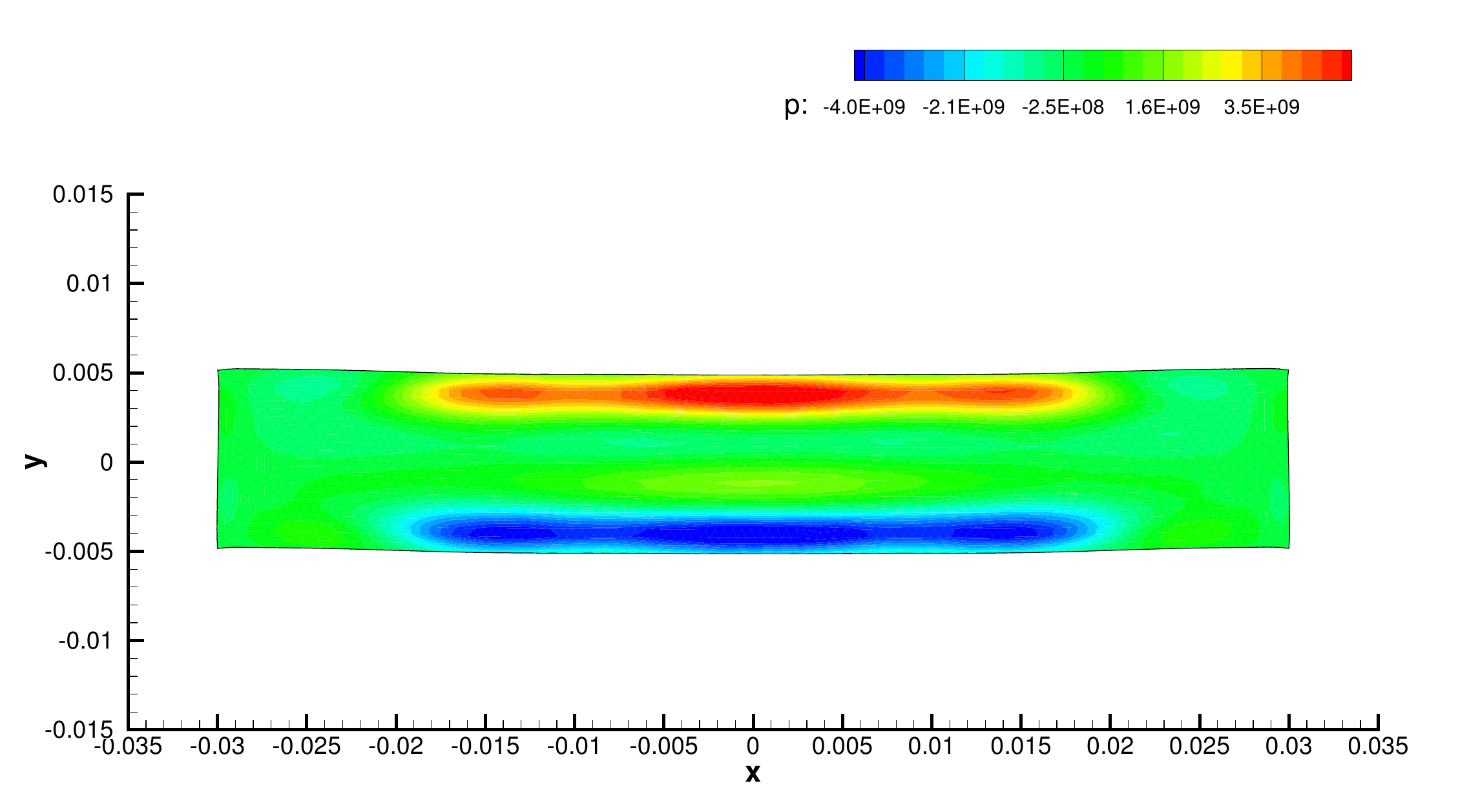}
			  &           
			\includegraphics[draft=false,width=0.47\textwidth]{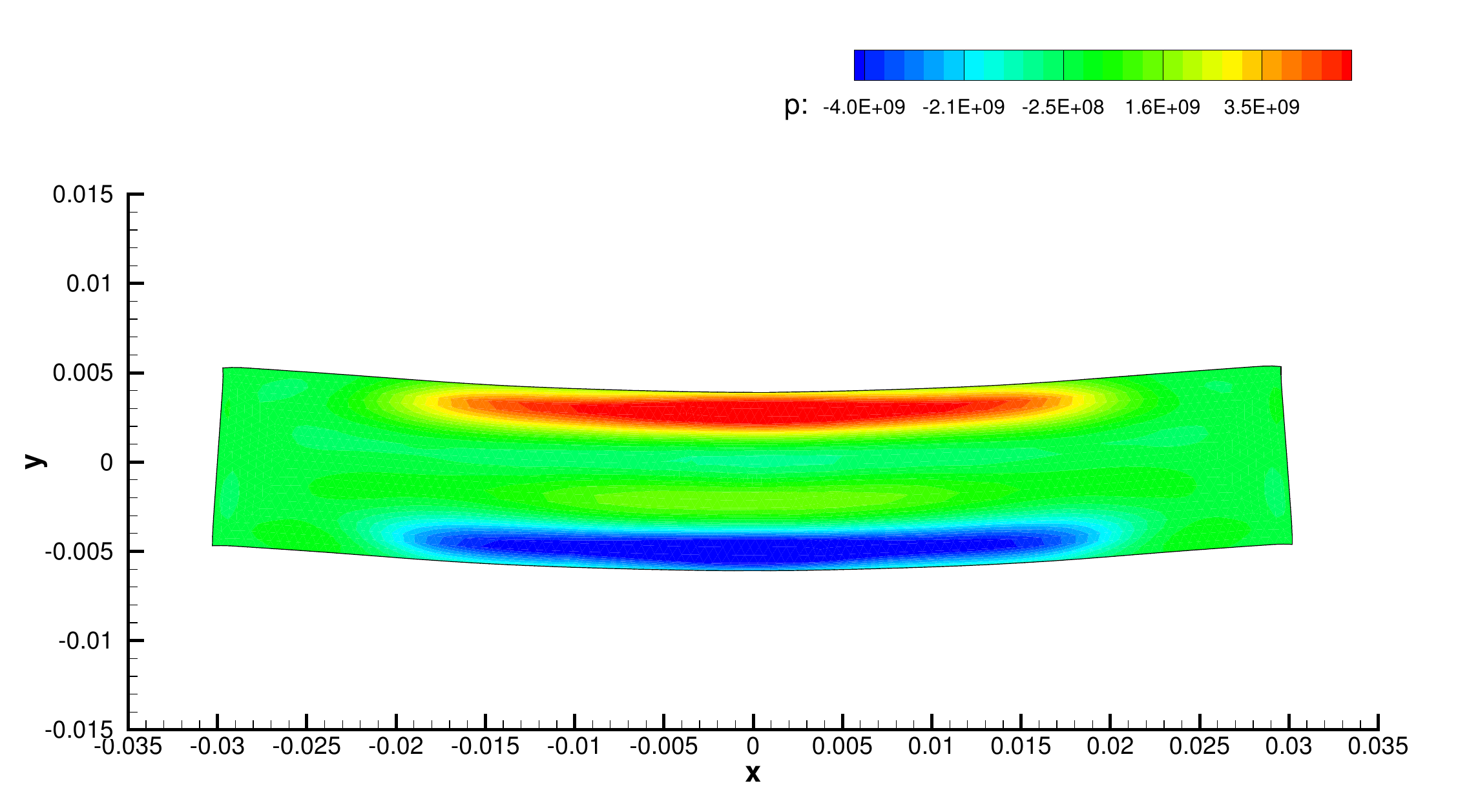}
			 \\
			\includegraphics[draft=false,width=0.47\textwidth]{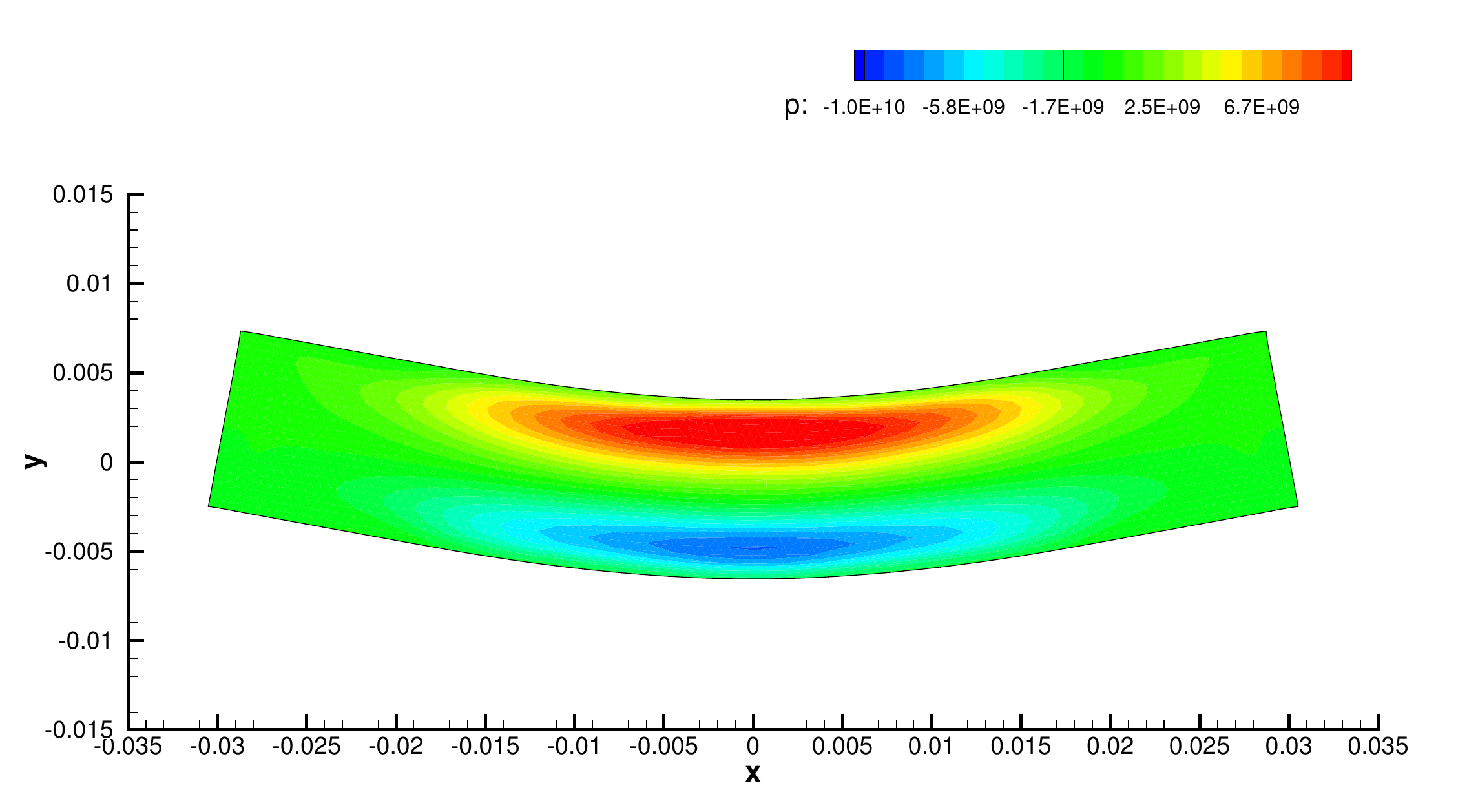}
			  &           
			\includegraphics[draft=false,width=0.47\textwidth]{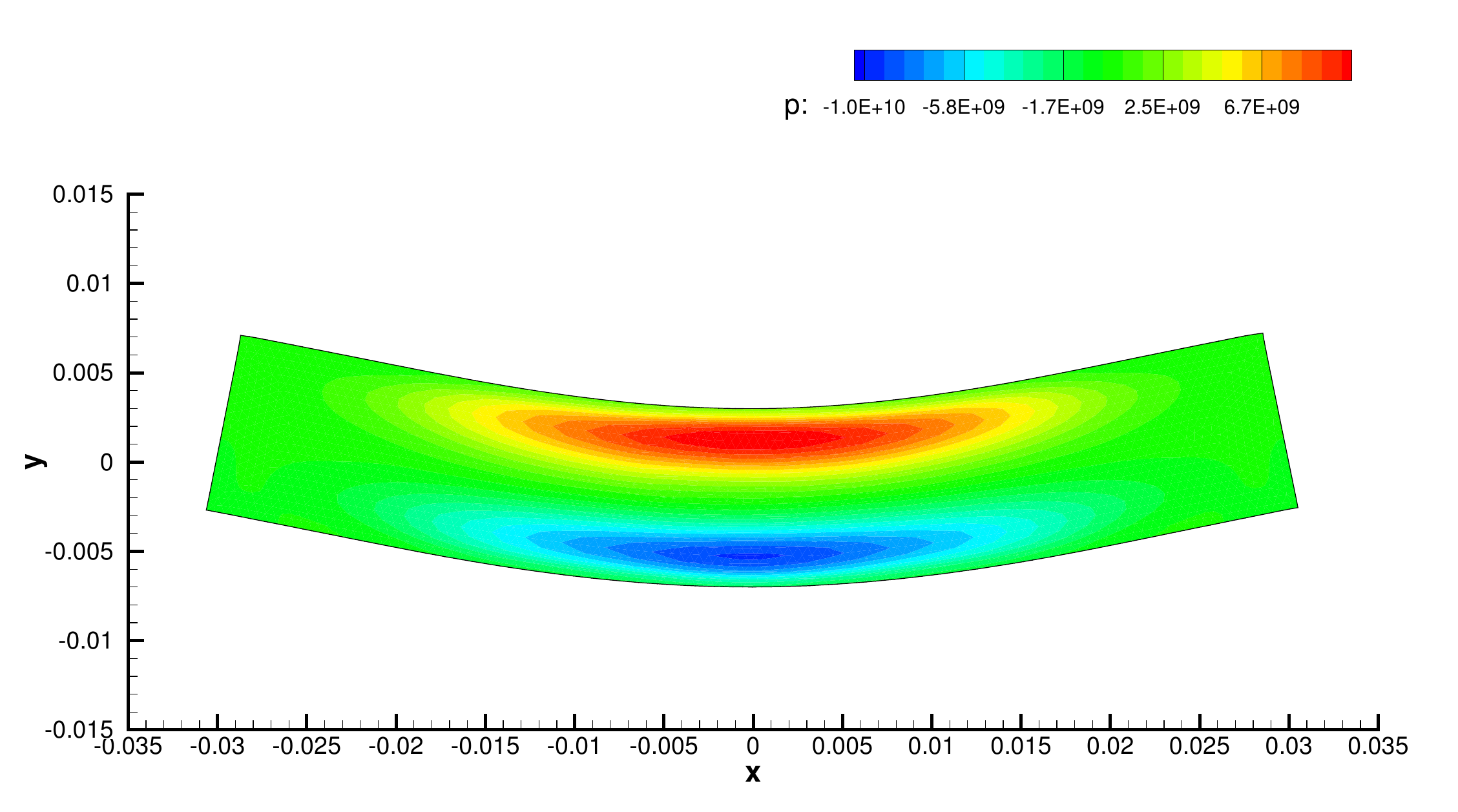}
			 \\
			\includegraphics[draft=false,width=0.47\textwidth]{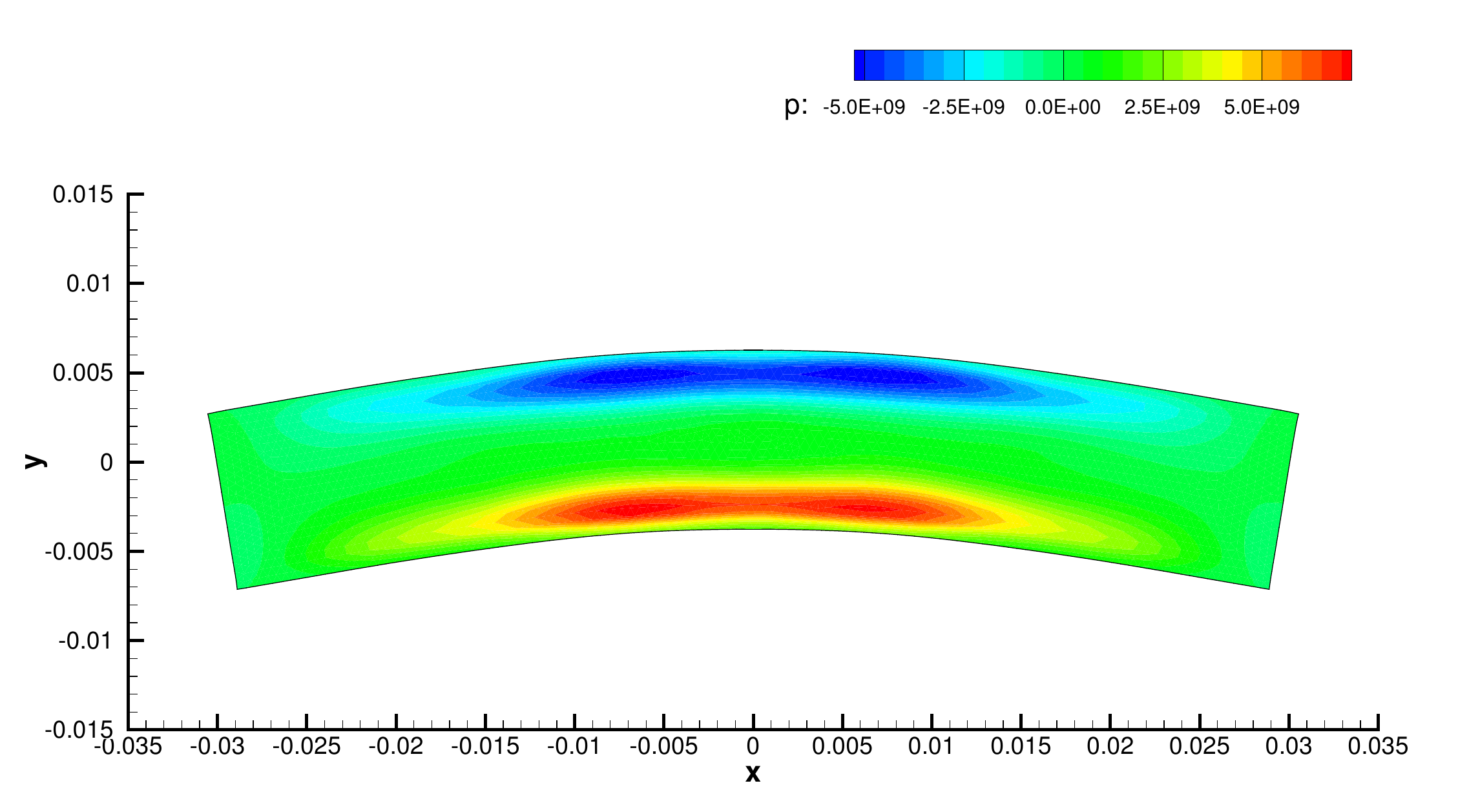}
			  &           
			\includegraphics[draft=false,width=0.47\textwidth]{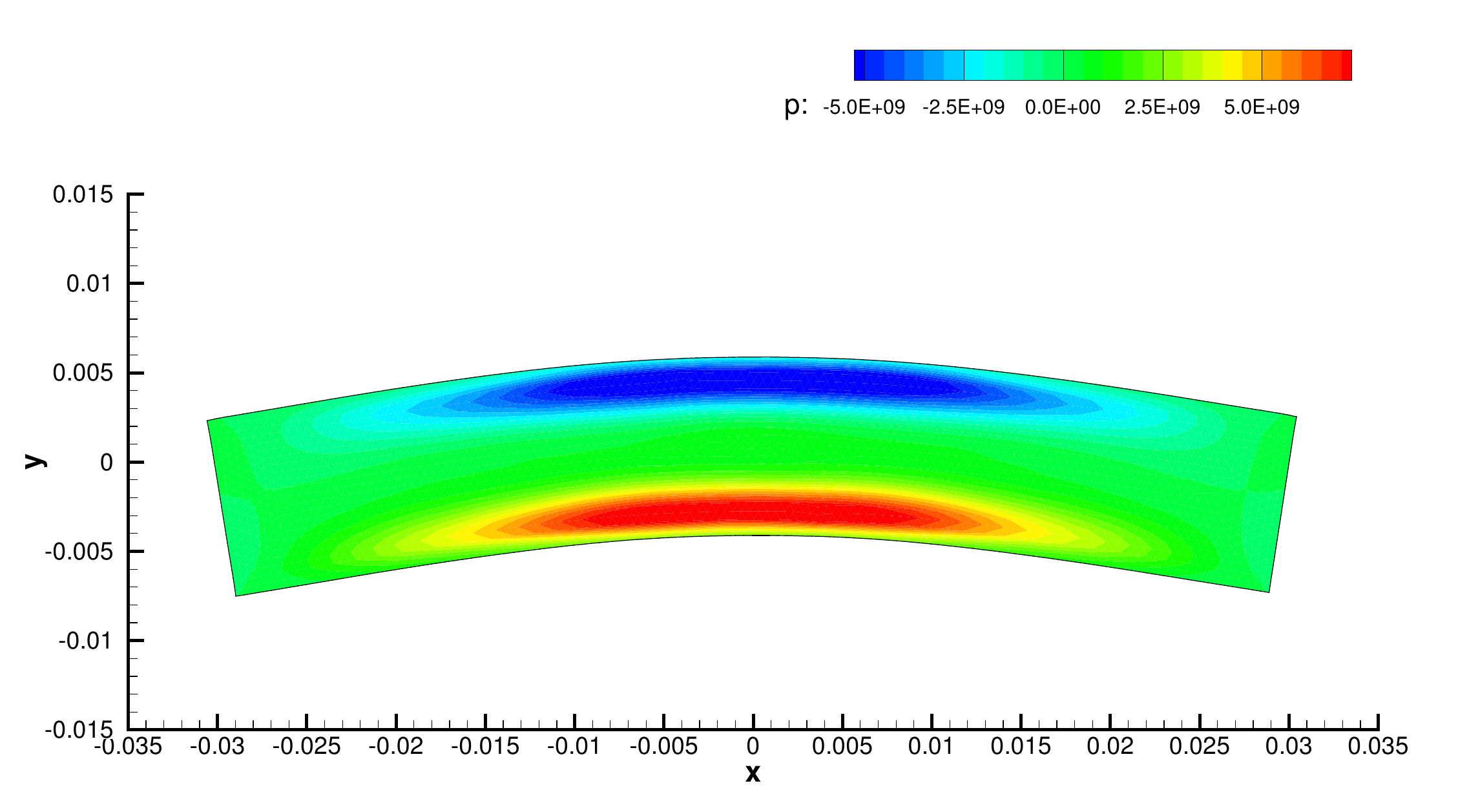}
			 \\
		\end{tabular} 
		\caption{Pressure contours obtained with ALE ADER-WENO third order schemes for the 
		oscillating Beryllium plate at output times $t=8\cdot10^{-6}$, $t=16\cdot10^{-6}$, 
		$t=24\cdot10^{-6}$ and $t=32\cdot10^{-6}$ (from top row to bottom row). Left column: 
		hypoelastic Wilkins model. Right column: hyperelastic \GPR model. } 
		\label{fig.Be_plate-pressure}
	\end{center}
\end{figure}

\begin{figure}[!htbp]
	\begin{center}
		\begin{tabular}{cc} 
			\includegraphics[draft=false,width=0.47\textwidth]{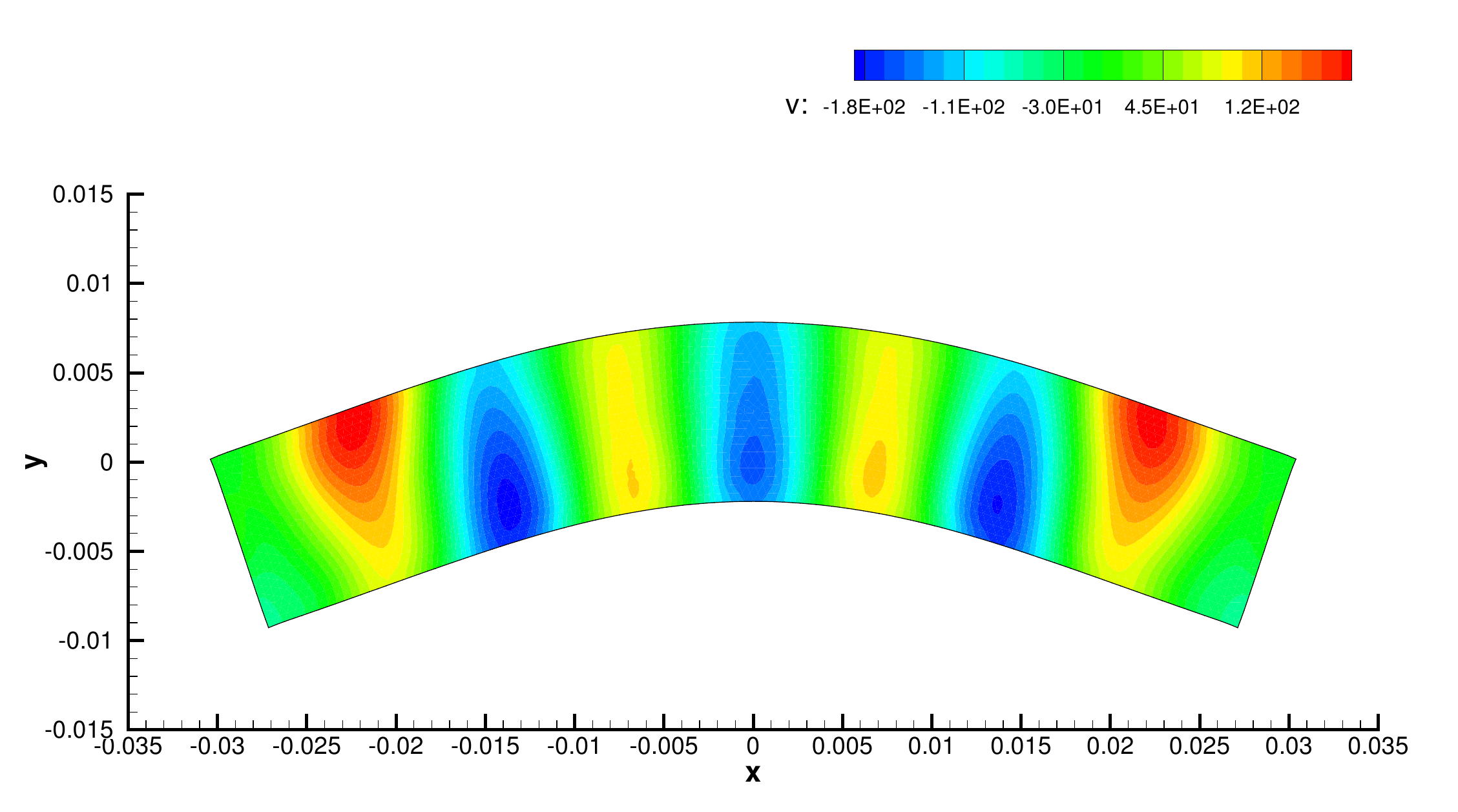}
			  &           
			\includegraphics[draft=false,width=0.47\textwidth]{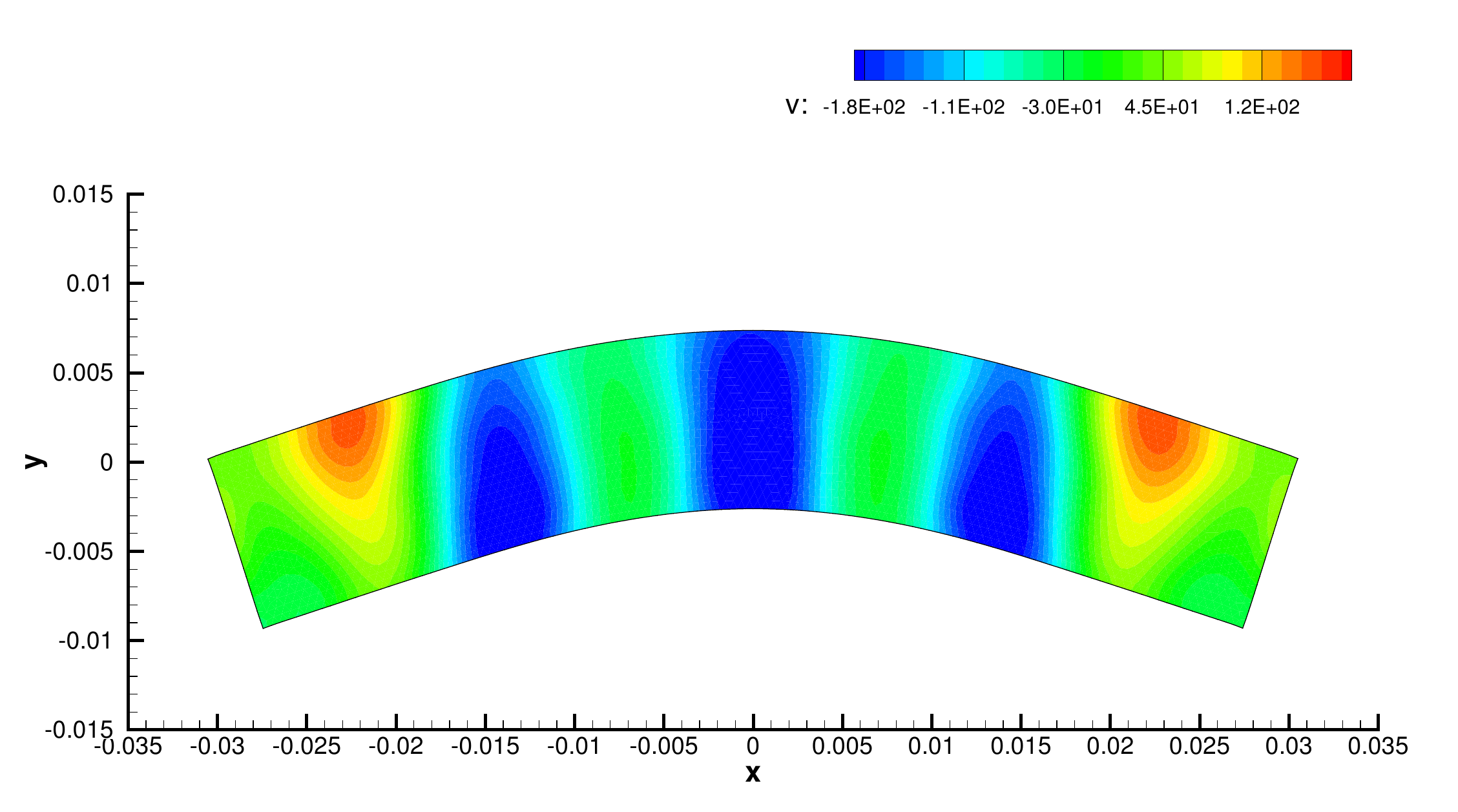}
			 \\
			\includegraphics[draft=false,width=0.47\textwidth]{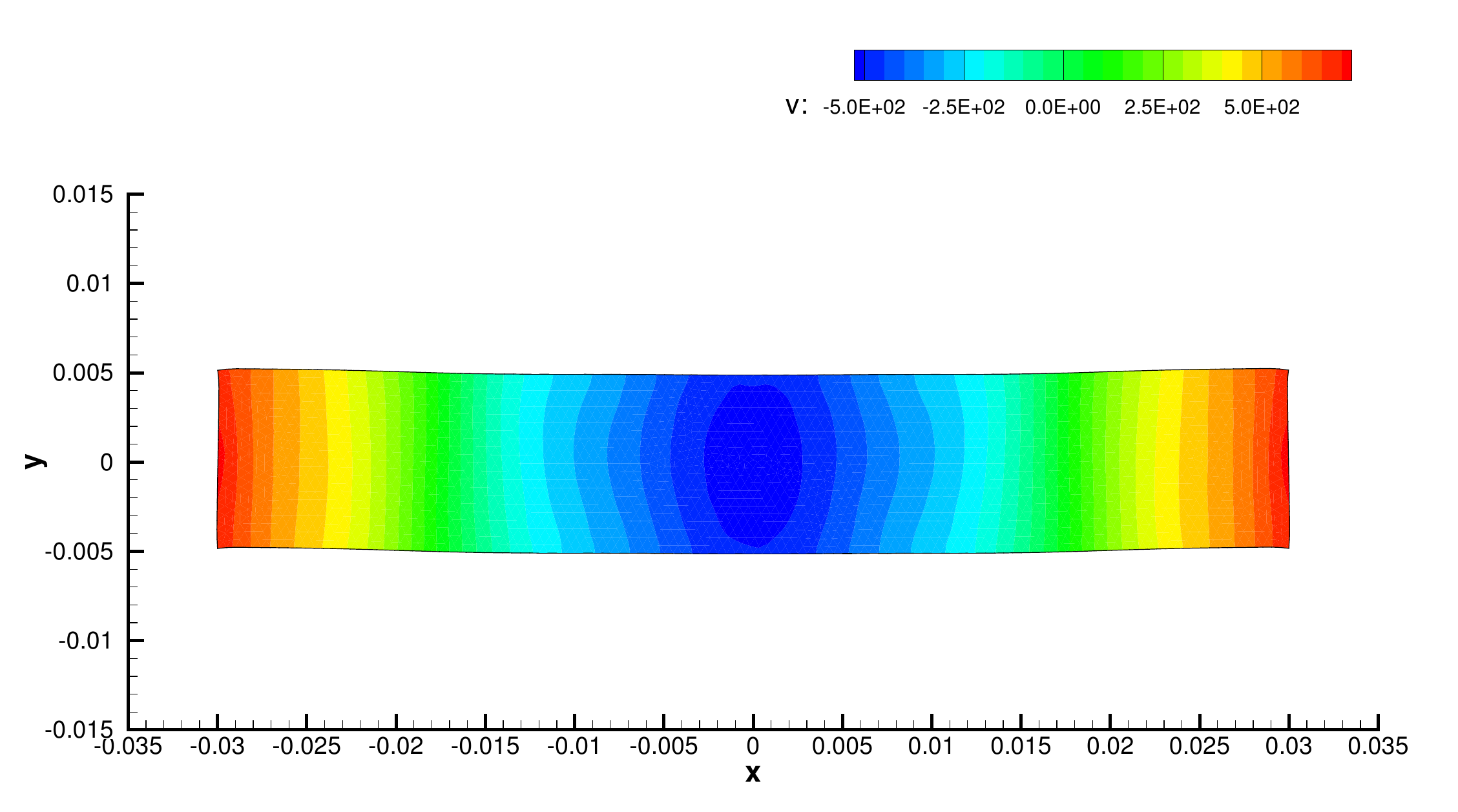}
			  &           
			\includegraphics[draft=false,width=0.47\textwidth]{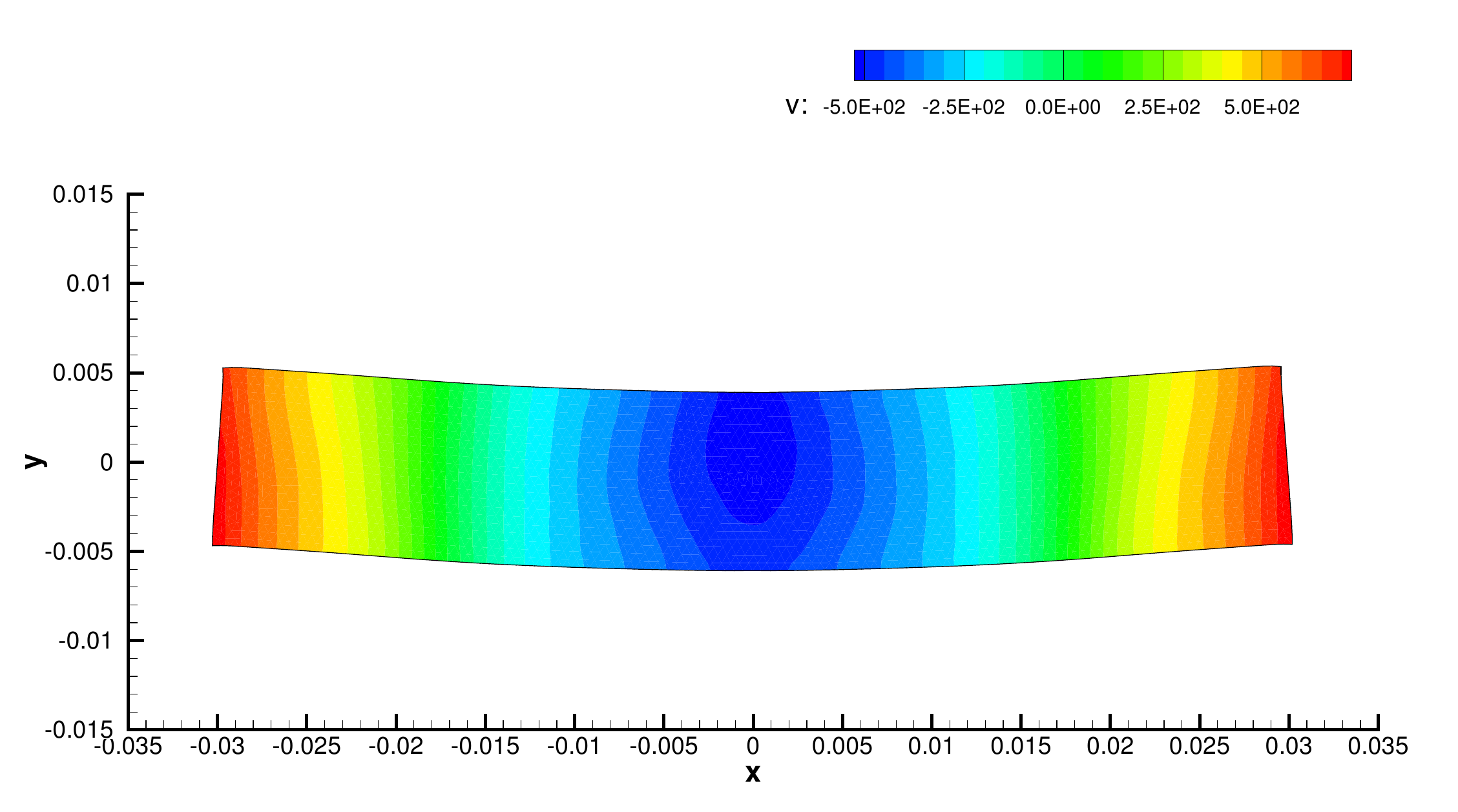}
			 \\
			\includegraphics[draft=false,width=0.47\textwidth]{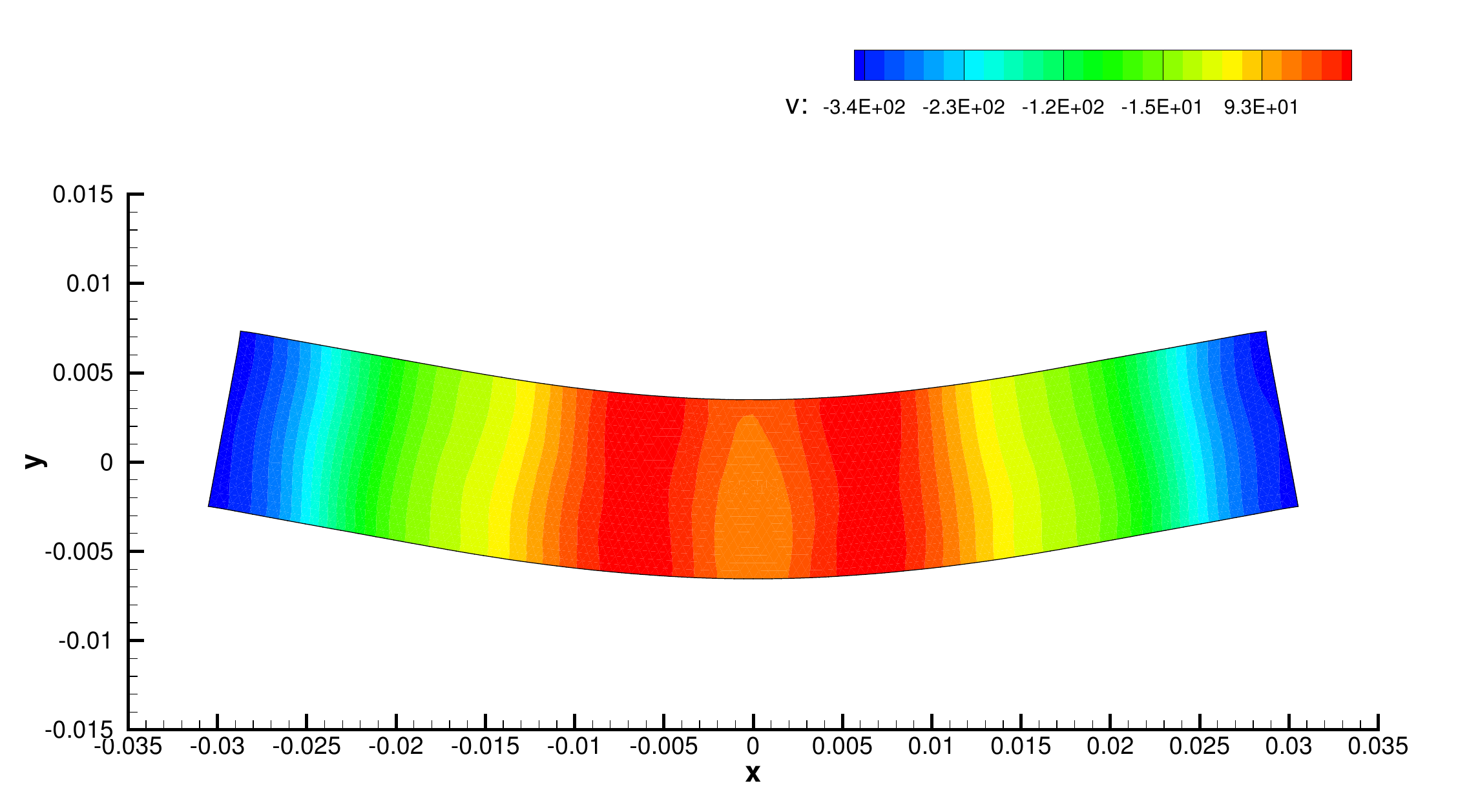}
			  &           
			\includegraphics[draft=false,width=0.47\textwidth]{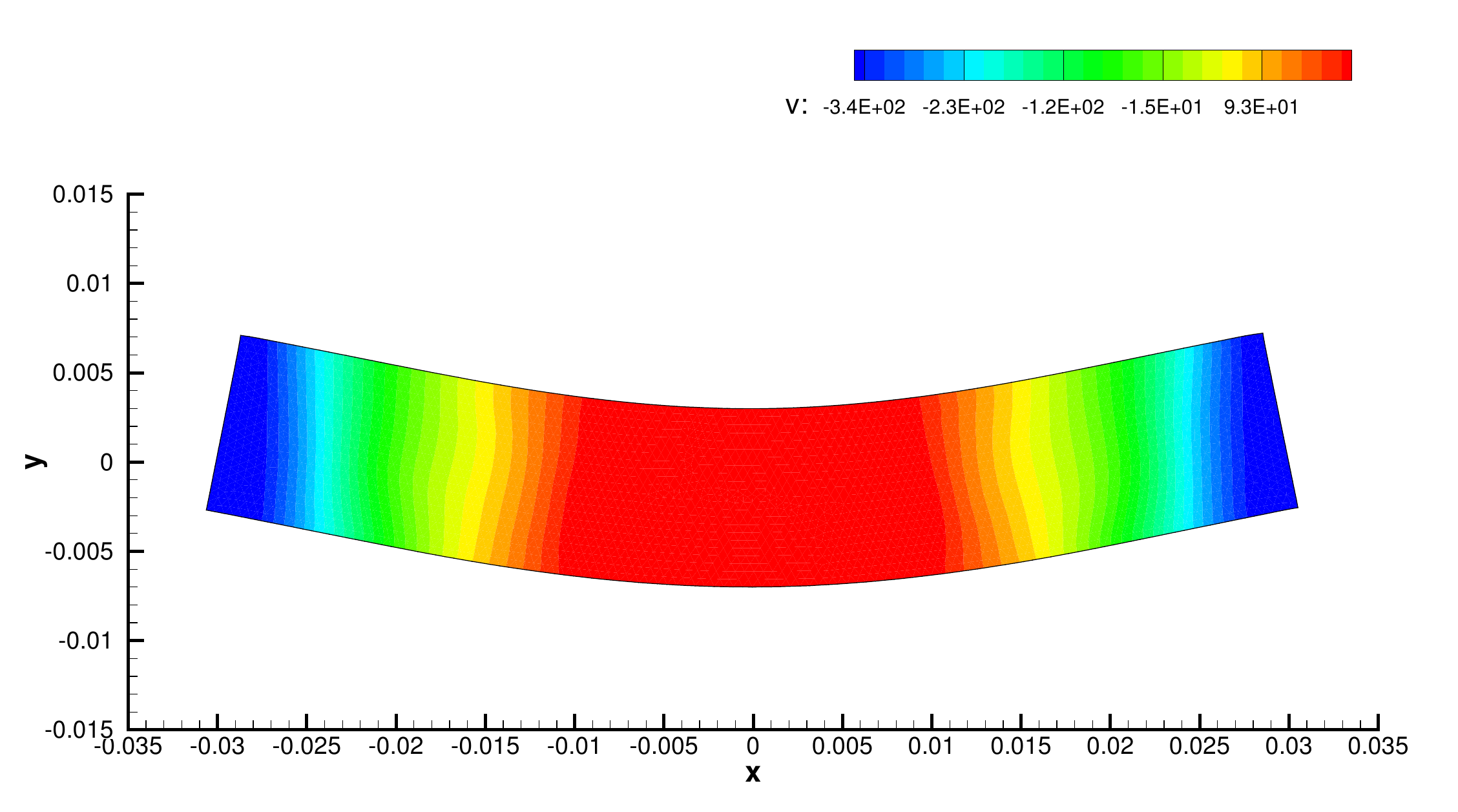}
			 \\
			\includegraphics[draft=false,width=0.47\textwidth]{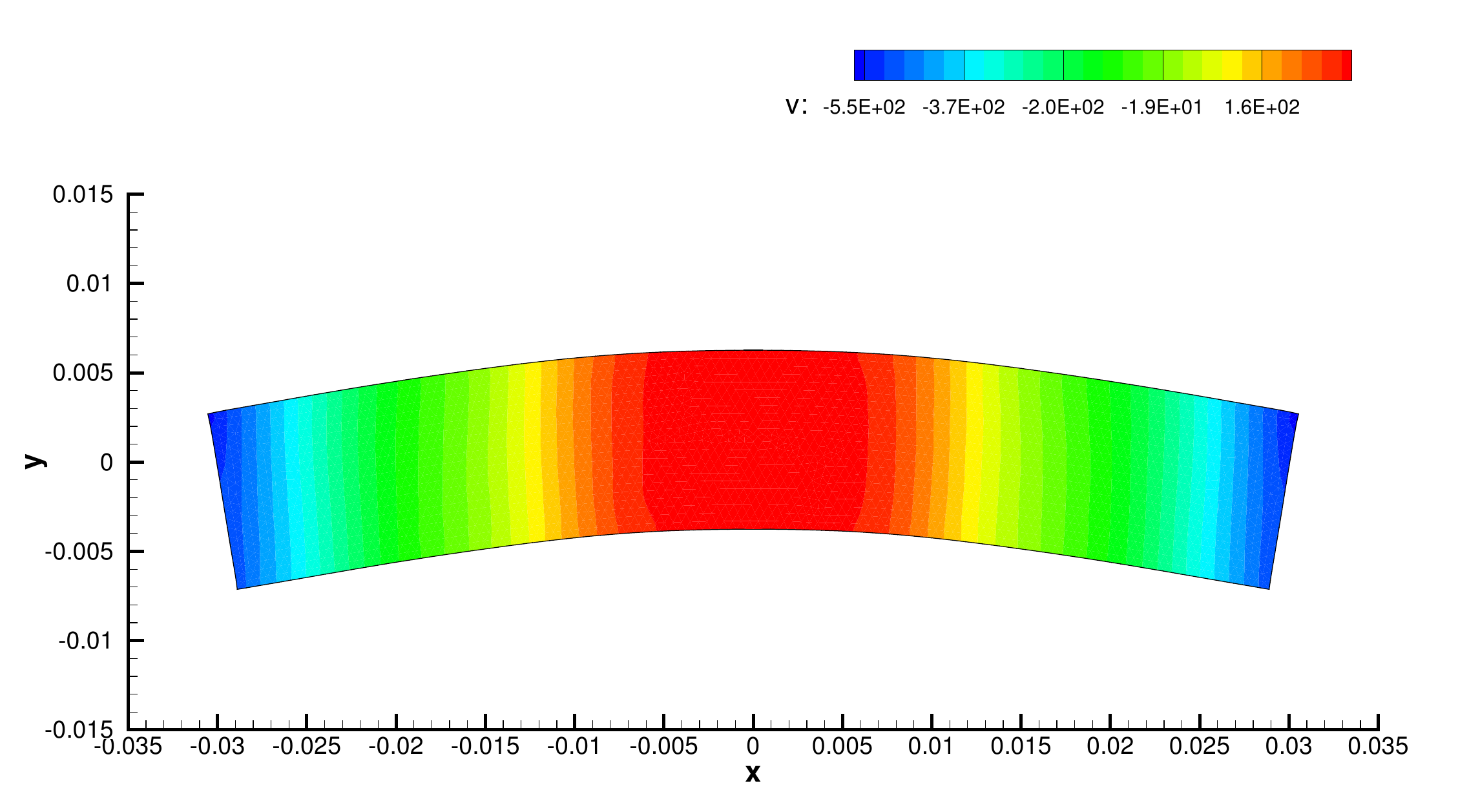}
			  &           
			\includegraphics[draft=false,width=0.47\textwidth]{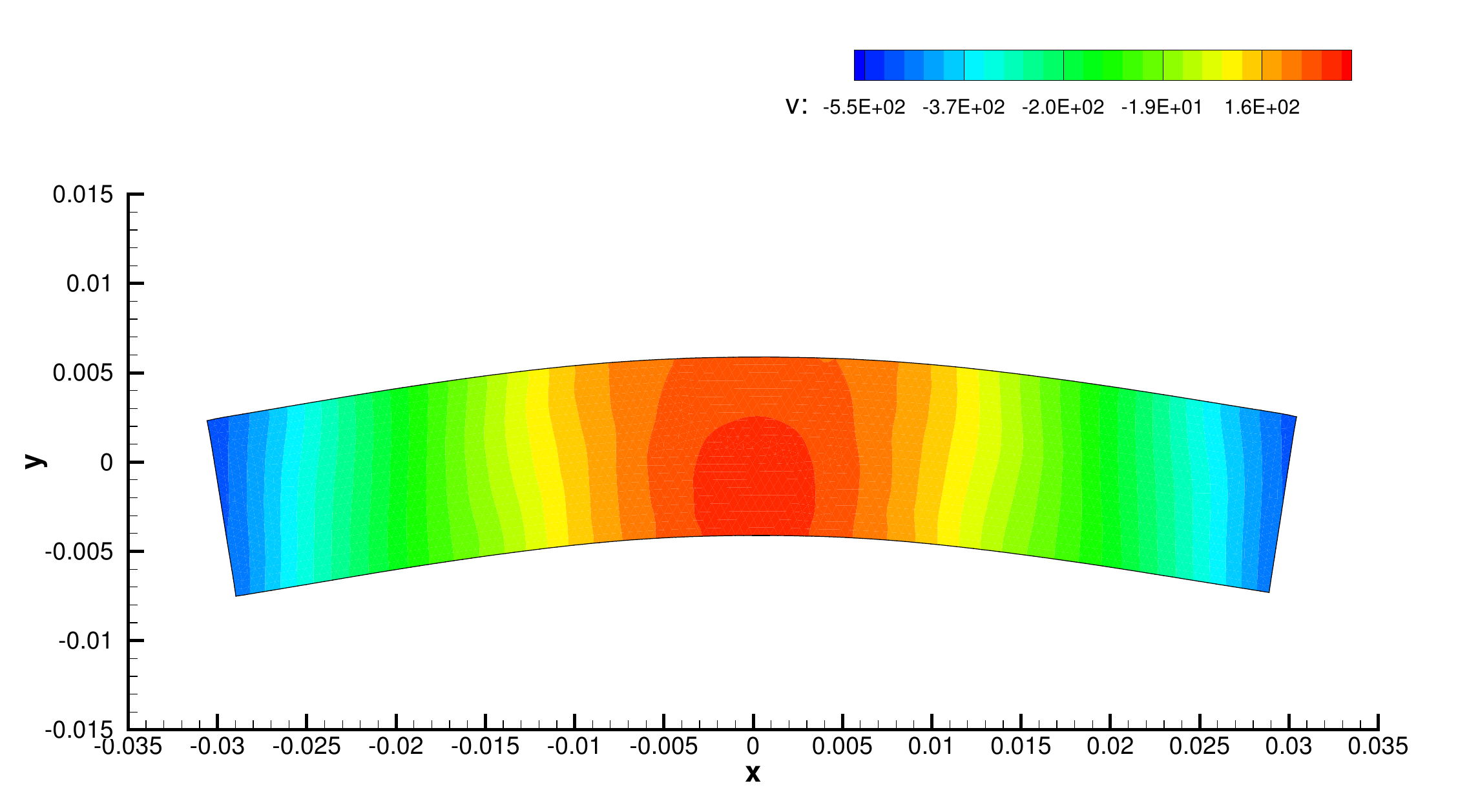}
			 \\
		\end{tabular} 
		\caption{Velocity contours obtained with ALE ADER-WENO third order schemes for the 
		oscillating Beryllium plate at output times $t=8\cdot10^{-6}$, $t=16\cdot10^{-6}$, 
		$t=24\cdot10^{-6}$ and $t=32\cdot10^{-6}$ (from top row to bottom row). Left column: 
		hypoelastic Wilkins model. Right column: hyperelastic \GPR model. } 
		\label{fig.Be_plate-velocity}
	\end{center}
\end{figure}

This problem simulates the purely elastic vibrations of a beryllium plate loaded by an initial 
velocity distribution that is not zero \cite{Burton2015,BoscheriDumbser2016}.
The computational domain is initially set to $\Omega(t=0)=[-3;3]\times [-0.5;0.5]$ and the mesh is 
made of $3210$ cells
with a characteristic mesh size of $h=0.0065$. 
Free traction boundary conditions are considered everywhere.
The plate is initialized with the reference density and pressure for beryllium
taken from Table~\ref{tab:dataEOS}. 
The distortion field $ \A $ of the hyperelastic model is initially set to identity while
 $\SSS=\bm{0}$ for the hypoelastic model.  
The velocity field is given by $\v=(0,v(x))$ where
\begin{equation}
v(x) = A \omega\left\{  
   C_1 \left( \sinh(\Omega(x+3))+\sin(\Omega(x+3)) \right) - S_1\left( \cosh(\Omega(x+3))+\cos(\Omega(x+3)) \right)
    \right\},
\end{equation}
with $\Omega=0.7883401241$, $\omega=0.2359739922$, $A=0.004336850425$, $S_1=57.64552048$ and 
$C_1=56.53585154$. The final time is set to $t_f=53.25$ according to \cite{Burton2015} such that it 
corresponds to two complete flexural periods $\omega$. 
  In Figs.~\ref{fig.Be_plate-pressure} and \ref{fig.Be_plate-velocity} we present the  pressure and 
  the vertical velocity component respectively for intermediate times $t=8 \time 10^{-6}$, 
  $t=15\time 10^{-6}$, $t=23\time 10^{-6}$ and $t=32\time 10^{-6}$ which cover approximately one 
  flexural period.  The color scales are identical to allow for a direct comparison of two models. 
Qualitatively the plate is behaving as expected and comparably for those two models. The 
oscillations decay only due to the numerical dissipation. Only tiny 
differences between the solutions can be observed
meaning that we are rather in the linear elastic regime.
Moreover, these third order accurate results visually compare well against known results from other 
Lagrangian schemes \cite{Sambasivan_13,Burton2015}.

\subsection{Cylindrical shell compression} \label{ssec:Shell}
\begin{figure}[!htbp]
	\begin{center}
		\begin{tabular}{cc} 
			\includegraphics[draft=false,width=0.47\textwidth]{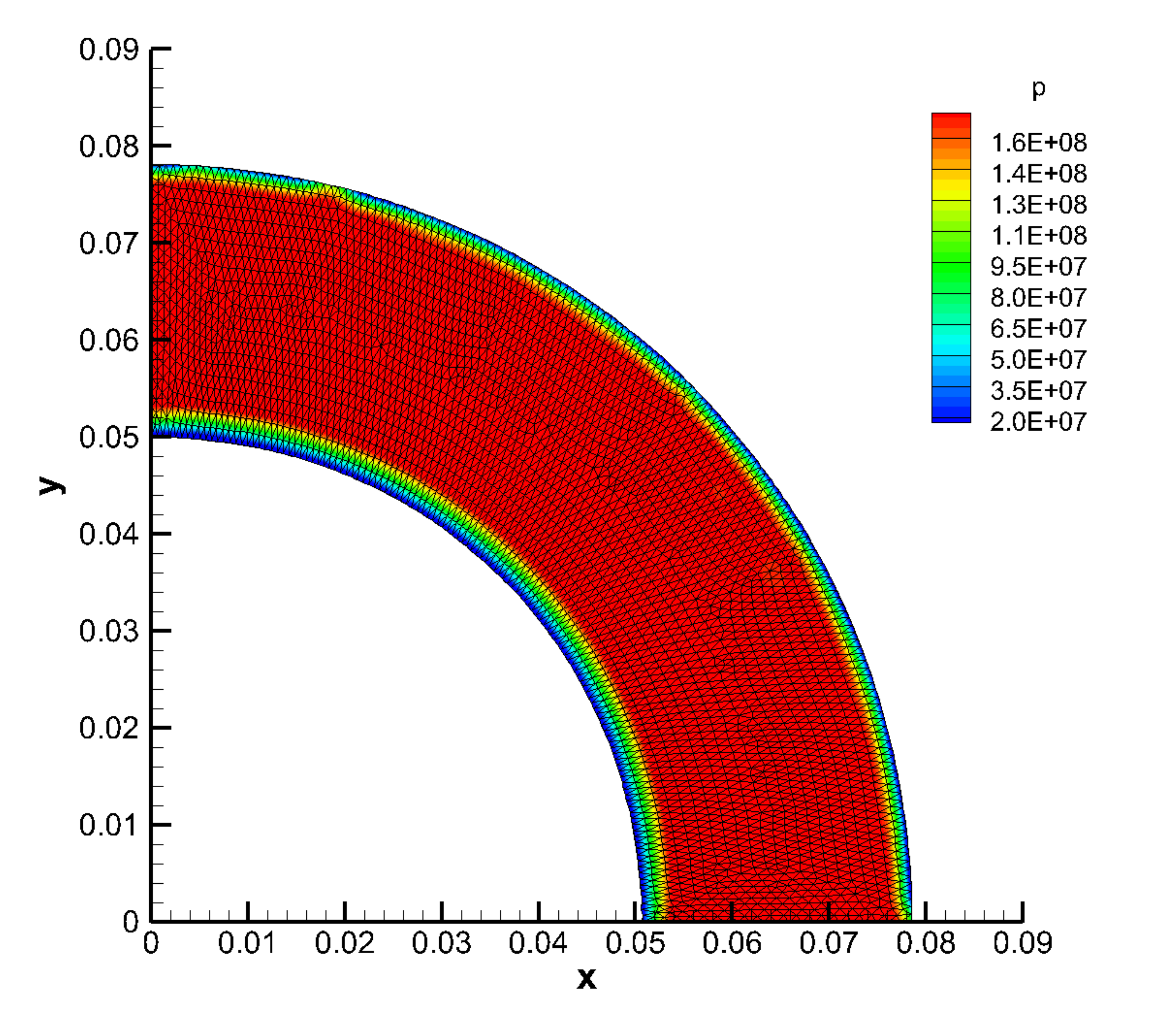}
			  &           
			\includegraphics[draft=false,width=0.47\textwidth]{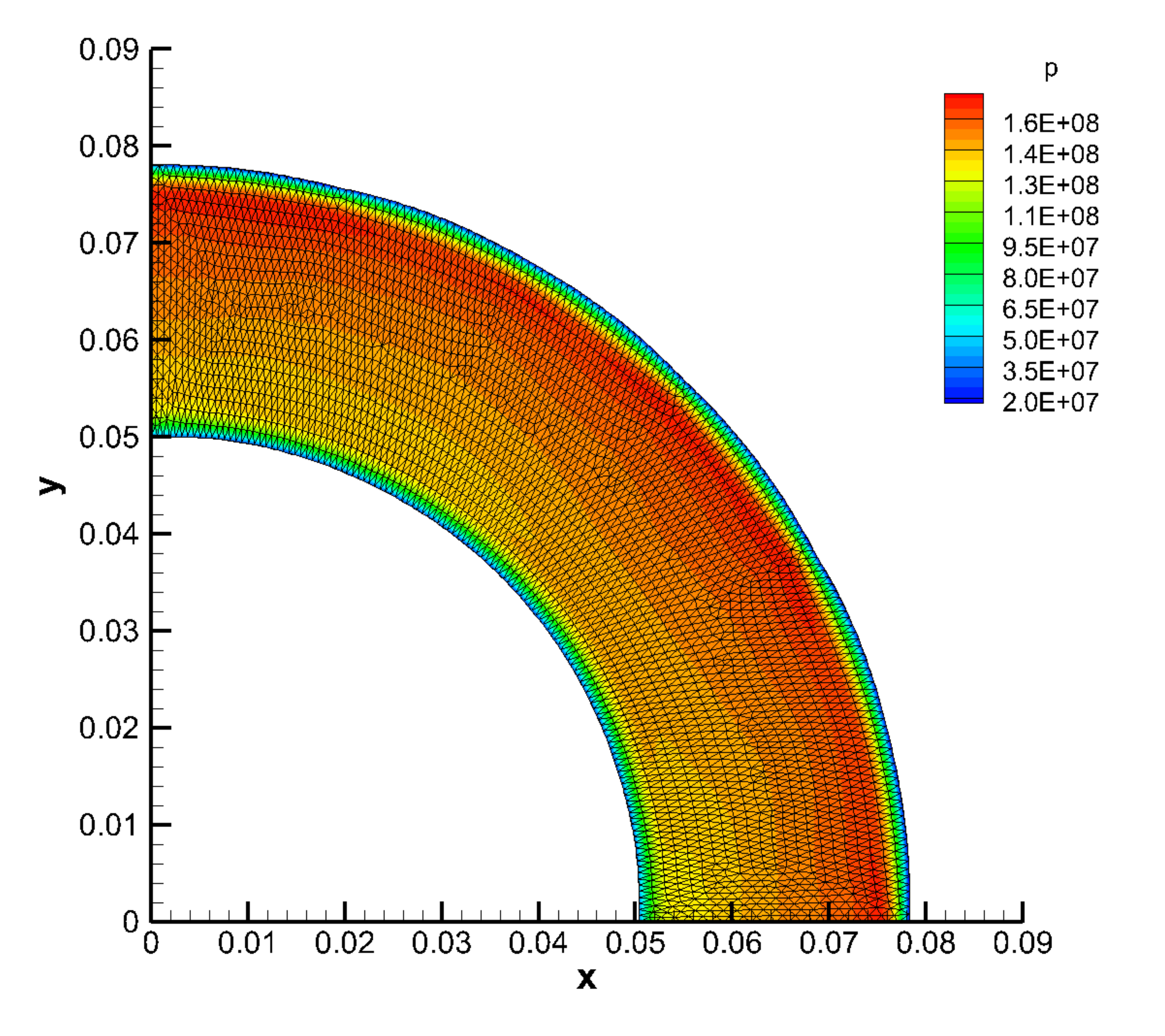}
			 \\
			\includegraphics[draft=false,width=0.47\textwidth]{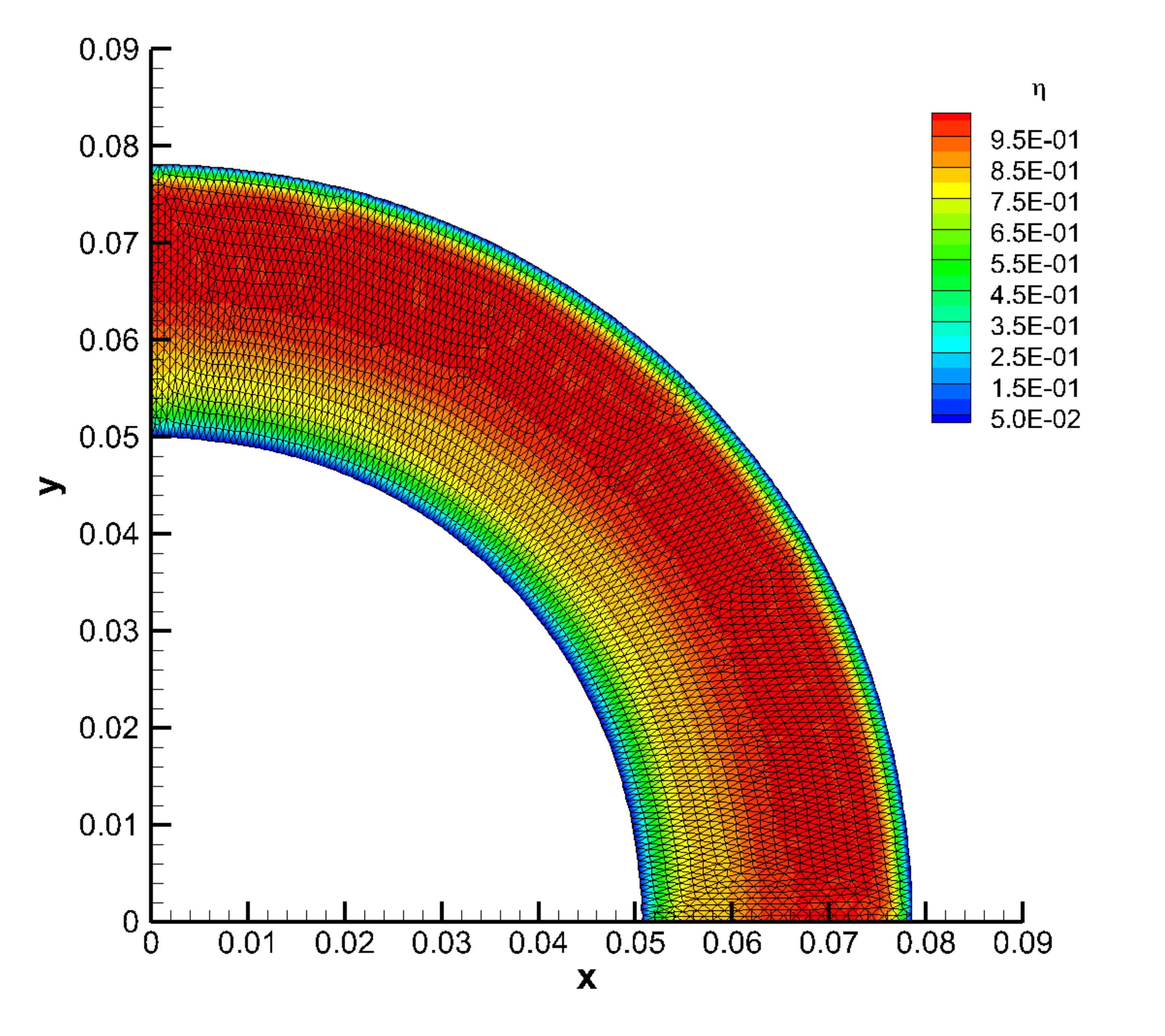}
			  &           
			\includegraphics[draft=false,width=0.47\textwidth]{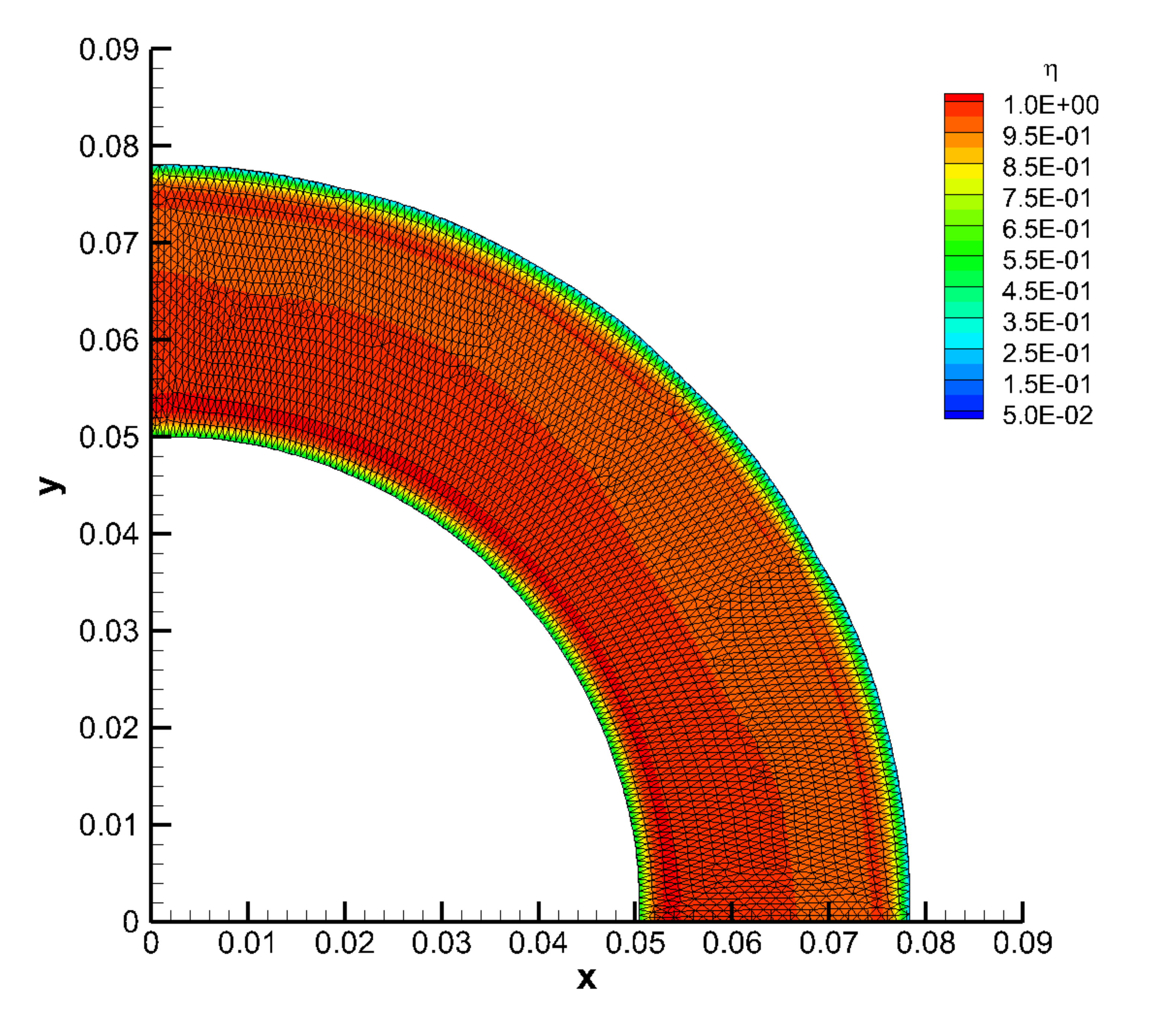}
			 \\
		\end{tabular} 
		\caption{Pressure contours (top row) and plasticity map (bottom row) obtained with ALE ADER-WENO third order schemes for the shell compression test case with characteristic mesh size $h=10^{-3}$.
			 Left column: hypoelastic Wilkins model. Right column: hyperelastic \GPR model. } 
		\label{fig.Shell}
	\end{center}
\end{figure}

\begin{figure}
  \begin{center}
  \begin{tabular}{cc}
	    \includegraphics[draft=false,width=0.47\textwidth]{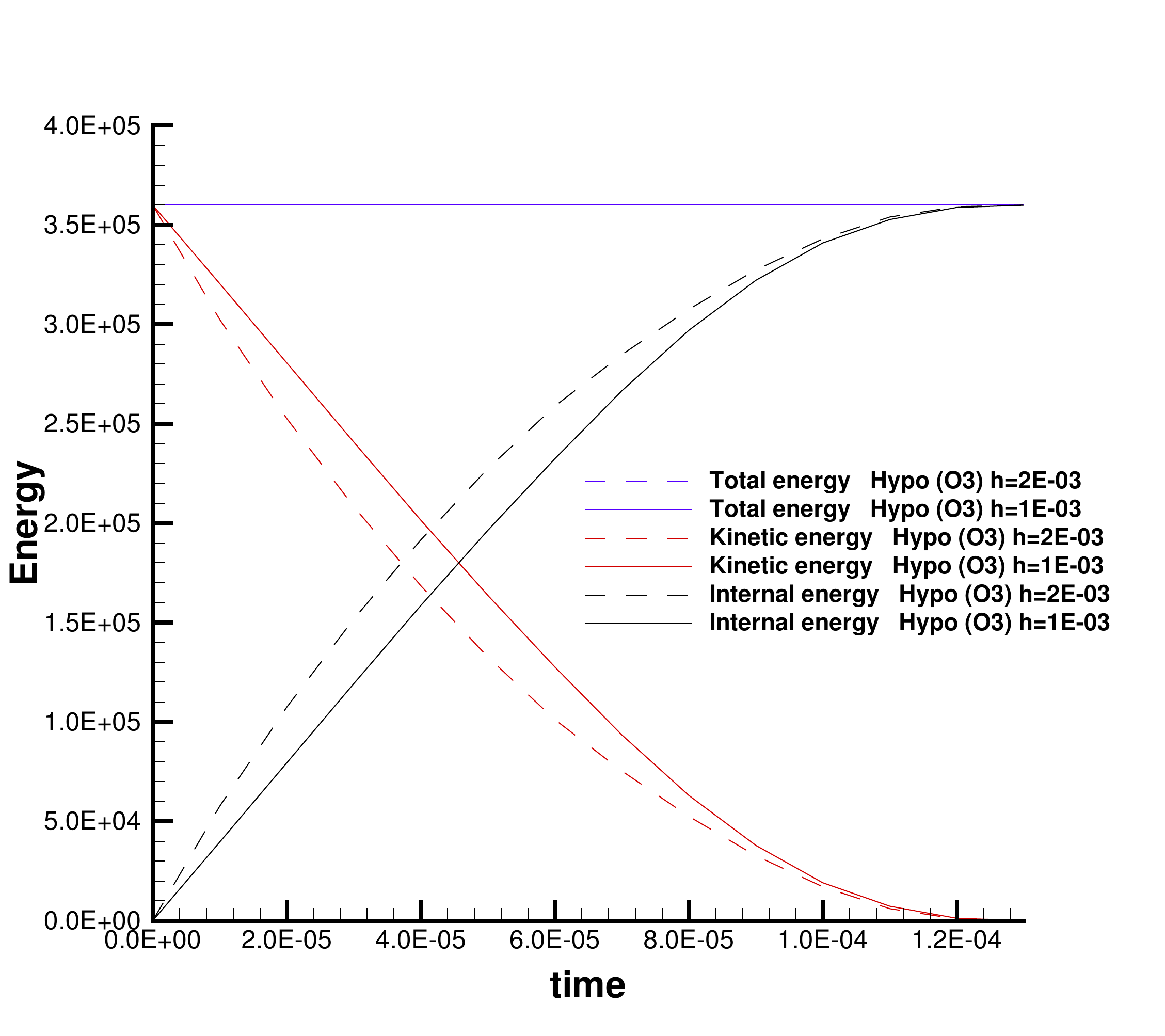}
	      &           
			\includegraphics[draft=false,width=0.47\textwidth]{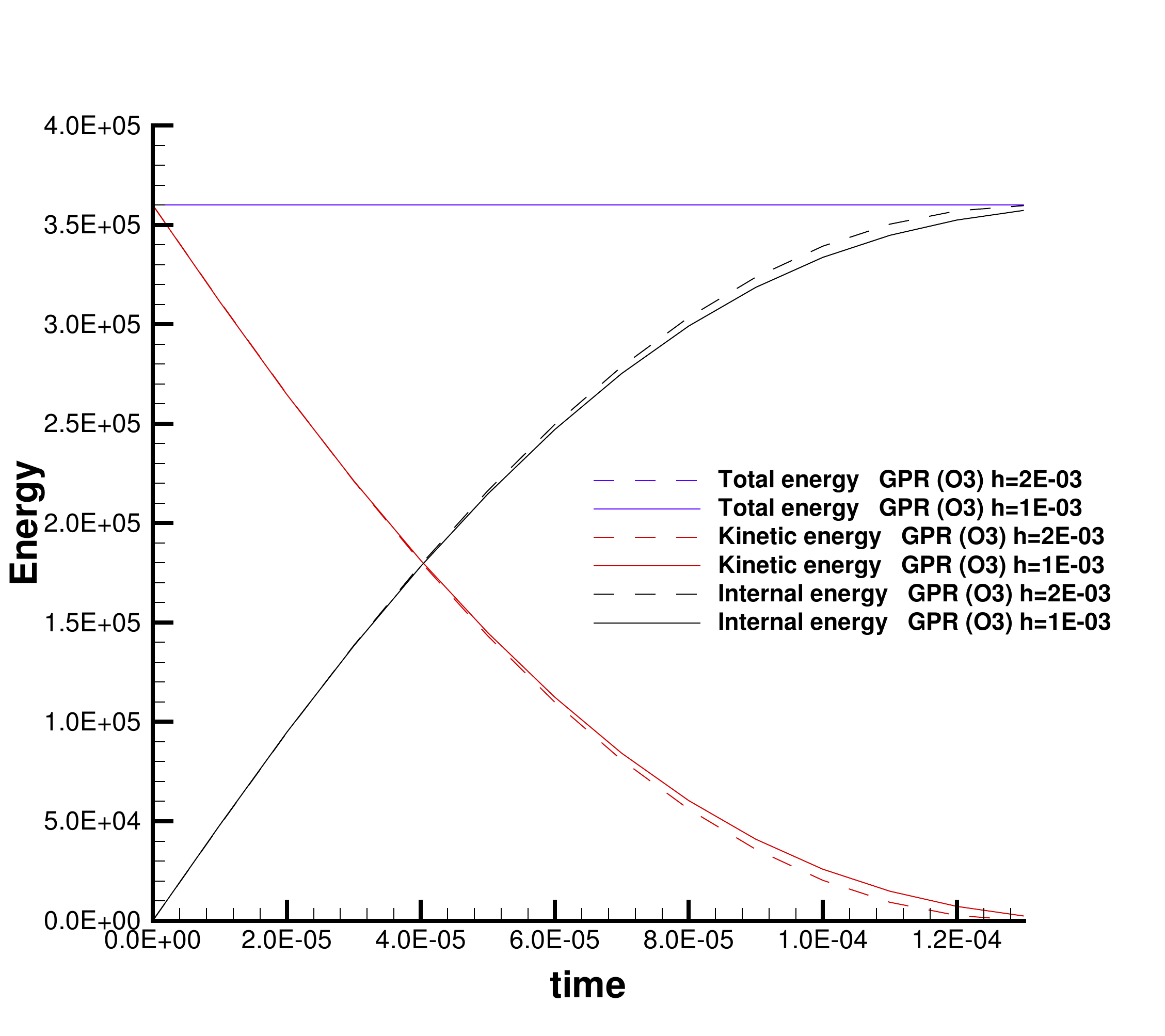}
			 \\
      \includegraphics[draft=false,width=0.47\textwidth]{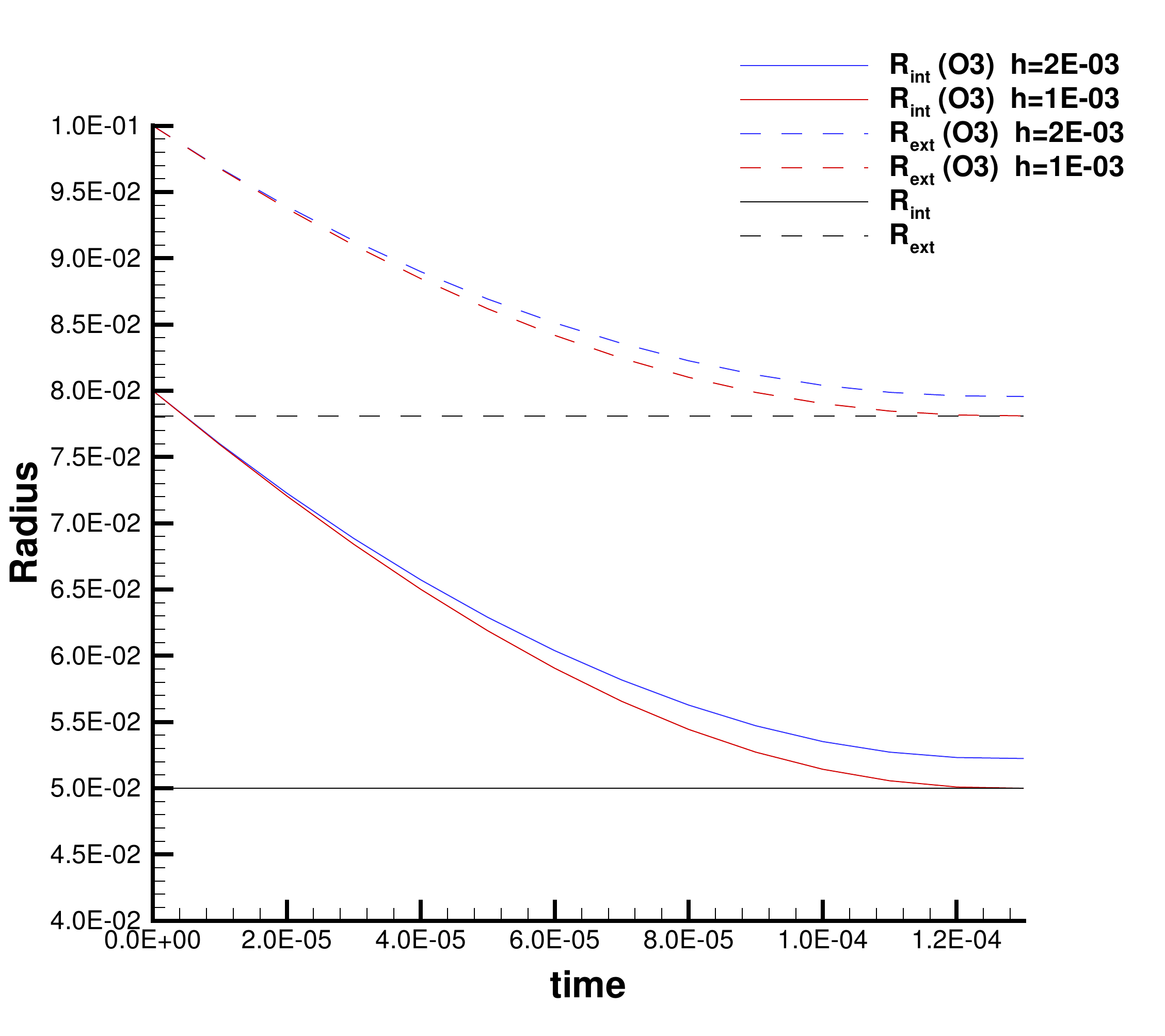}
        &           
			\includegraphics[draft=false,width=0.47\textwidth]{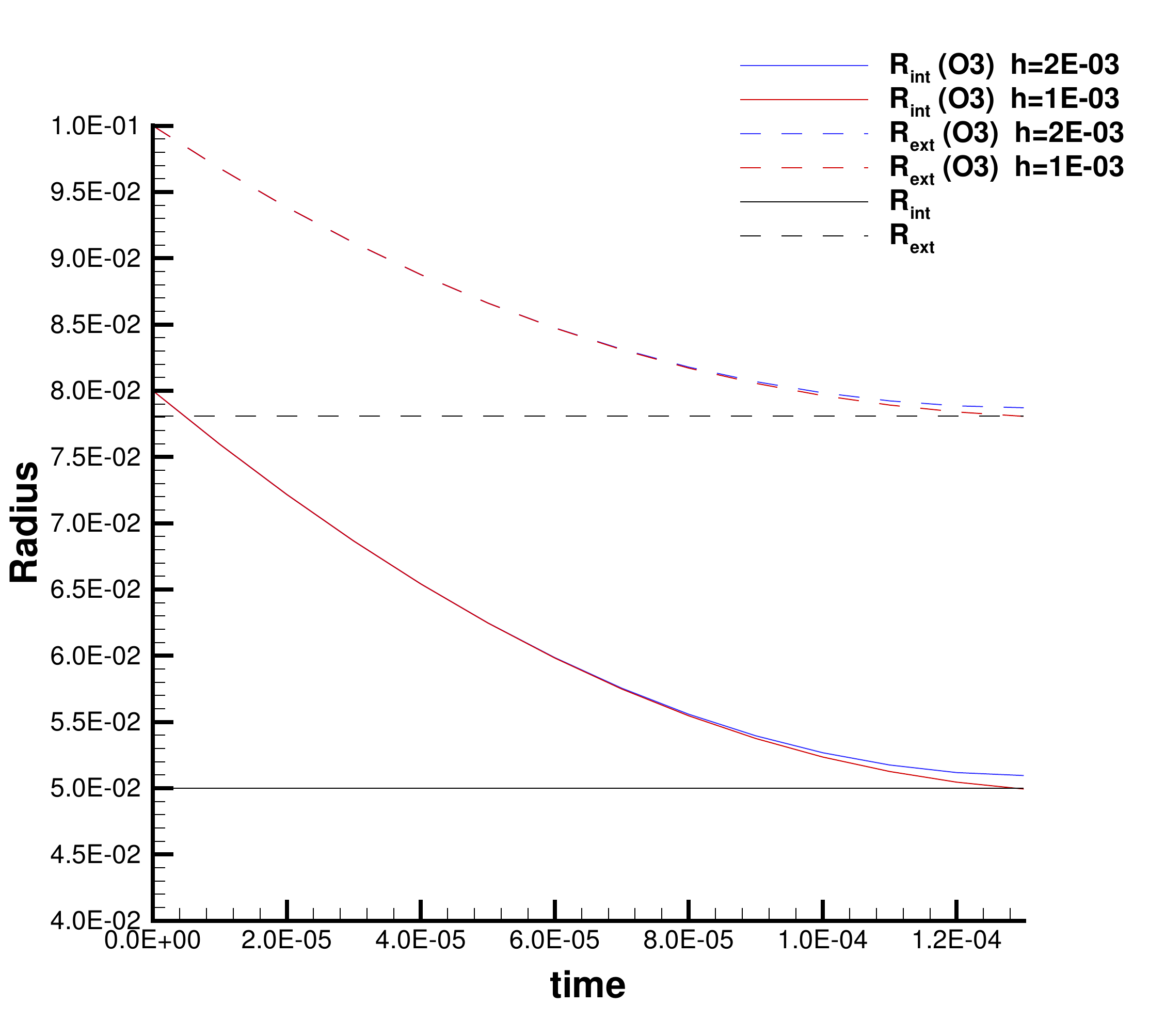}
			 \\
  \end{tabular}
  \caption{Top: analysis of energy conservation for the shell collapse test problem for both computational meshes and governing equations. Bottom: time evolution of the internal and external radius for both computational meshes and governing equations.}
  \label{fig.Shell-analysis}
\end{center}
\end{figure}

The next test case is known as the Verney problem \cite{Verney,KammLANL08,Verney2} but we consider 
the problem statement as it is proposed by Howell and Ball in \cite{Howell2002}. It
consists in a cylindrical beryllium shell which collapses responding to an initial inwardly 
directed radial velocity field. 
The initial setup is taken from \cite{KammLANL08}.
The shell $\Omega$ with inner radius $R_i=8\times 10^{-2}$m and outer radius
$R_o=10\times10^{-2}$m is made of beryllium under Mie-Gruneisen equation of state.
In this test case, the material experiences elastoplastic deformations.
Initial pressure is set to $p_0=10^{-6}$Pa and the radial velocity magnitude is given by 
$v_0(r)=-V_0\left(\frac{R_i}{r}\right)^2$, where we have chosen
$V_0=417.10$m/s. 
The initial kinetic energy due to the velocity distribution is entirely dissipated by the plastic deformation 
of the material leading to a deceleration of the shell. 
At the end of the simulation, set to $t_{\text{final}}=125\mu$s, the shell arrives at a complete 
rest state
for a value of the inner/outer radii equal to $r_i=5 \times 10^{-2}$m and $r_o=7.81 \times 
10^{-2}$m. This exact solution has been derived by Howall and Ball in \cite{Howell2002} under the 
ideal plasticity assumption.
Only one quarter of the shell is considered in Cartesian geometry and its associated 
unstructured mesh has a characteristic mesh size of $h=10^{-3}$ or $h=2\times 10^{-3}$, 
leading to about $98 500$ and $197 000$ cells, respectively.
Free traction boundary conditions are considered on the inner and outer radii,
 while symmetric boundary conditions are used for points on the axis $x=0$ and $y=0$. 
The results of the pressure contours (top) and the plasticity map (bottom) 
are displayed in Fig.~\ref{fig.Shell} at the final instant of time 
for both models; \GPR on left panels, Wilkins on the right ones. It can be observed that 
while qualitatively equivalent, these results differ in the field representation while maintaining a good radial
symmetry.
Next on the top panels of Fig.~\ref{fig.Shell-analysis} we present the classical plots of energy 
balance where we
can observe the transfer of kinetic energy into internal energy with conserved total energy.
We observed that the hypo-elastic model of Wilkins presents a shift in chronometry when the mesh is refined,
meaning that one has not reached the mesh convergence for this model. 
Contrarily the hyper-elastic \GPR model does not present such shift.
At last the bottom panels of the same figure illustrate the evolution of the mean inner/outer radii of the shell
and their convergence towards their exact values (black lines) for both resolutions and models. A 
little bit faster convergence with respect to the mesh refinement can be observed for the \GPR 
model. Overall, these third order accurate results visually compare well against known results from 
other Lagrangian schemes \cite{Sambasivan_13}. The distribution of the  norm of the 
deviatoric stress $\sigma=\sqrt{3 \tr(\dev{\bm{\sigma}}^2)/2} $ is plotted in 
Fig.~\ref{fig.Shell-analysis} versus radius. From these plots, one can judge about the degree of 
symmetry violation in our simulations. Thus, the distribution of $ \sigma $ looks perfectly 
symmetric (all points lie on a single line) for the \GPR model, while it is less symmetric for the 
Wilkins model.

\subsection{Taylor bar impact on a wall} \label{ssec:Taylor}
\begin{figure}[!htbp]
	\begin{center}
		\begin{tabular}{cc} 
			\includegraphics[draft=false,width=0.47\textwidth]{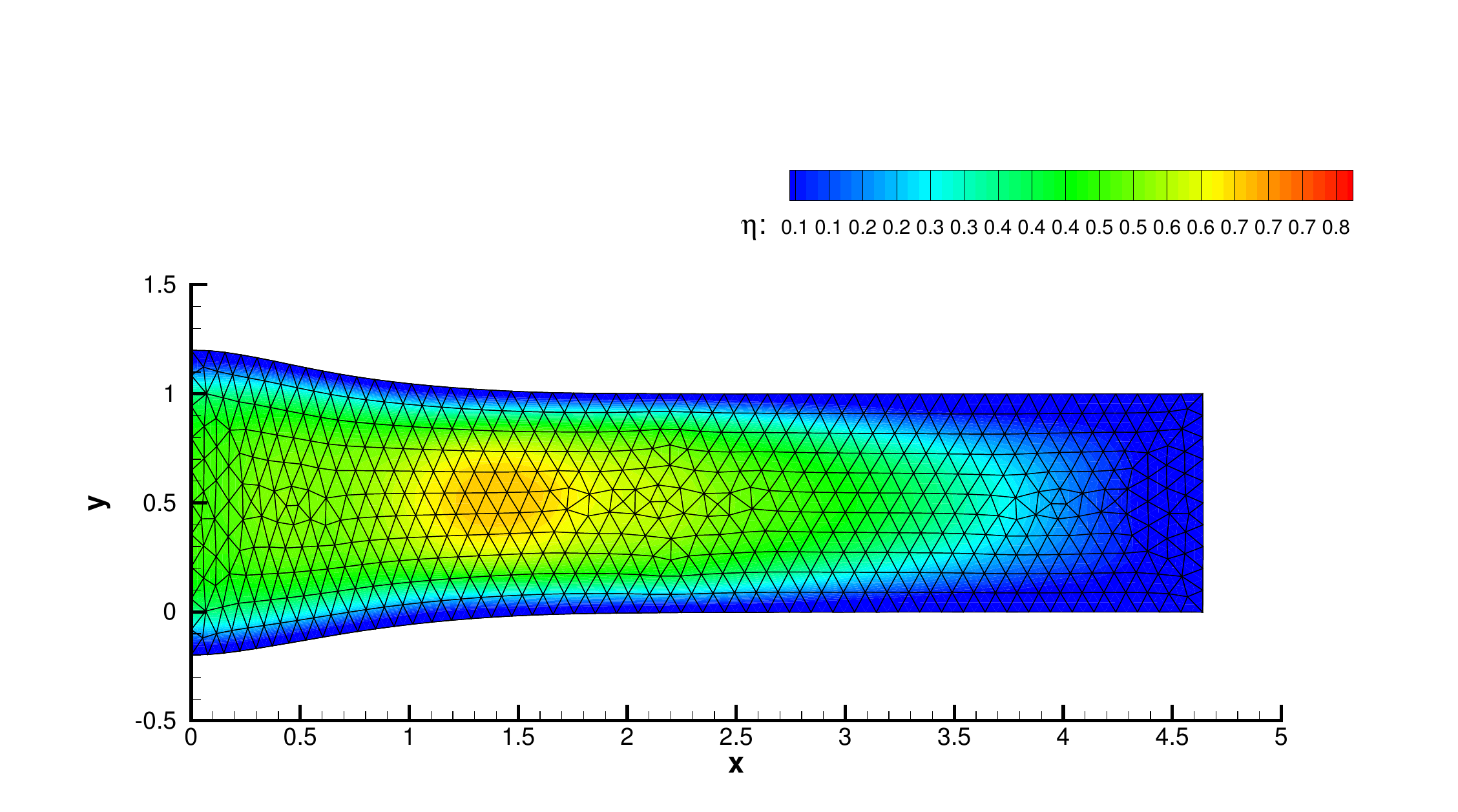}
			  &           
			\includegraphics[draft=false,width=0.47\textwidth]{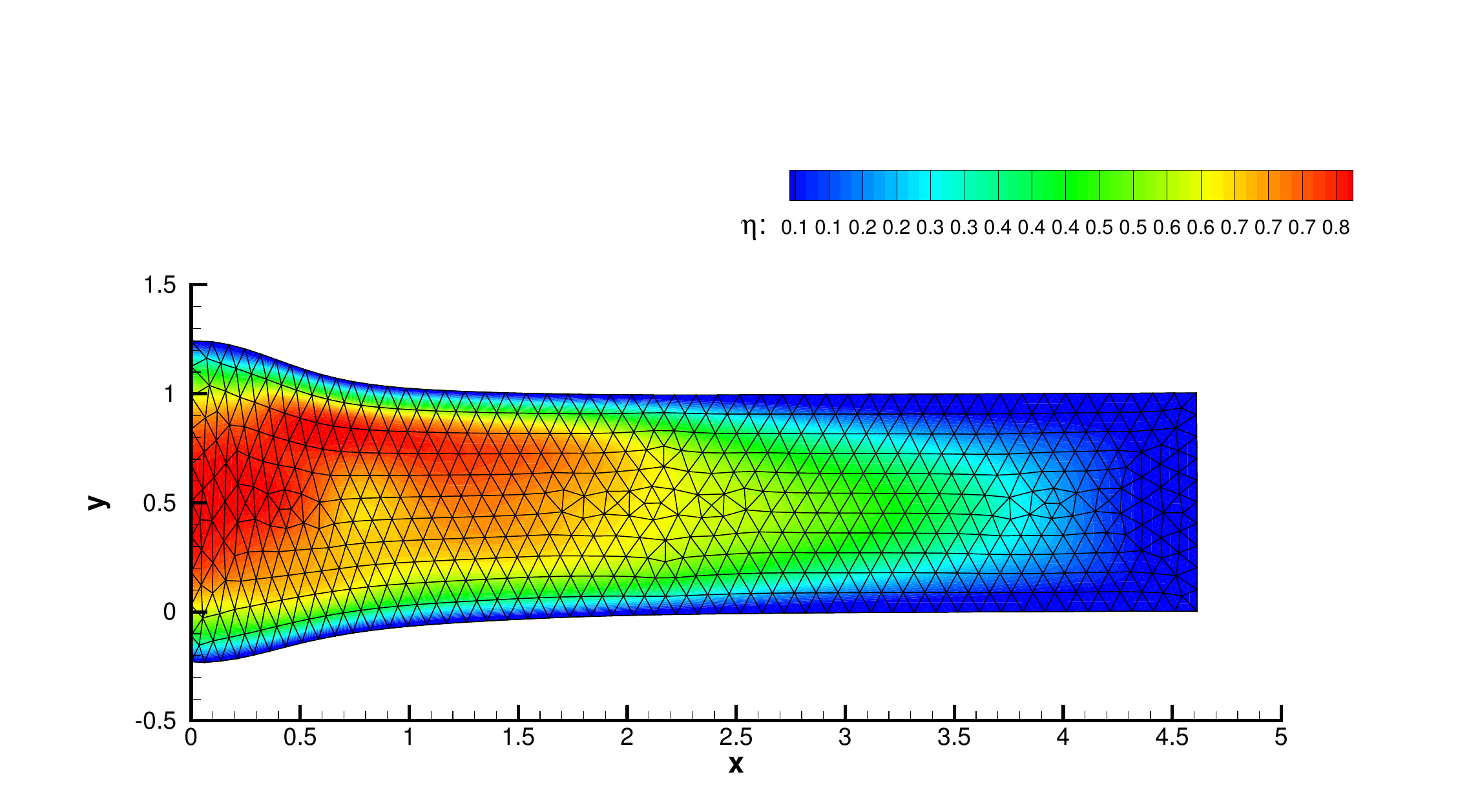}
			 \\
			\includegraphics[draft=false,width=0.47\textwidth]{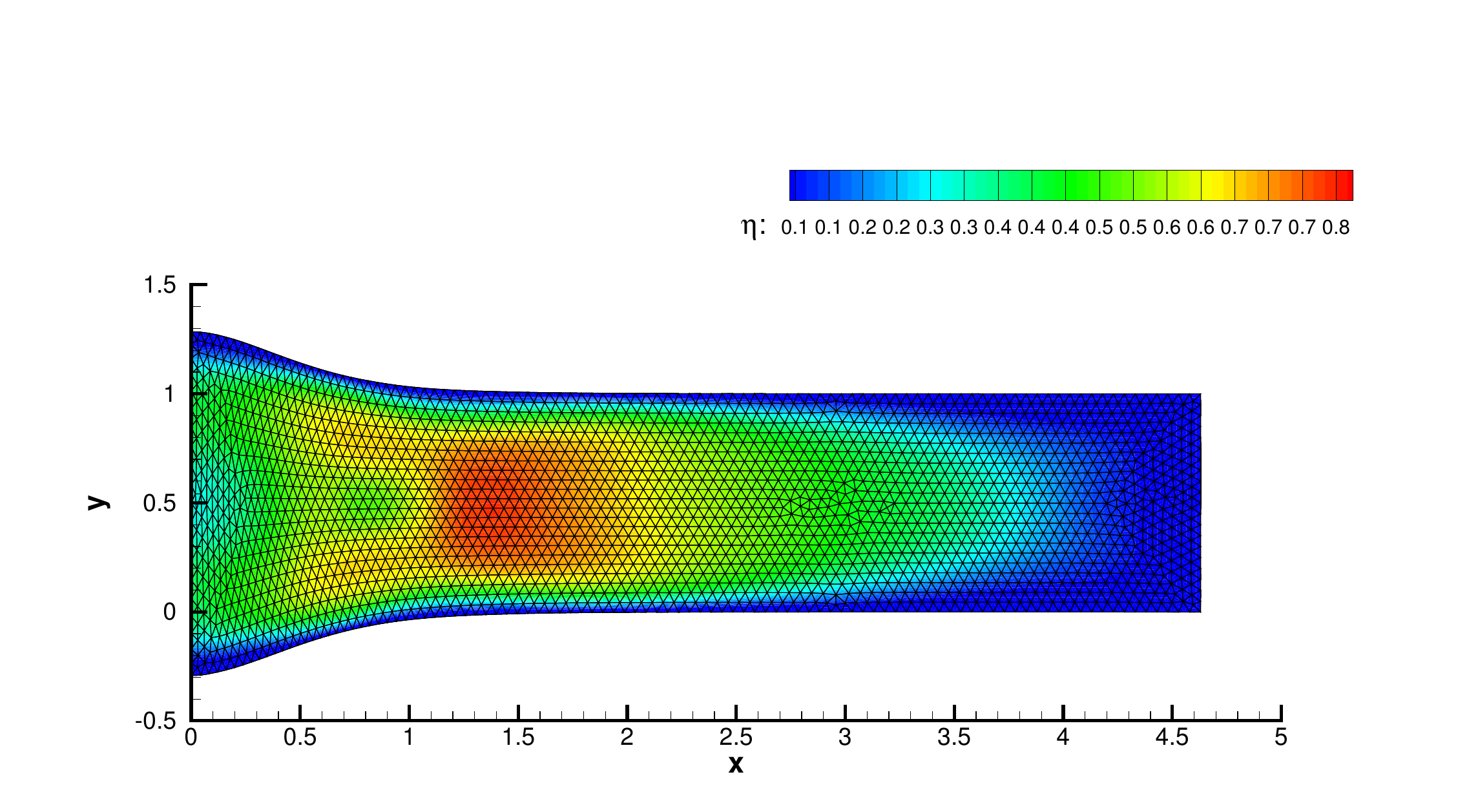}
			  &           
			\includegraphics[draft=false,width=0.47\textwidth]{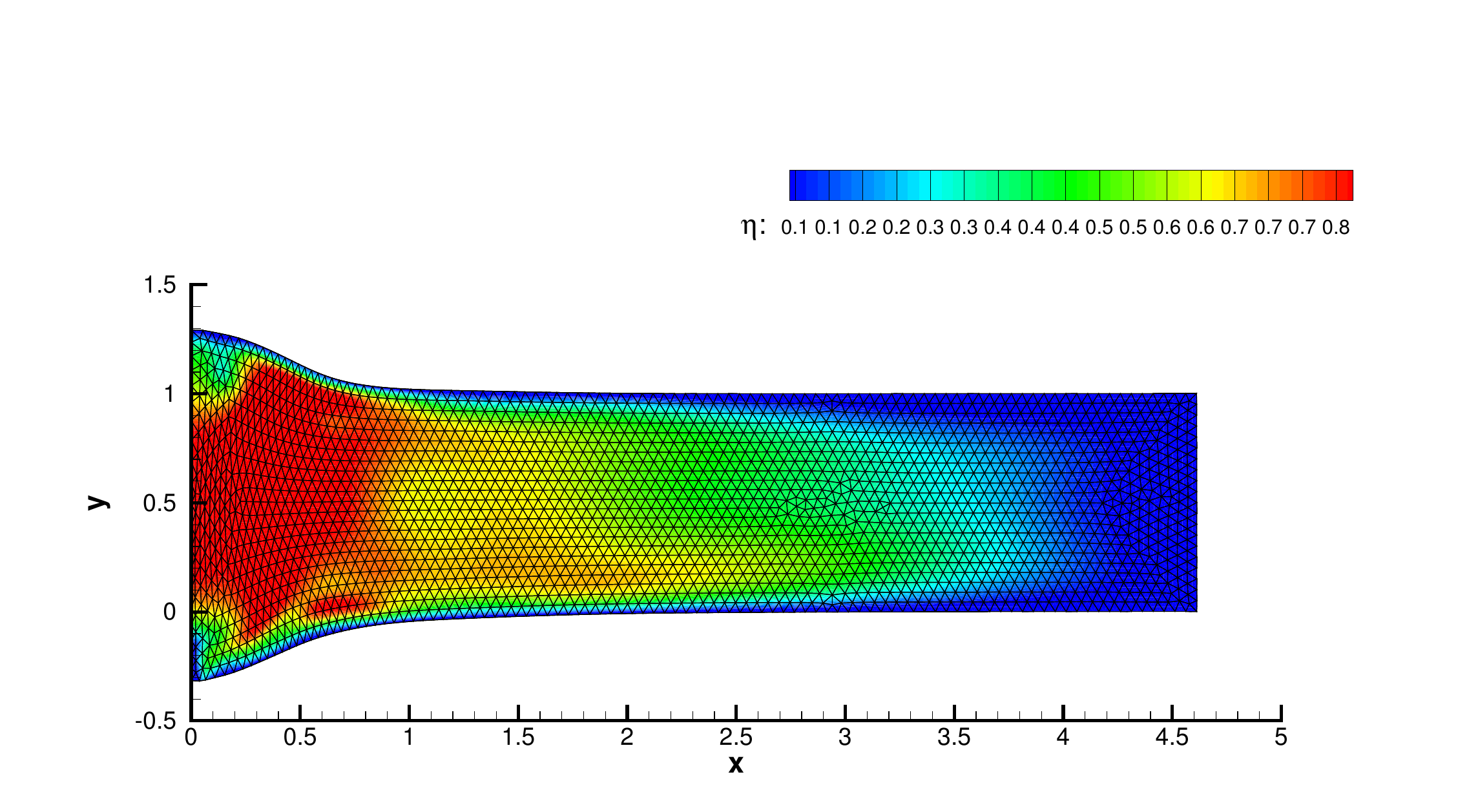}
			 \\
		\end{tabular} 
		\caption{Plasticity color map of the Taylor rod test problem at the final time 
		$t_{\text{final}}=0.005$ computed on an unstructured grid of characteristic mesh size 
		$h=0.10$ (top row) and $h=0.05$ (middle row) with  ADER-WENO third order schemes. Left 
		column: hypoelastic Wilkins model results. Right column: hyperelastic \GPR model results. } 
		\label{fig.TaylorRod}
	\end{center}
\end{figure}

\begin{figure}
  \begin{center}
  \begin{tabular}{cc}
	    \includegraphics[draft=false,width=0.47\textwidth]{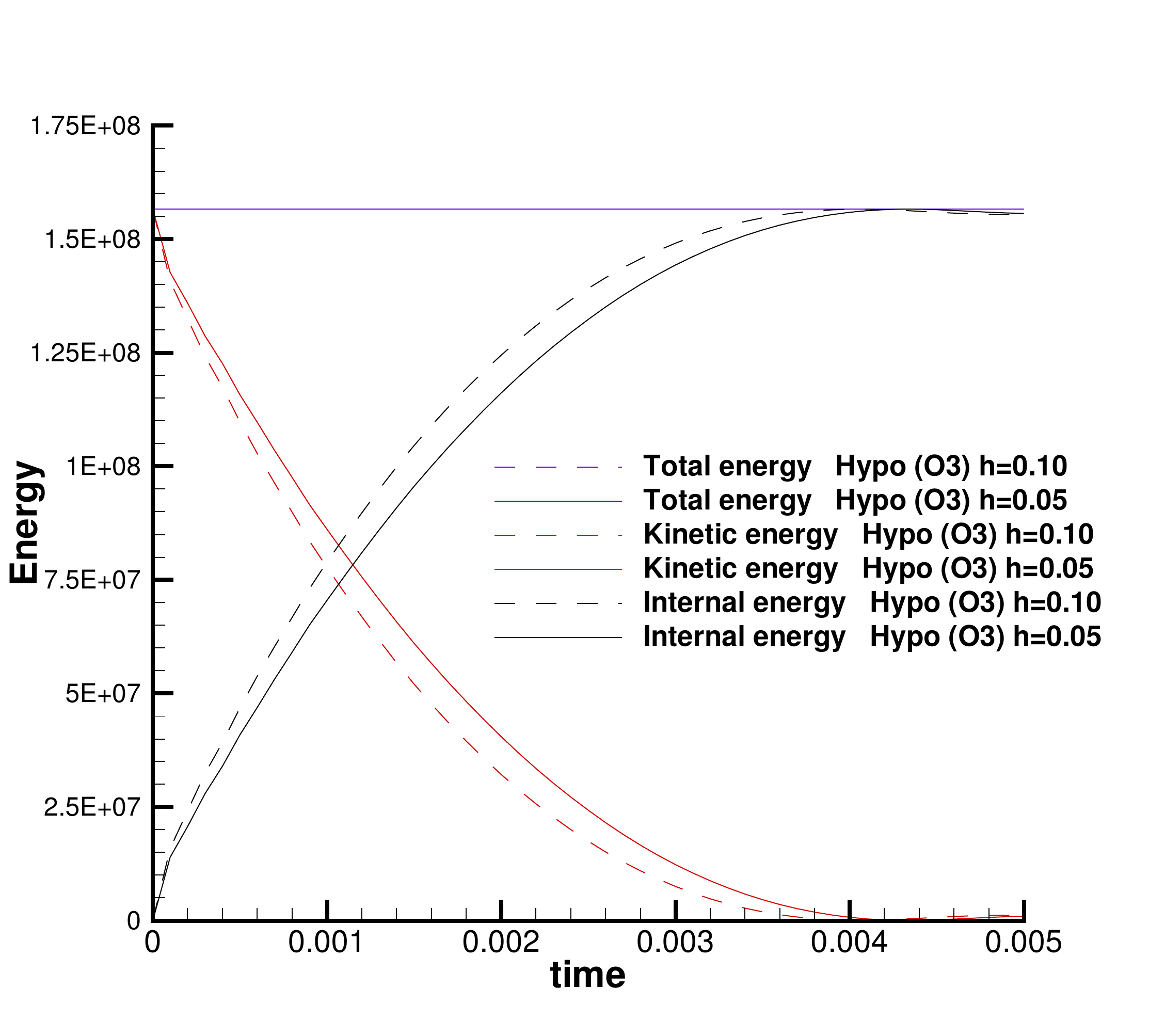}
	      &           
			\includegraphics[draft=false,width=0.47\textwidth]{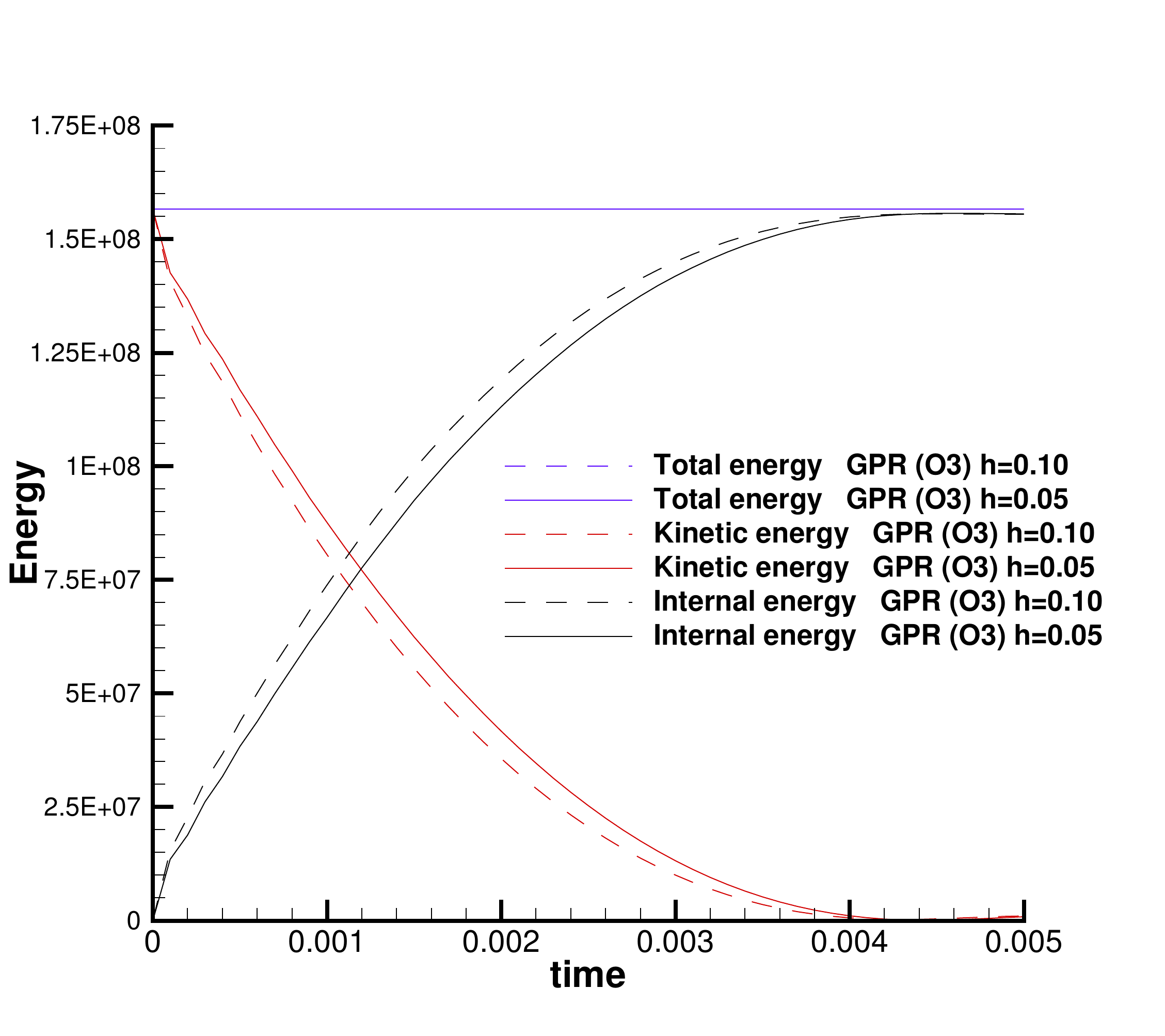}
			 \\
      \includegraphics[draft=false,width=0.47\textwidth]{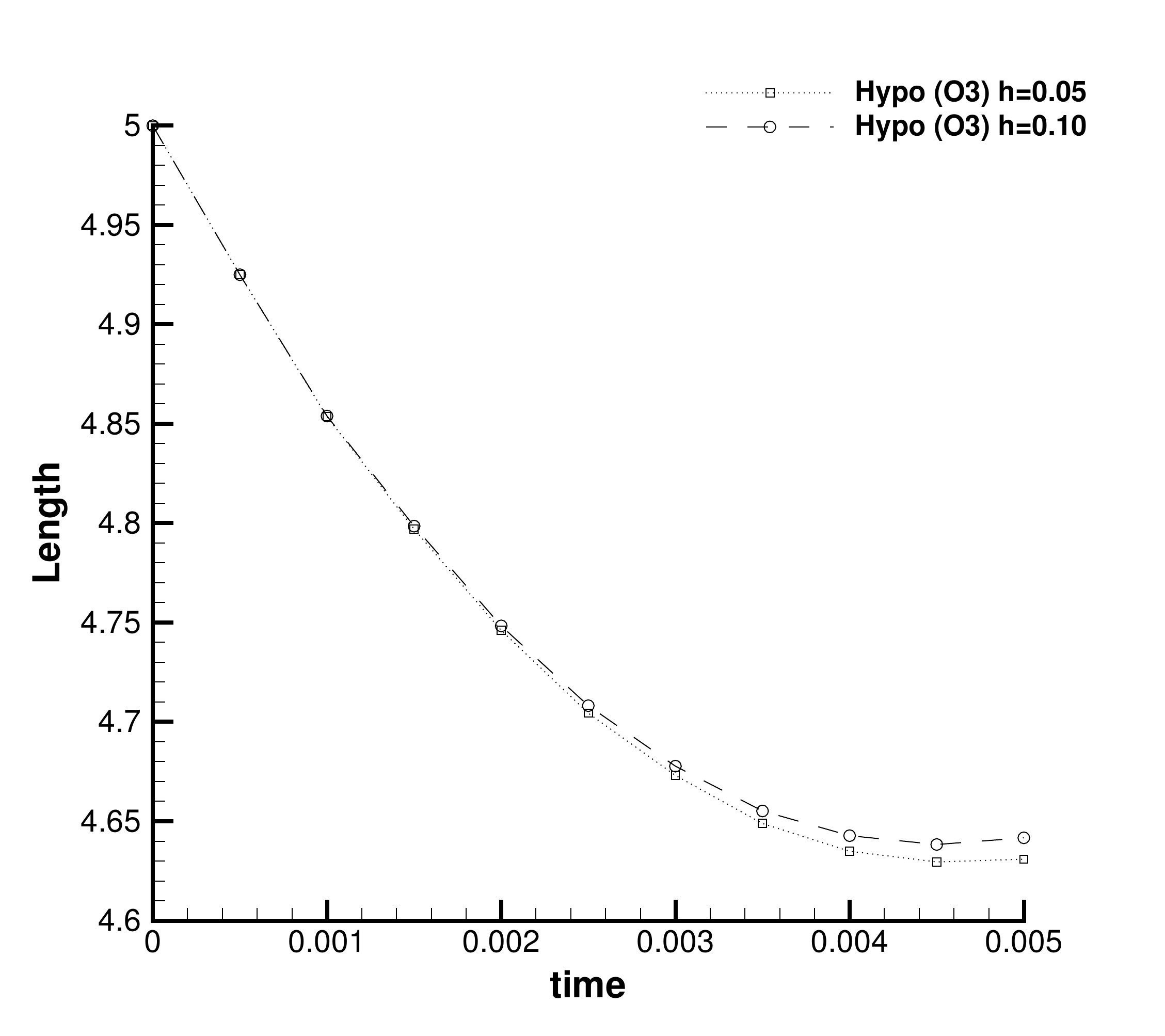}
        &           
			\includegraphics[draft=false,width=0.47\textwidth]{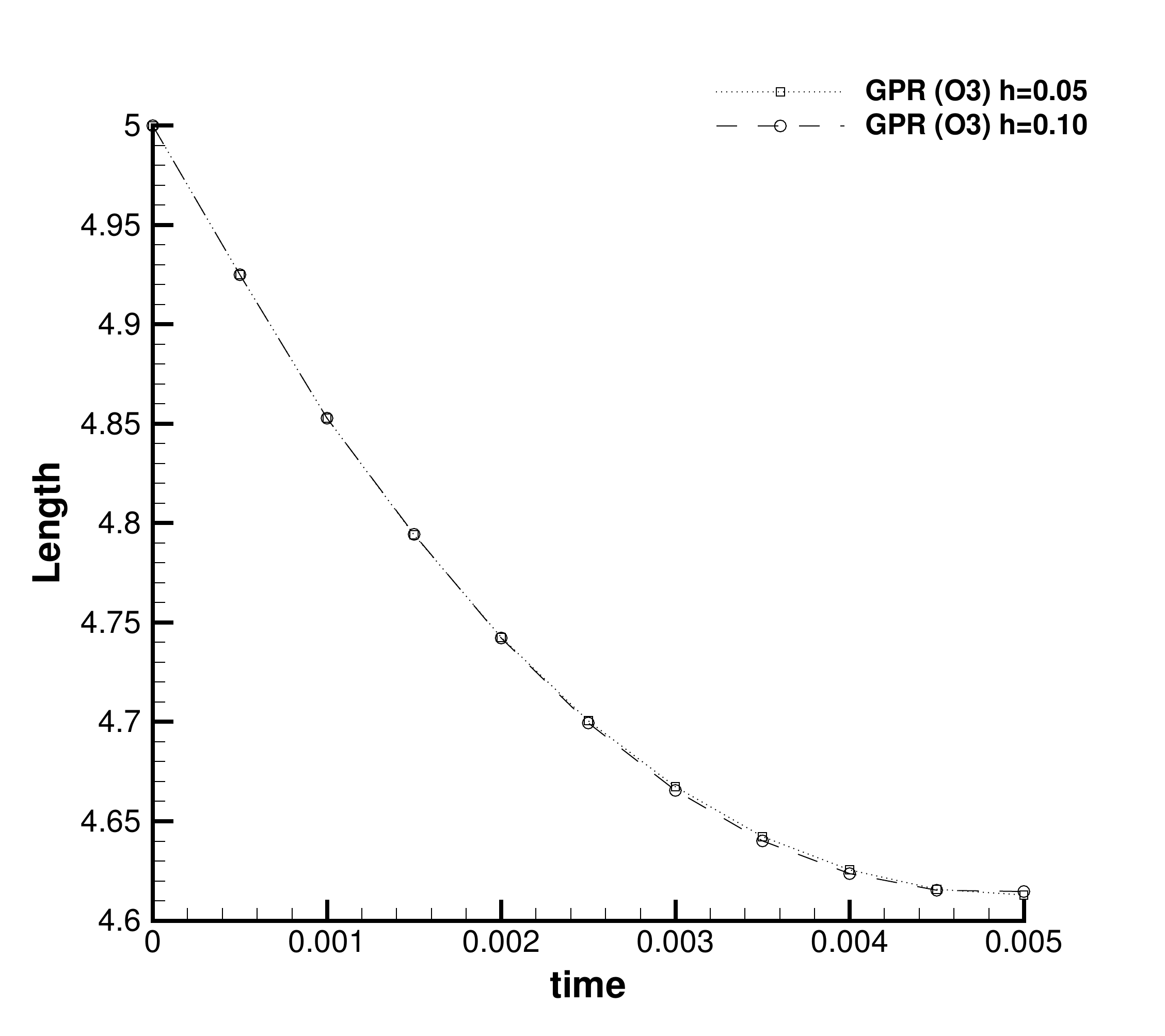}
			 \\
  \end{tabular}
  \includegraphics[draft=false,width=0.5\textwidth]{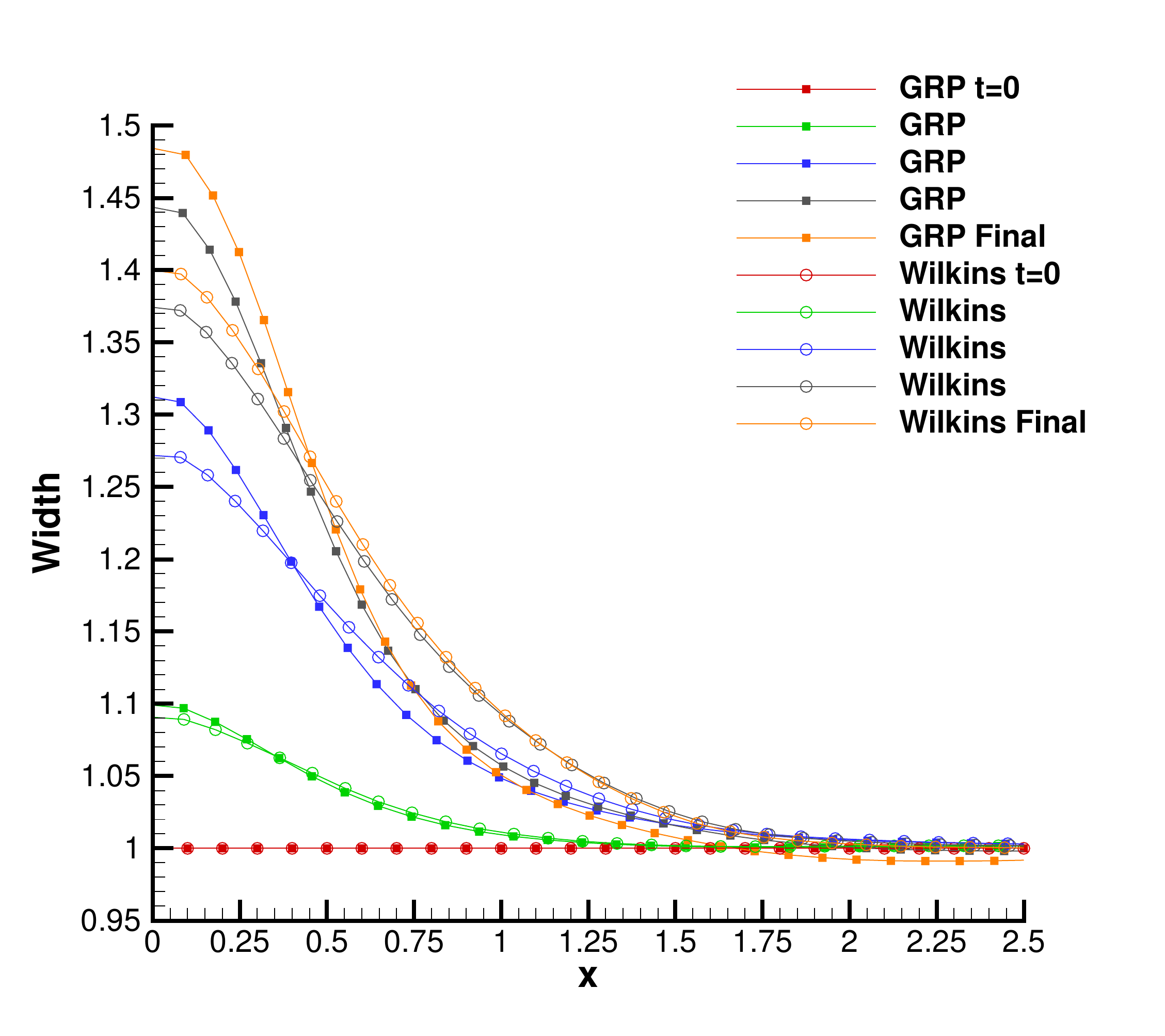}
   \\
  \caption{Top: analysis of energy conservation for the Taylor rod test problem for both computational meshes and governing equations. Middle: time evolution of the target length for both computational meshes and governing equations.
  Bottom: diagnostics on the bar width for $5$ time instants.}
  \label{fig.TaylorRod-analysis}
\end{center}
\end{figure}

This problem consists of the impact of a two-dimensional aluminum bar impacting on a
rigid wall. Initially proposed as a cylindrical bar by Taylor \cite{Taylor}, a planar 
geometry 
counterpart
has been designed and simulated in \cite{Kluth2010,Maire2013}. This later setup is considered
in this paper. 
The computational domain is the initial projectile $\Omega(t=0)=[0:5]\times [0:1]$ and the mesh is made
 of about $8960$ ($h=0.10$) or $35872$ ($h=0.05$) cells. 
The final time is set to $t_{\text{final}}=0.005$.
We consider aluminum material with the   initial reference data from Table~\ref{tab:dataEOS}. Only 
the initial velocity is set to $\vv=(-150,0)$.
The boundary conditions are free traction for all boundaries except the left boundary which is a wall type.
Although there exists no exact solution for this problem it is nonetheless employed for robustness and accuracy
benchmarking. During the impact the kinetic energy is entirely converted into the internal energy 
through plastic dissipation. 
The results are displayed in Fig.~\ref{fig.TaylorRod} where we have plotted the meshes and the norm 
of the deviatoric stress $\sigma=\sqrt{3 \tr(\dev{\bm{\sigma}}^2)/2} $ 
in colors for a coarse (top) and
refined mesh (bottom panel) for the hypoelastic model of Wilkins (left column) and the hyperelastic 
\GPR model (right column).
Since the same color scale is used, we can observe that the two models produce different 
distribution of $ \sigma $ even if the general behavior 
of the rod is alike. 
In Fig.~\ref{fig.TaylorRod-analysis}, we present the internal, kinetic and total energies for 
Wilkins model 
on the left and \GPR model on the right for the coarse (dashed line) and fine (straight line) mesh 
results.
As expected for both models, the total energy is conserved and the kinetic energy is decreasing to 
the benefit of the 
internal energy up to the time at which the bar stops moving. At this time the whole kinetic energy 
has been converted into the internal energy.
On the middle panels of Fig.~\ref{fig.TaylorRod-analysis}, we also plot the evolution of the length 
of the bar. It seems that for the 
hyperelastic \GPR model the convergence towards its limit is mildly dependent on the mesh 
resolution unlike
the hypoelastic model of Wilkins. Note that such mesh resolution dependency has already been 
observed in \cite{Maire2013}. 
At last, the bottom panel of Fig.~\ref{fig.TaylorRod-analysis} presents the position comparison of 
the bar side boundary as a function of $x$.
Five different times are plotted for both models. We can observe that the shape of the bar is truly different
already at the third time corresponding to the middle of the simulation. The differences do 
increase in time leading to rather different final shapes.

Finally, we remark that the plasticity models which are based on the ideal plasticity constitutive 
law, like the Wilkins model, usually require an incorporation of an extra strain-hardening scalar 
parameter in order to fit 
experimental data. On the other hand, the strain hardening due to the 
finiteness of the rate of the dislocation nucleation is, in fact, accounting for in the rate 
dependent character of the \GPR model and does not require the introducing of extra state variables,  
and a good agreement with the experimental data can be achieved~\cite{Barton2012,Hank2017} with a 
proper model for the strain dissipation time $ \tau $. 

\subsection{Elastic shear layer test} \label{ssec:shear}
\begin{figure}
  \begin{center}
  \begin{tabular}{cc}
	    \includegraphics[draft=false,width=0.47\textwidth]{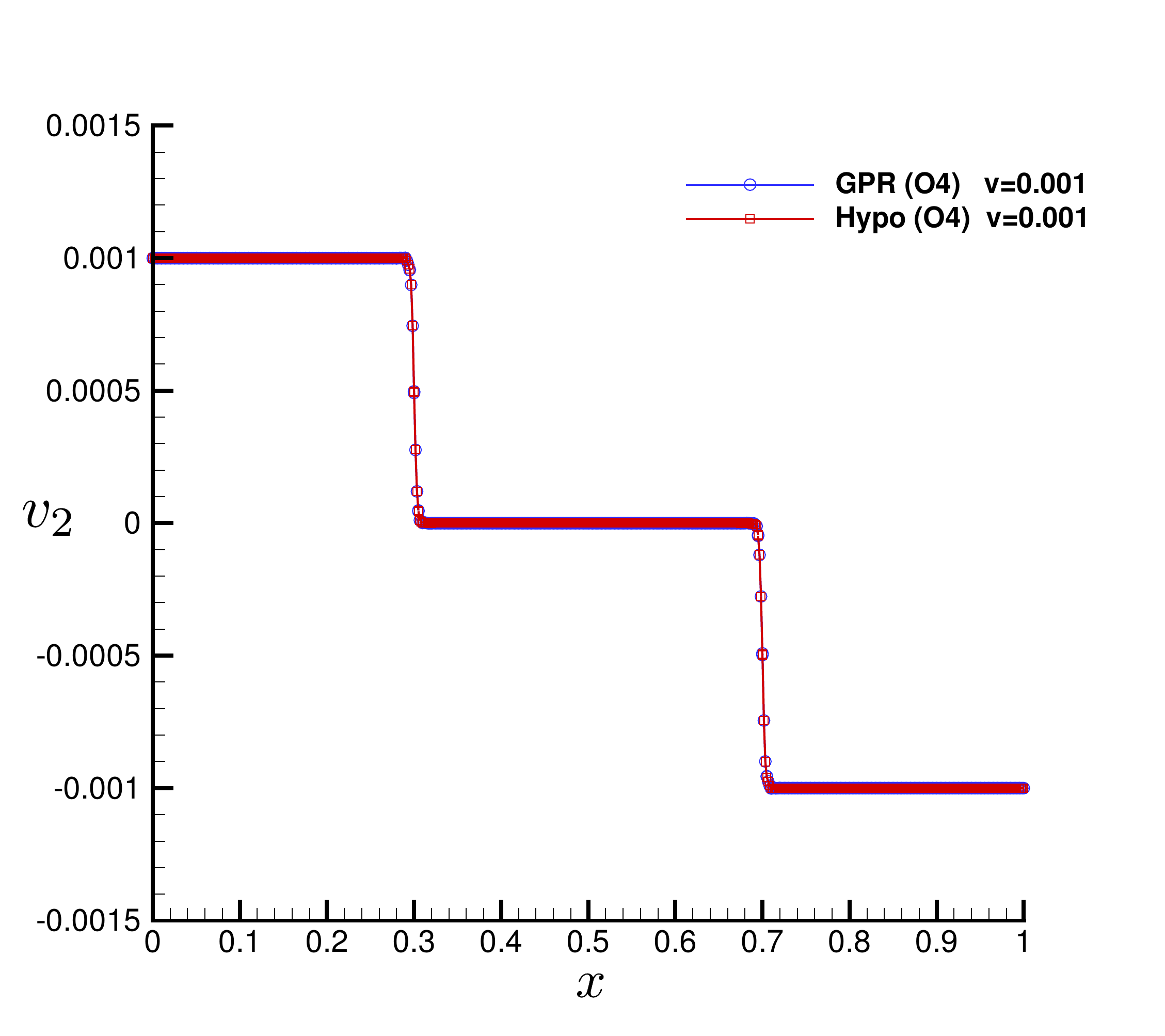}
	      &  
			\includegraphics[draft=false,width=0.47\textwidth]{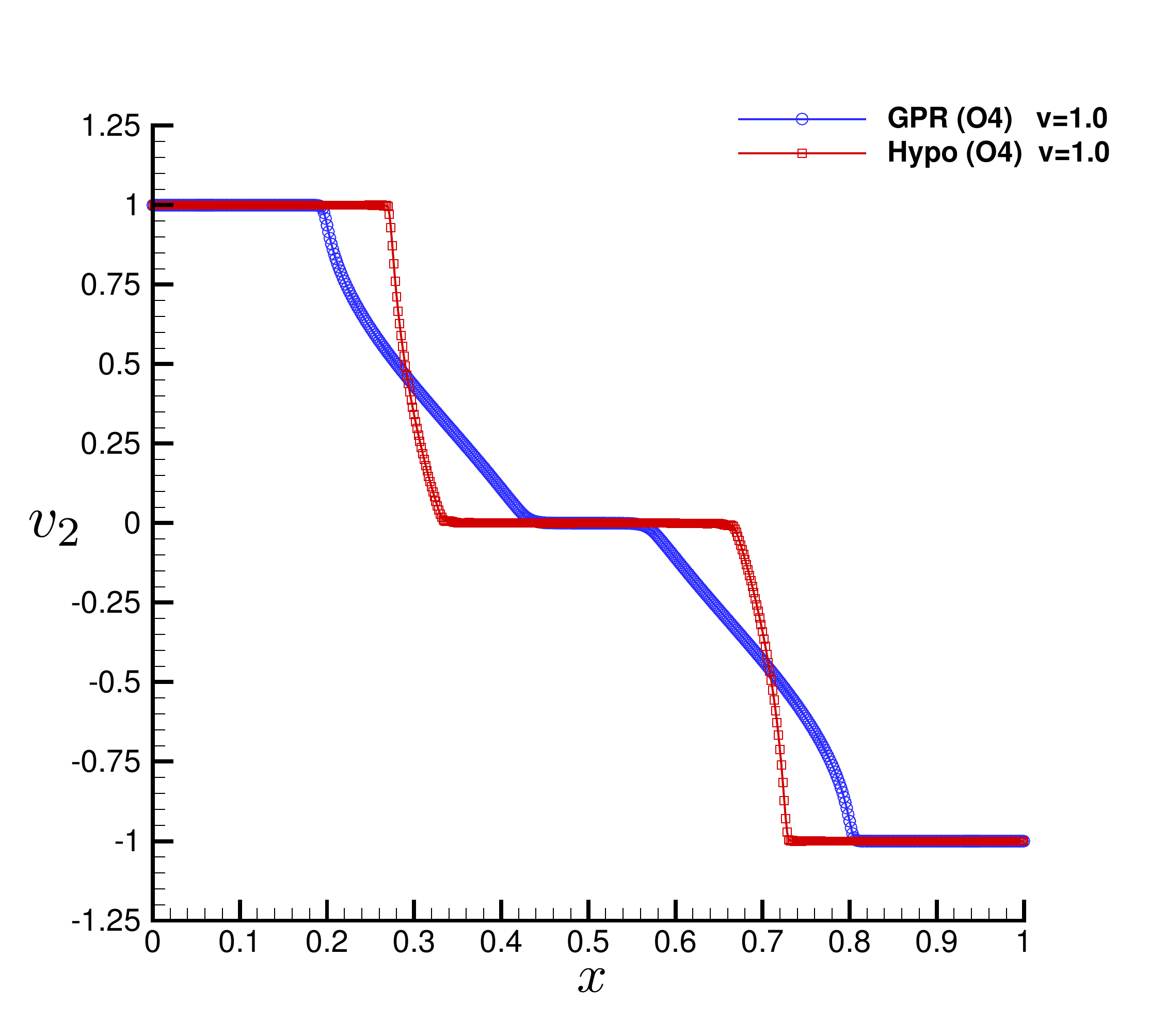}  
			      \\				
			\includegraphics[draft=false,width=0.47\textwidth]{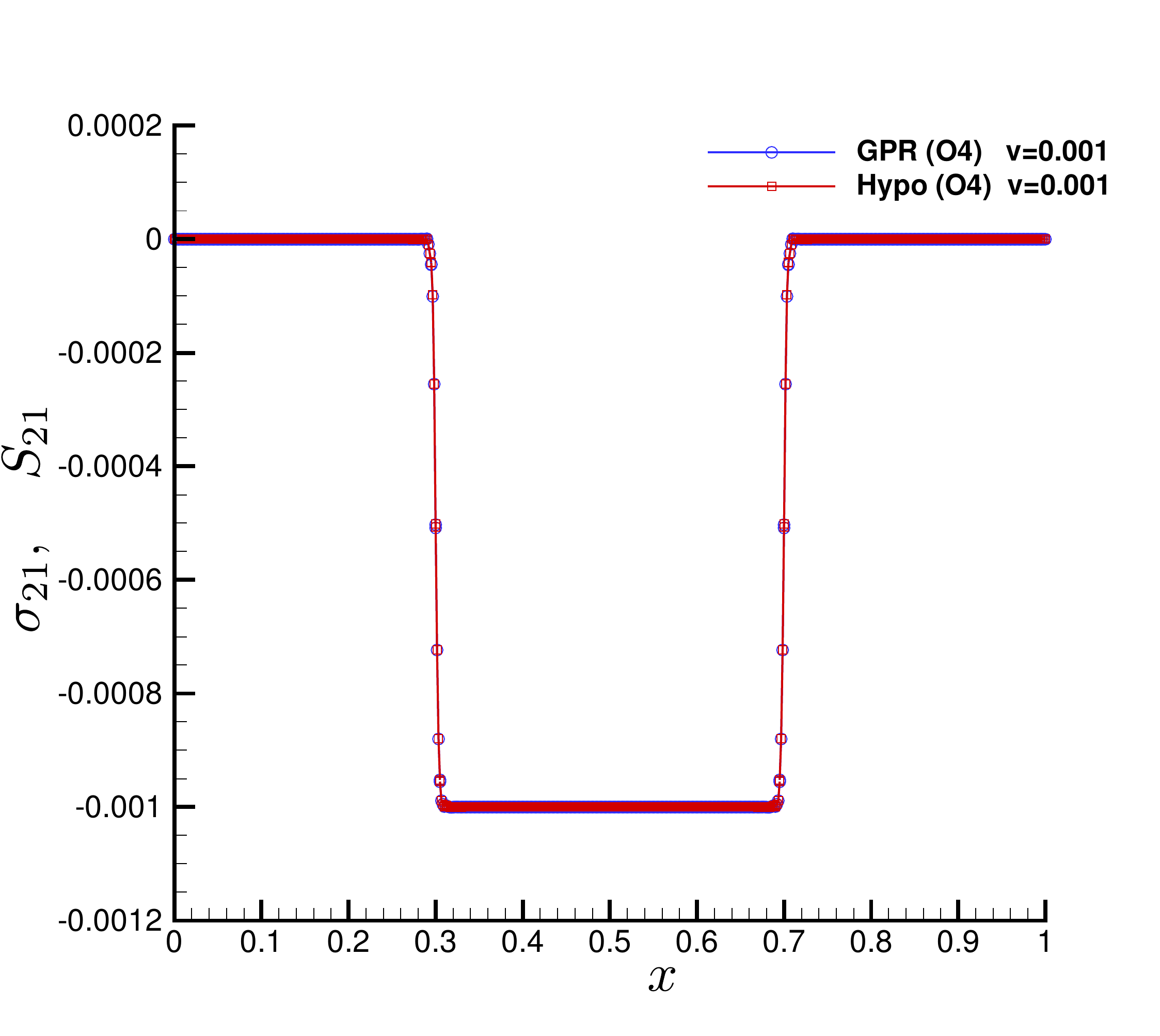}
			 &
			\includegraphics[draft=false,width=0.47\textwidth]{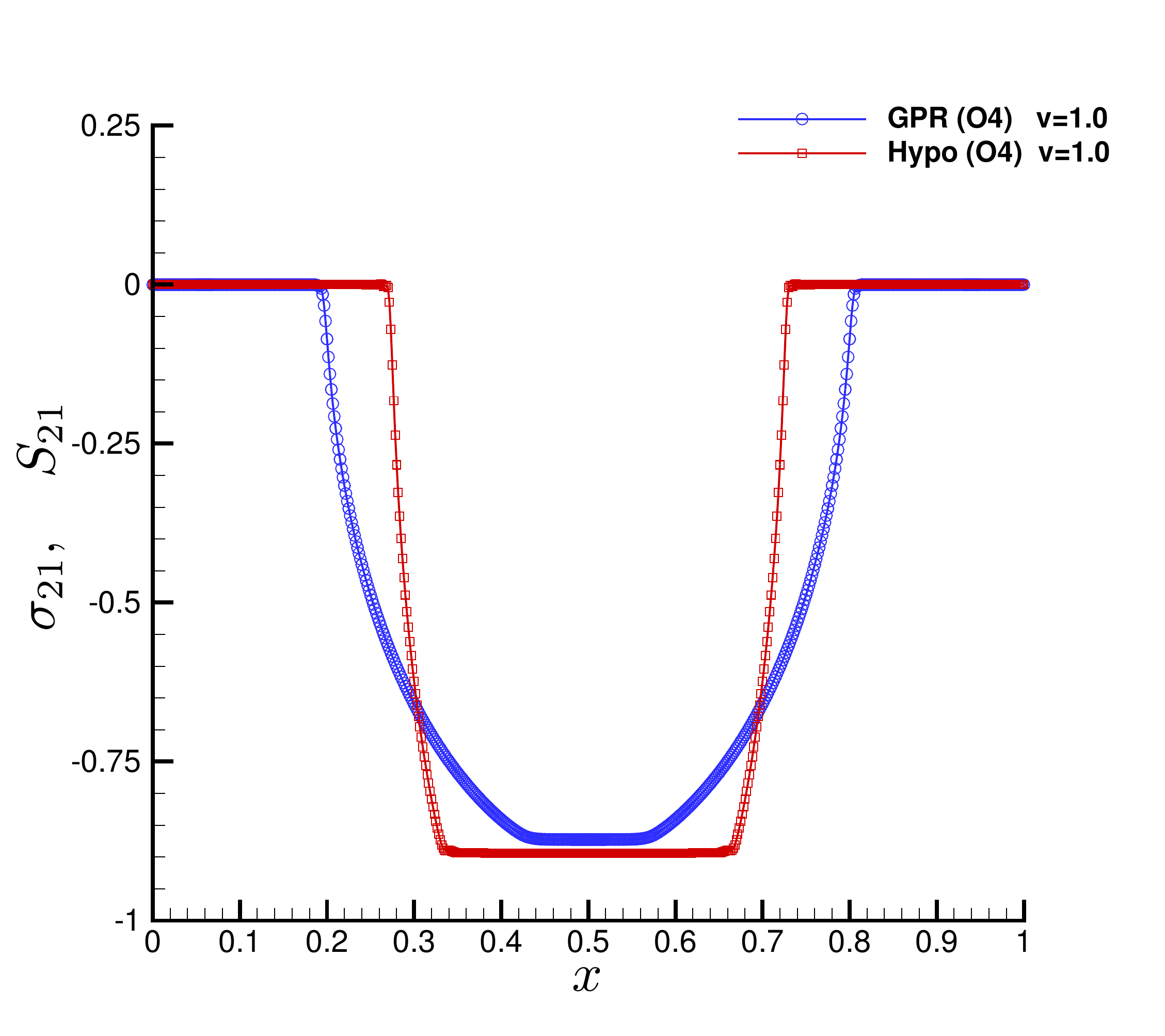}
			 \\
			\includegraphics[draft=false,width=0.47\textwidth]{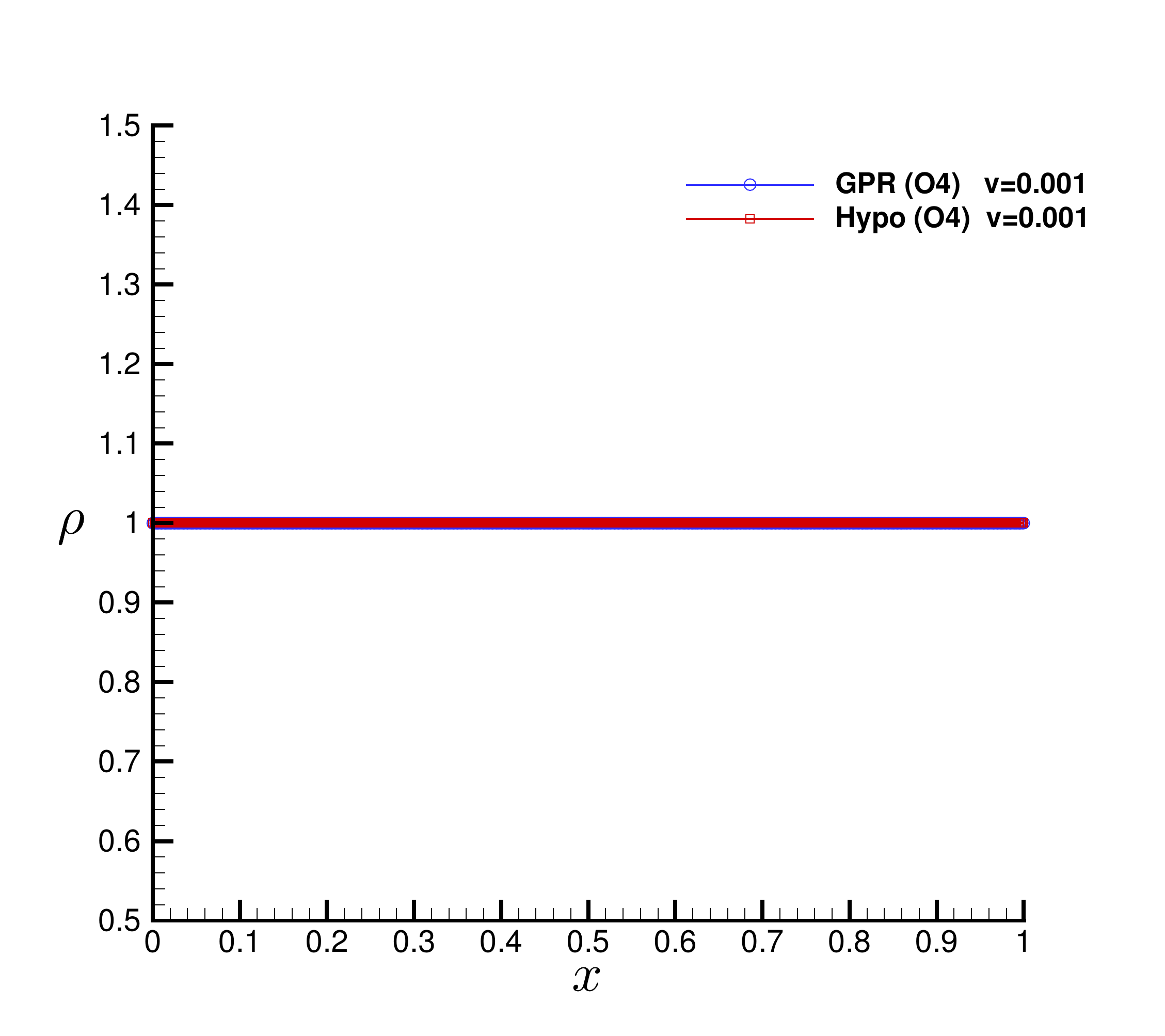}
			 &
			\includegraphics[draft=false,width=0.47\textwidth]{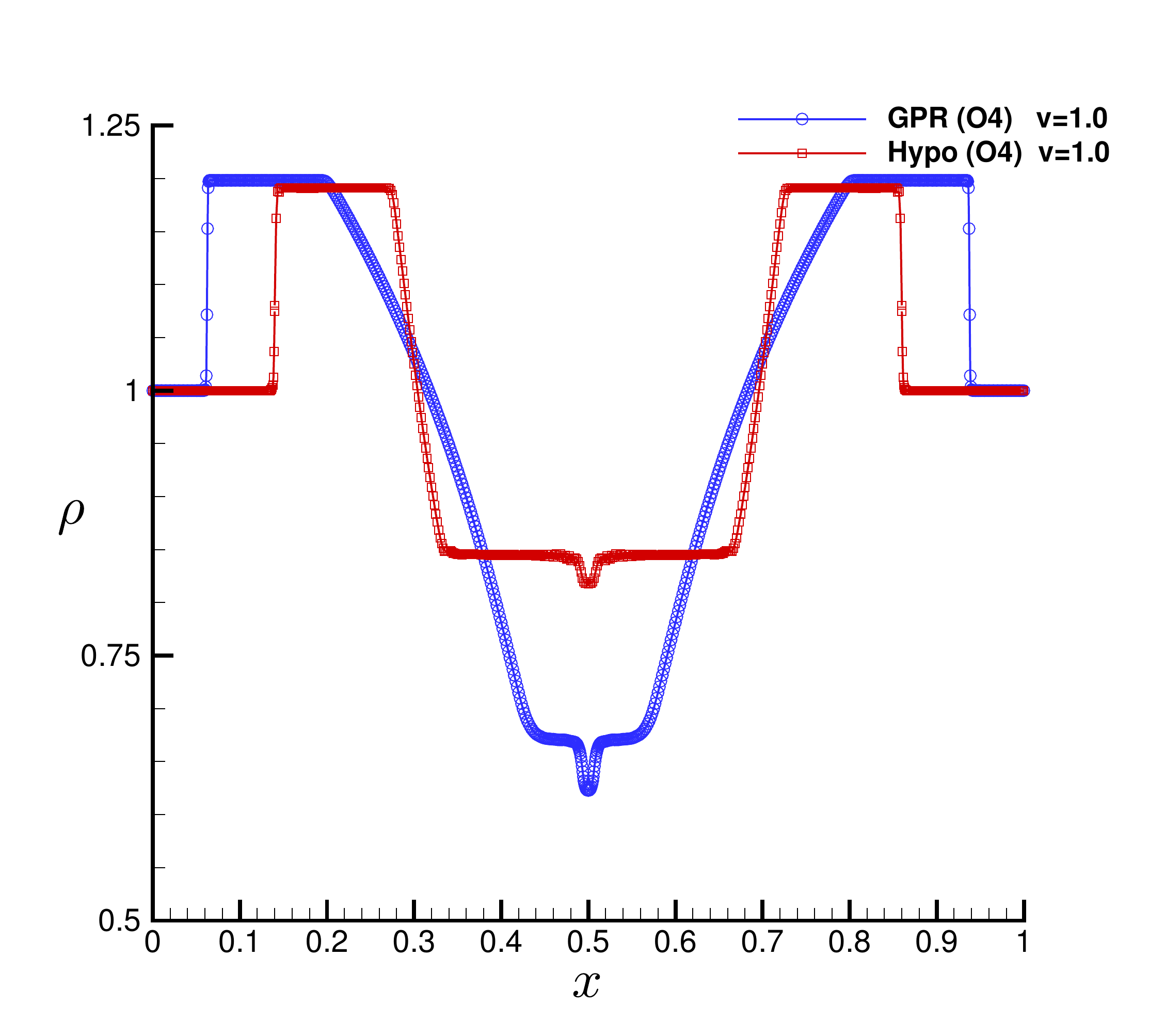}
			 \\
  \end{tabular}
  \caption{Shear layer test case results computed with ADER-DG fourth order scheme with initial 
  shear velocity $v^s_1=10^{-3}$ (left column) or $v^s_2=1$ (right column) for both models are 
  shown: velocity along the 
  $y-$axis (top row), shear stresses $ \sigma_{21} $ and $ S_{21} $ (middle row) and density 
  (bottom row). }
  \label{fig.shear}
\end{center}
\end{figure}

Here, we propose to solve a very simple test case which aims at showing the different responses of 
the hypoelastic and the hyperelastic materials in the non-linear elastic regime. The rectangular 
computational domain $\Omega(t=0)=[0:1]\times [0:0.1]$ is paved with $200 \times 10$ square cells. 
The material is characterized by $\rho_0=1$, $c_v=c_0=c_s=1$ and the static yield strength 
$\sigma_Y=10^{20}$ is set to a very large value to avoid plastic deformations to occur. The 
stiffened gas 
equations of state is used with $\gamma=1.4$ and $\pi_\infty=0$. The initial conditions are given 
in 
terms of a classical one-dimensional Riemann problem where a discontinuity is located at $x_d=0.5$ 
and 
separates the right state $\mathbf{U}_R=(\rho_R, u_R, v_R, p_R)=(\rho_0, 0, v^s, 1)$ from the left 
one $\mathbf{U}_L=(\rho_L, u_L, v_L, p_L)=(\rho_0, 0, -v^s, 1)$. For the \GPR model the distortion 
tensor is initialized with the identity matrix, i.e. $\A=\II$, while the shear velocity is 
assigned two different values, namely $v^s_1=10^{-3}$ and $v^s_2=1$. The 
numerical results have been computed using a fourth order ADER Discontinuous 
Galerkin discretization (ADER-DG), supplemented with the finite volume subcell limiter presented in 
\cite{DGLimiter1,DGLimiter2,ALEDG} and they are depicted in Fig.~\ref{fig.shear}. When the shear 
stress is mild, then the linear elasticity equations are recovered from both models, and, as 
expected, the results are almost identical, see left panels. On the contrary, the hyperelastic 
model computes a totally different solution (in terms of location, amplitude and shape of the waves) 
when the material is subjected to strong shear. This is of course not surprising, because the 
Wilkins model was designed for the regimes when the elastic deformations are small, while its 
extension to large non-linear elastic deformations is rather problematic and requires retaining 
more terms in the general hypoelastic time evolution equation \eqref{stress.rate.gen} and, what 
is more challenging, it requires solution dependent elastic moduli in \eqref{stress.rate.gen}, 
\eg see~\cite{Kim2016}. This is why hypoelastic models are only very rarely used for modeling 
large elastic deformations (occurring for example in rubber). On the contrary, the hyperelastic 
models, and in particular the \GPR model, provide a more flexible framework for modeling 
linear and non-linear elastic and elastoplastic deformations. 

\subsection{Elastic rotor problem} \label{ssec:rotor}

\begin{figure}
  \begin{center}
  \begin{tabular}{cc}
	    \includegraphics[draft=false,width=0.47\textwidth]{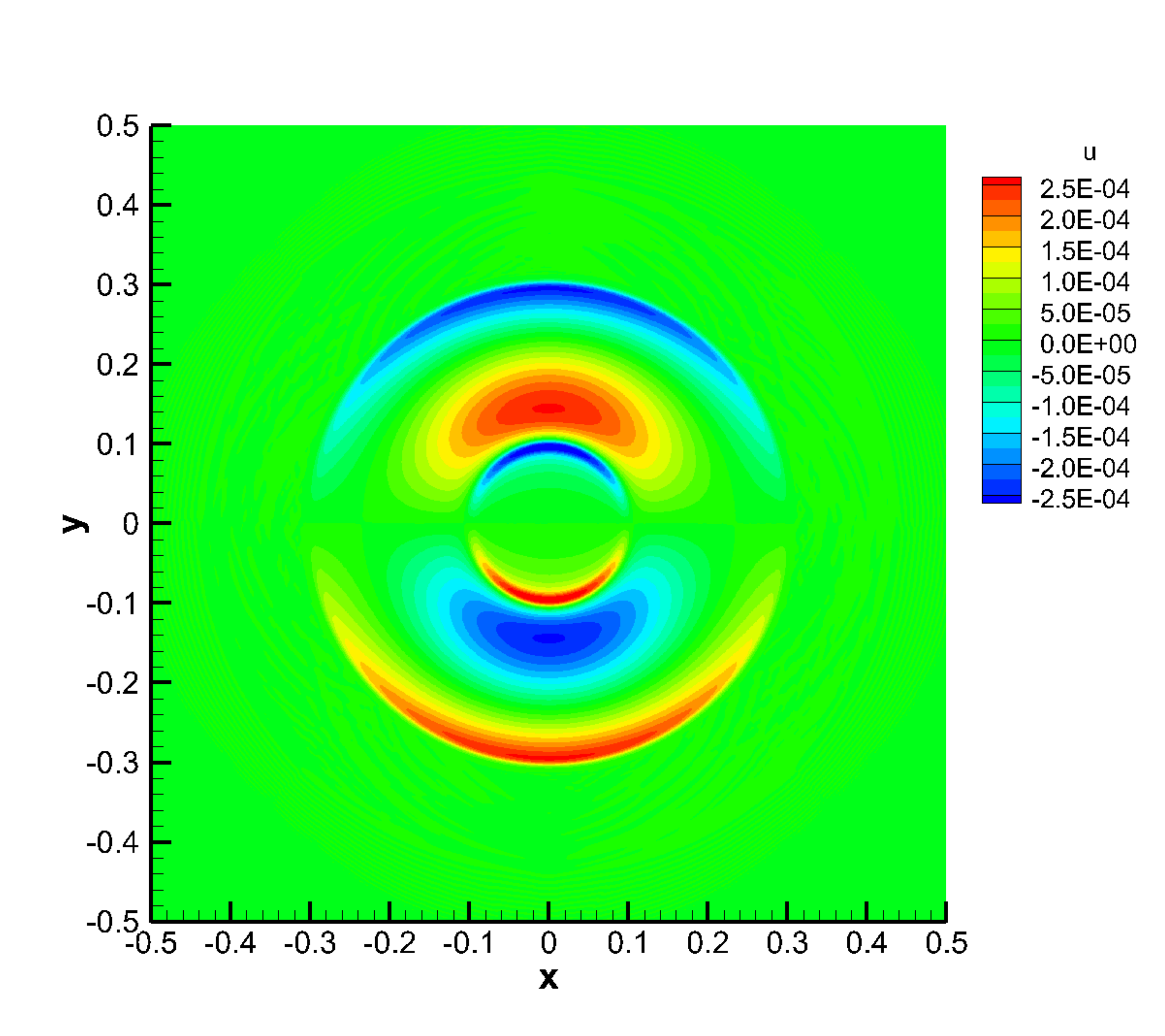}
	      &           
			\includegraphics[draft=false,width=0.47\textwidth]{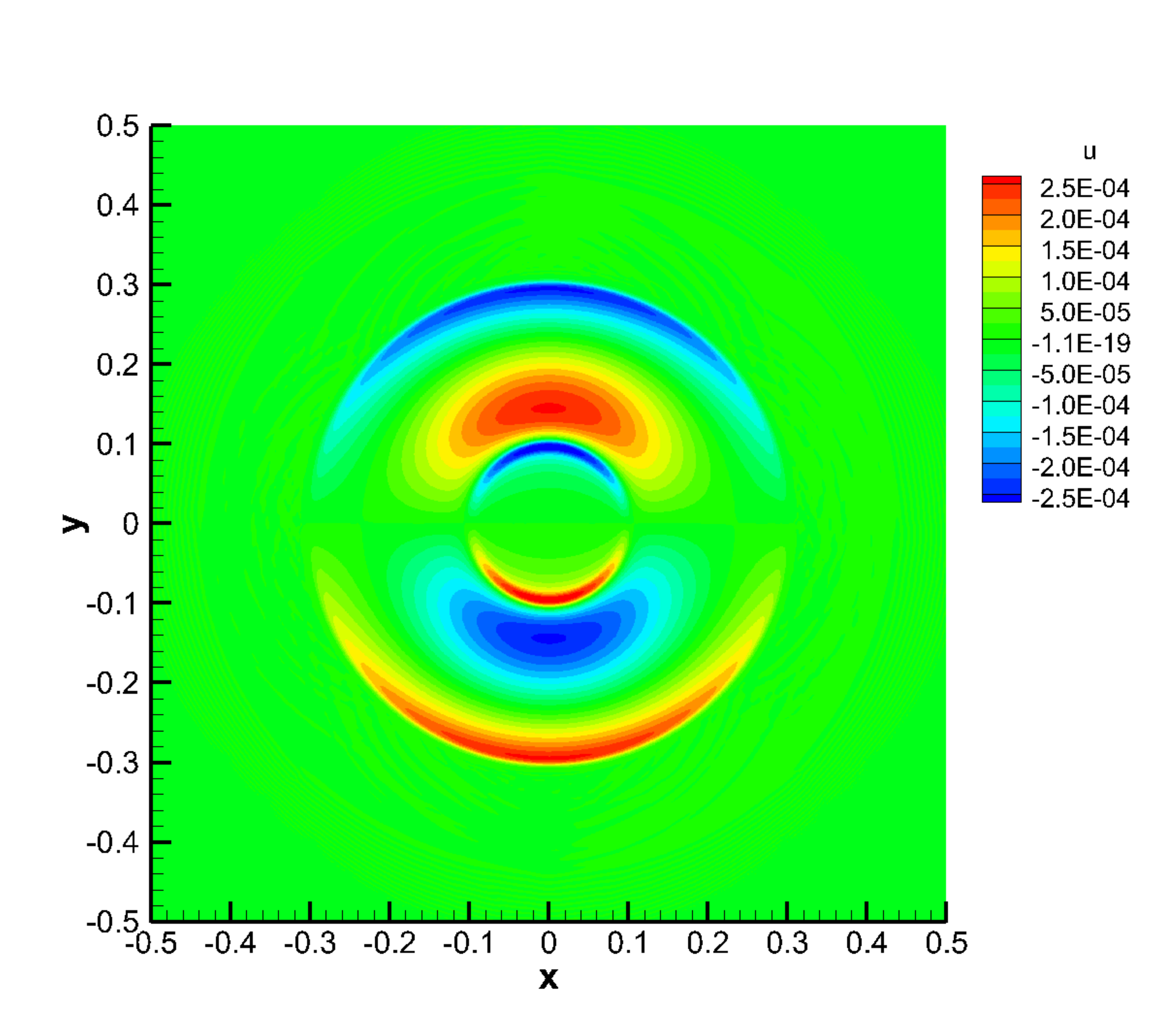}
			 \\
      \includegraphics[draft=false,width=0.47\textwidth]{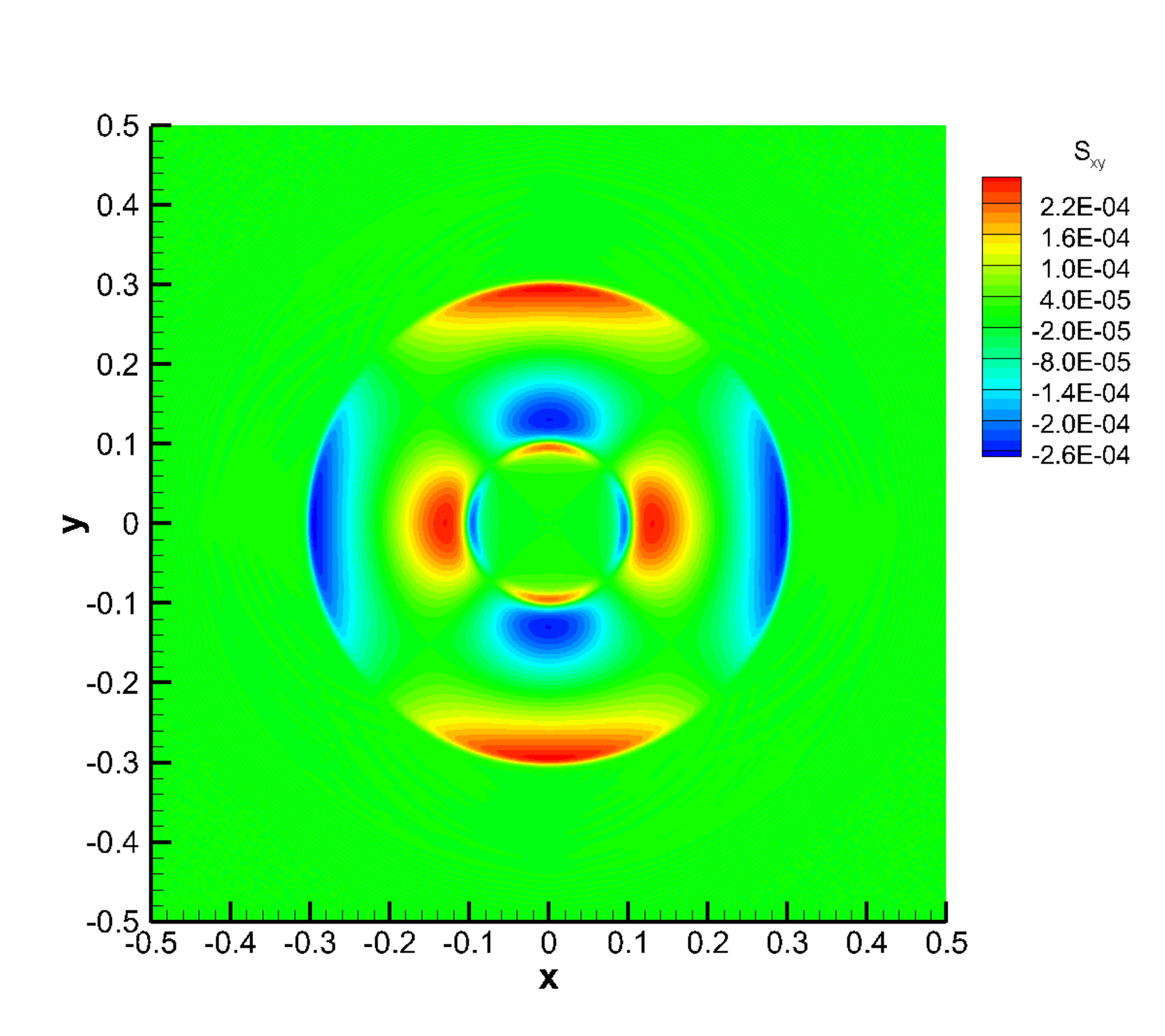}
        &           
			\includegraphics[draft=false,width=0.47\textwidth]{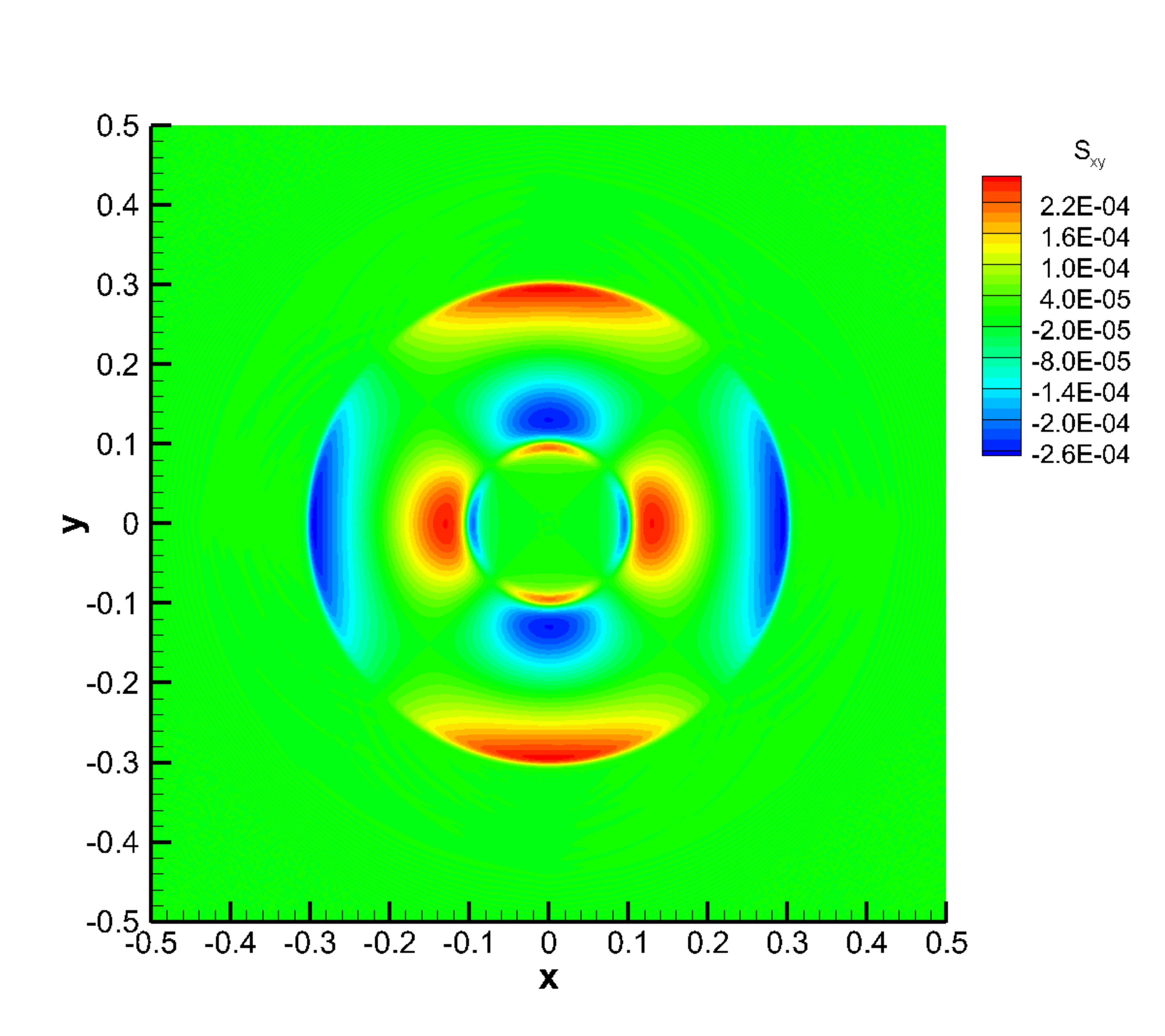}
			 \\
  \end{tabular}
  \caption{Results for the solid rotor problem obtained with ADER-DG fourth order scheme with 
  initial angular velocity of the rotor of $\vv\times \omega=0.001$. Horizontal velocity (top row) 
  and stress tensor component $ \sigma_{21} $ and $S_{21}$ (bottom row) are displayed.
  The Wilkins model results are displayed on the left panels, \GPR model ones on the right panels.}
  \label{fig.rotor-small}
\end{center}
\end{figure}

\begin{figure}
  \begin{center}
  \begin{tabular}{cc}
	    \includegraphics[draft=false,width=0.47\textwidth]{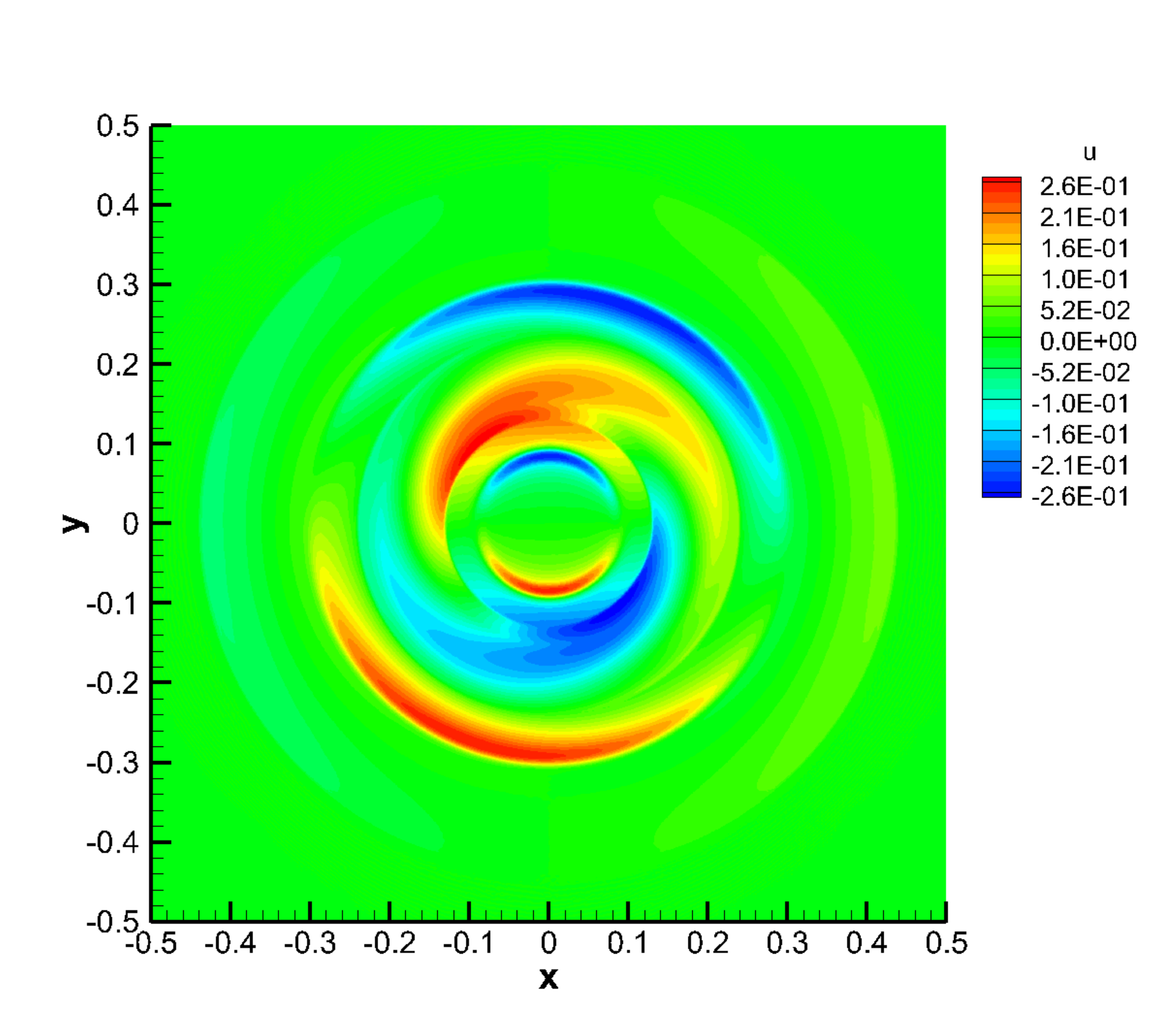} 
	     &           
			\includegraphics[draft=false,width=0.47\textwidth]{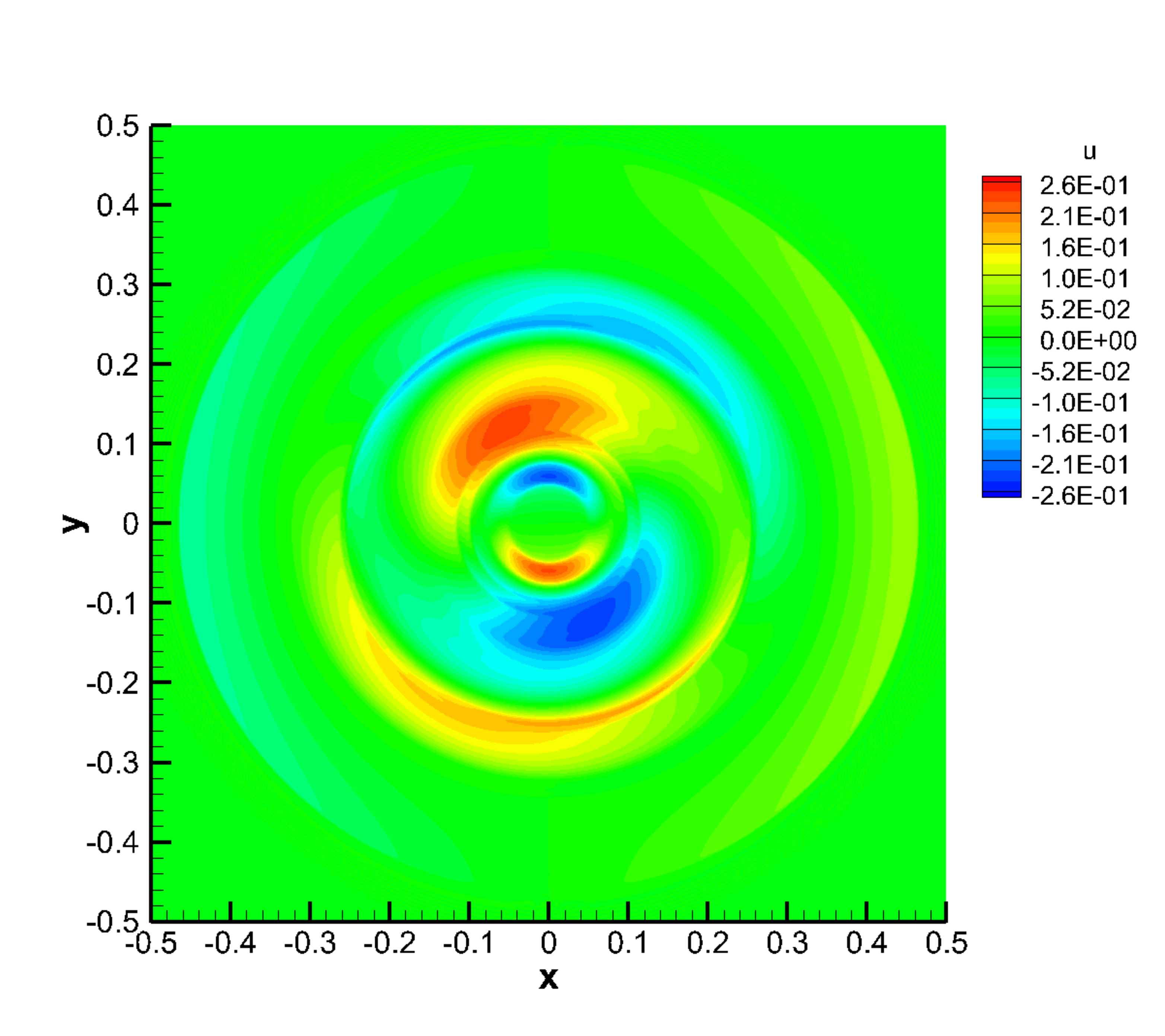}
			 \\
      \includegraphics[draft=false,width=0.47\textwidth]{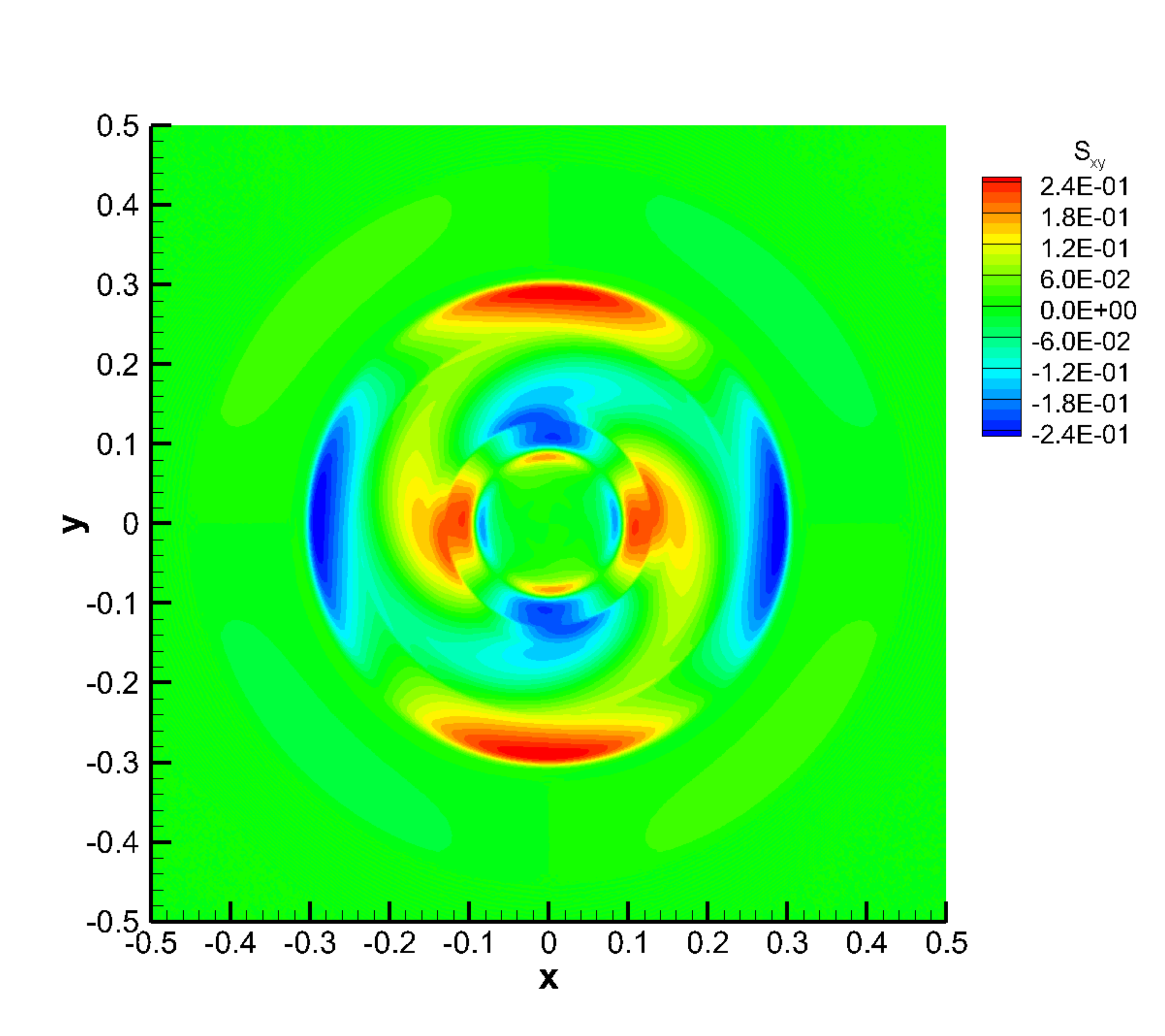} 
       &           
			\includegraphics[draft=false,width=0.47\textwidth]{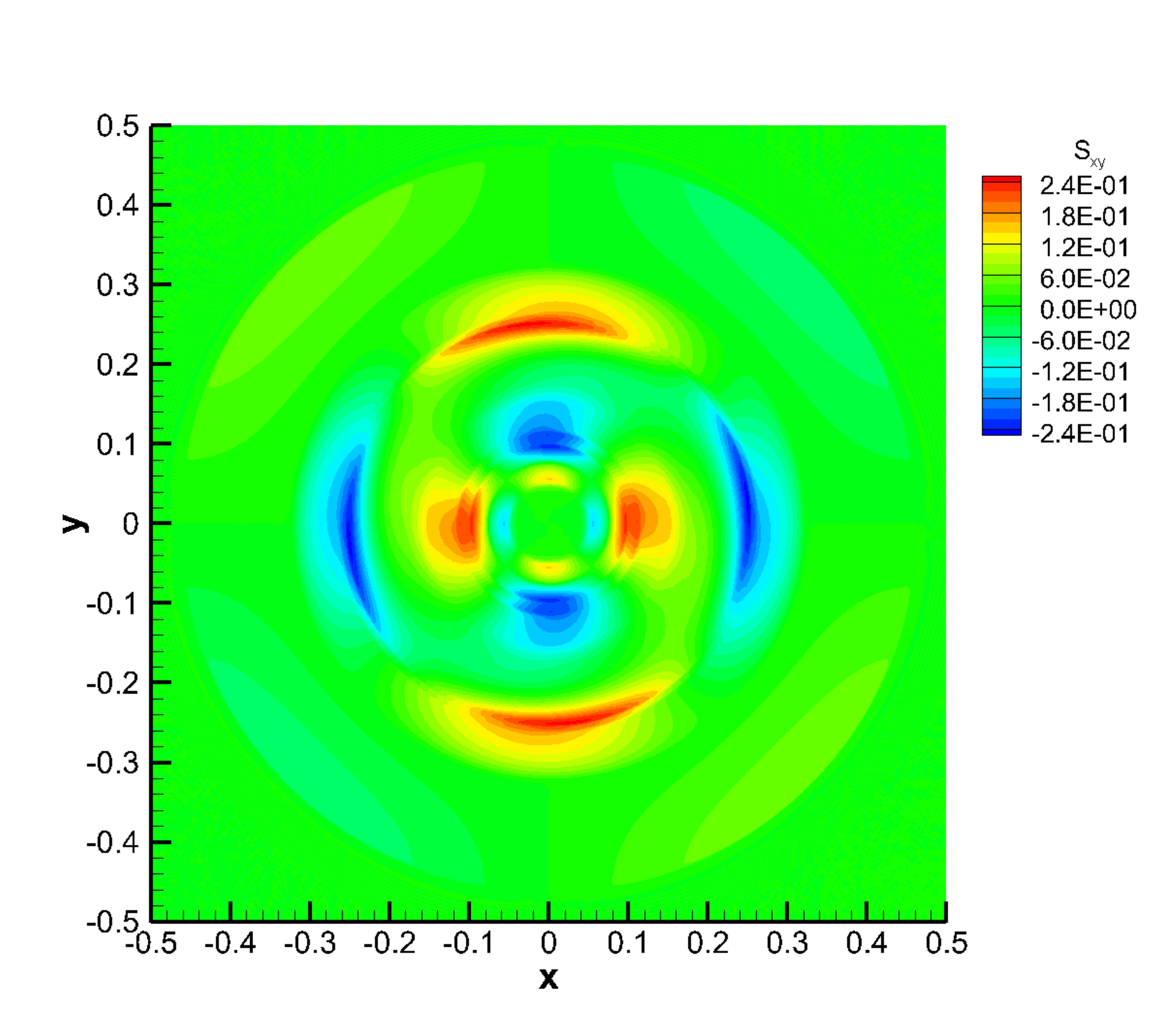}
			 \\
			\includegraphics[draft=false,width=0.47\textwidth]{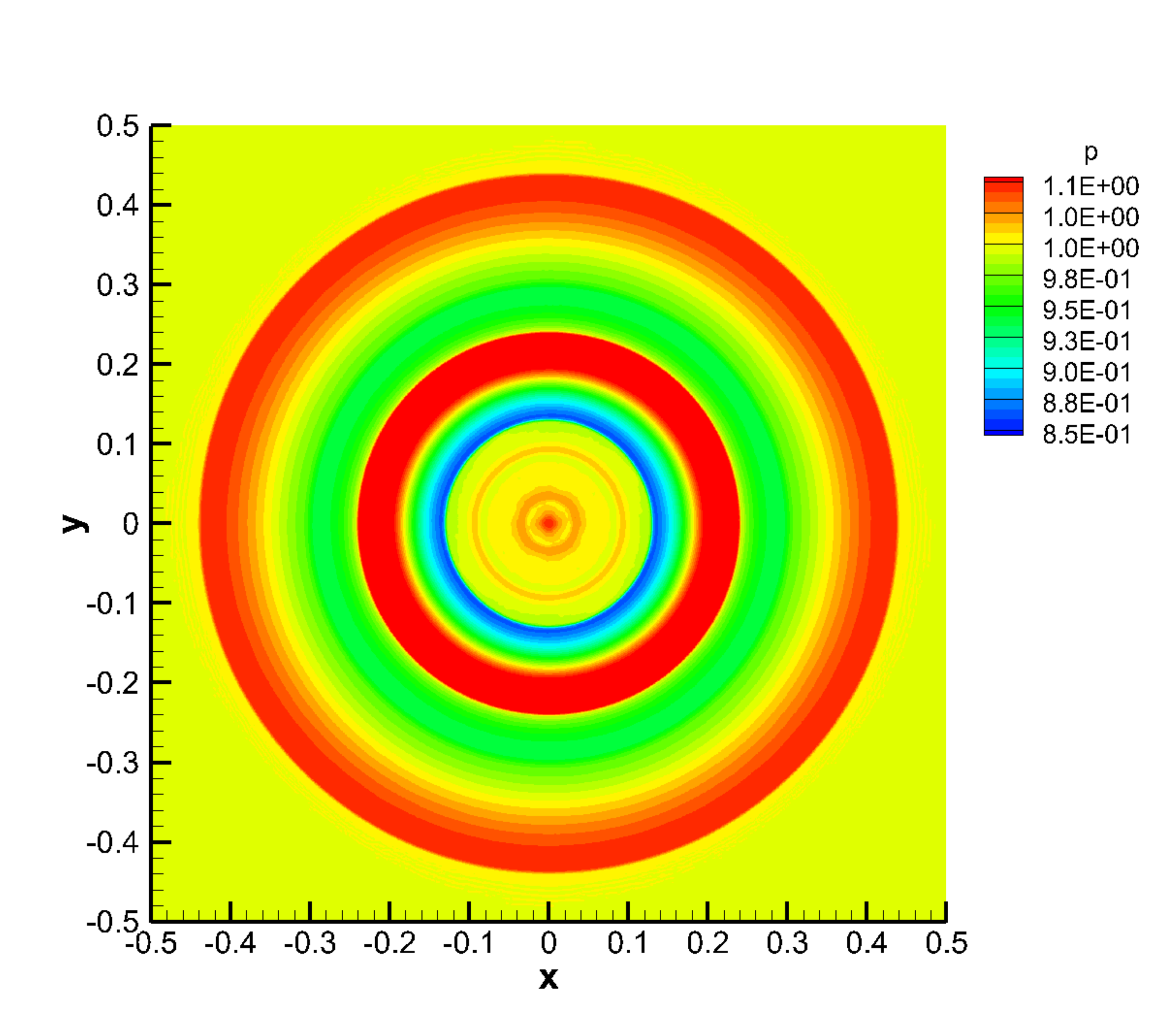}
			  &           
			\includegraphics[draft=false,width=0.47\textwidth]{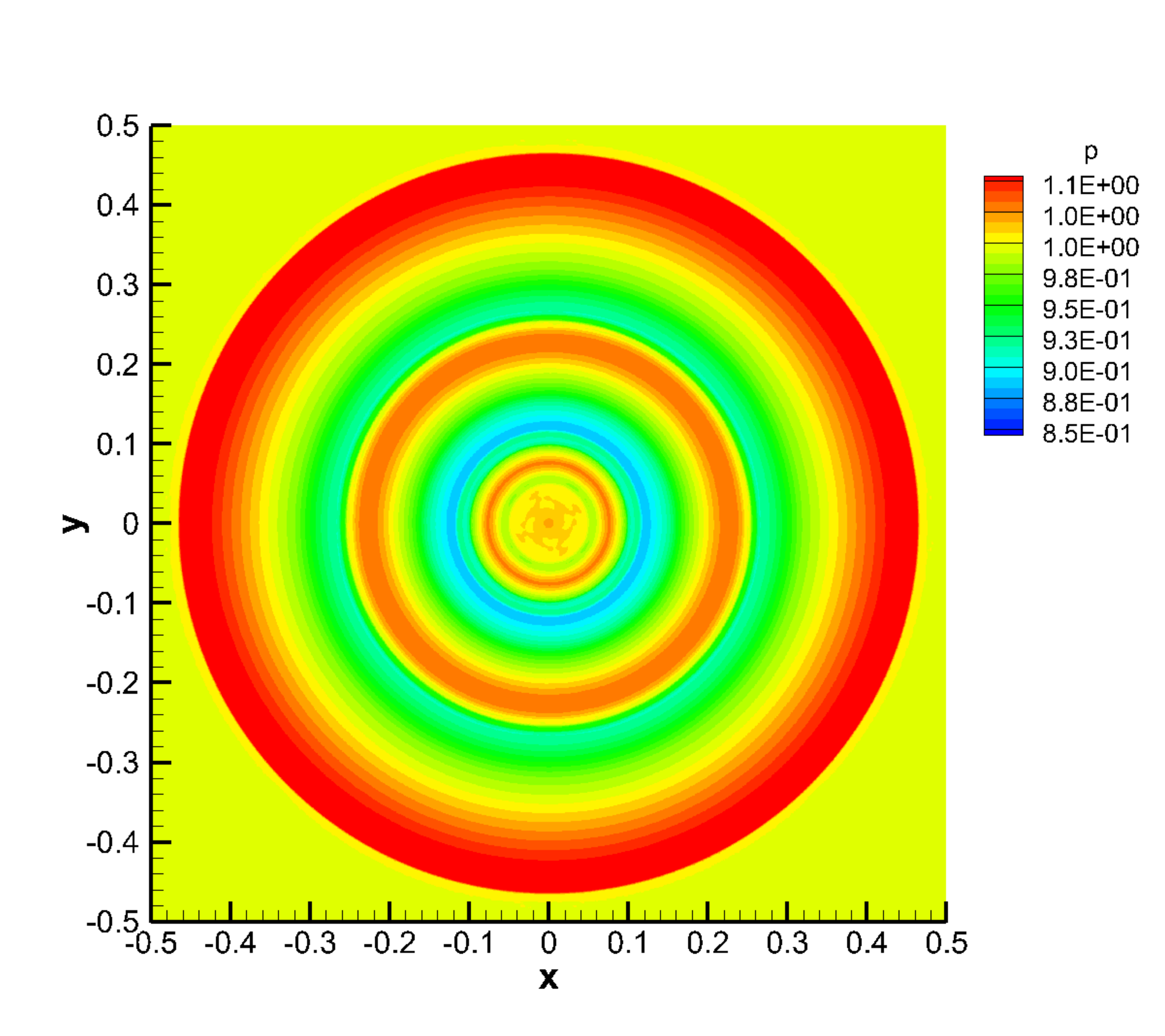}
			 \\
  \end{tabular}
  \caption{Results for the solid rotor problem obtained with ADER-DG fourth order scheme with 
  initial angular velocity of the rotor of $\vv \times \omega=1.0$. Horizontal velocity (top row), 
  stress tensor component $ \sigma_{21} $ and $S_{21}$ (middle row) and pressure $p$ (bottom row) 
  are displayed.
  The Wilkins model results are displayed on the left panels, \GPR model ones on the right panels.}
  \label{fig.rotor}
\end{center}
\end{figure}

Finally, both physical models are used to simulate the solid rotor problem proposed in 
\cite{Dumbser2008} for the equations of nonlinear hyperelasticity. This test 
case consists of a homogeneous elastic material with a circular part subject  to a sudden  
velocity impulse with the angular velocity $\omega$. It might be interpreted as a highly 
simplified model for the inner (rotating) and the outer (fixed) part of a bearing which are 
suddenly attached together via spontaneous welding by friction. The computational domain is the 
square $\Omega(t=0)=[-0.5:0.5]\times [-0.5:0.5]$ which is discretized with a Cartesian mesh of 
characteristic mesh size $h=0.02$. Transmissive boundary conditions are set everywhere and the 
final time is chosen to be $t_{\text{final}}=0.005$. The rotor radius is $R_i=0.1$ while 
$r=\sqrt{x^2+y^2}$ denotes the generic radial coordinate. We select two different angular 
frequencies, namely $\omega_1=10^{-2}$ and $\omega_2=10$, so that the tangential velocity of the 
rotor is $v_{t,1}=10^{-3}$ and $v_{t,2}=1$ at $r=R_i$, respectively. The material parameters are 
the same as the ones used for the shear layer test problem from  the previous section. The 
simulations are performed using a fourth order ADER-DG scheme and the numerical results are shown 
in Figs.~\ref{fig.rotor-small} and \ref{fig.rotor} for $v_{t,1}=10^{-3}$ and $v_{t,2}=1$ 
respectively. 
Shear waves traveling into both the rotor and the stator as well as pressure waves can be observed. 
Even in this case, if the shear stresses are not too strong like in Fig.~\ref{fig.rotor-small} with 
$v_{t,1}=10^{-3}$, the hypoelastic Wilkins model and the hyperelastic \GPR model show almost the 
same behavior. This is expected because, in this case, they both reduce to the equations of linear 
elasticity. 
However, when the deformations are larger, see Fig.~\ref{fig.rotor} with $v_{t,2}=1$, the 
solutions of the models are different which, as in the previous example, can be explained by the 
inability of Wilkins model to deal with large reversible non-linear elastic deformations, and its extension to finite strains is a rather complicated process.

\section{Conclusion and Perspectives} \label{sec:conclusion}

The goal of this paper was to compare the \GPR hyperelastic model against the well-known 
hypoelastic model of Wilkins. 
While the former model is derived in the framework of symmetric hyperbolic and thermodynamically 
compatible equations and somewhat more complex, the latter is apparently simpler but thermodynamically inconsistent and suffers from modeling 'choices' that are sometimes difficult to justify. 
We have implemented both models into the same simulation codes based either on a high order direct ALE framework on moving unstructured meshes, or a high order ADER-Discontinuous Galerkin scheme with subcell finite volume limiter on fixed Cartesian grids. 
We have elaborated the differences between the two models from the modeling point of view. 
Moreover, we have performed a systematic comparison of their numerical behavior on a set
of well known test cases.

A large set of benchmark problems in 1D and 2D have been simulated. 
In this work the materials have been described by a Mie-Gr\"uneisen or stiffened gas equation of state for the hydrodynamic pressure. In the \GPR model, a simple equation of state based on the 
invariants of the trace-free part of the metric tensor $\mathbf{G} = \A^T \A$ has been used.  
The numerical results of both models  have been reported to assess their differences, and, also, to 
show some circumstances where the models behave alike.
We have observed that on the most simple elastic tests, both models produce very much resembling 
results. More differences are observed when transient plastic deformations occur since the 
inelasticity is incorporated in a fundamentally different way in these two models, in a  
rate-dependent manner in the \GPR model and in the rate-independent ideal plasticity manner in the 
Wilkins model. This fact was already expected from the analysis of the models. 
Since the numerical framework used for the comparison is the same, we can therefore clearly 
attribute those difference to the intrinsically different properties of the models. 

Even more differences have been observed in the case when the material undergoes finite elastic 
deformations. In this case, as it was expected, the hypoelastic Wilkins model has to be modified 
by adding extra differential terms to the evolution equation of the stress deviator with solution 
dependent elastic modulus, which is not a trivial task. On the contrary, due to the hyperelastic 
character of the \GPR model, no new differential terms should be applied to the model, but only the 
equation of state should be adjusted, which is more convenient from the practical viewpoint than modifying the evolution equation for the stress deviator.

 We would like to mention that although Wilkins model  is thermodynamically inconsistent, 
 this model is extremely simple, in particular when considering complex boundary conditions in multi-dimensions.
Contrarily the \GPR model, appealing from the modeling point of view, is more complex to implement
and demands extra care when boundary conditions are involved.
This will be the subject of a forthcoming work.
  
%
%
%
In the future, we plan to investigate the hyperelastic \GPR model for different equations 
of state and to apply it also to non-Newtonian fluids, in order to widen its range of practical
applicability. Moreover, we plan to compare the numerical results against available laboratory 
experiments. 
Other directions of research will be dedicated to  enriching the model capabilities by the ability 
to  deal with complex physics of inelastic 
deformations in order to account for damaging and dislocation dynamics via an explicit inclusion of 
the evolution equation of the damage order parameter~\cite{Resnyansky2003,Romenskii2007} and of the 
Burgers 
tensor~\cite{SHTC-GENERIC-CMAT}, respectively.



\section*{Acknowledgments}
The authors would like to thank M. Shashkov and J. Kamm for the inspiring discussion which brought 
the seminal idea for the research presented in this work.

The research contained in this paper has been financed by the European 
Research Council (ERC) under the
European Union's Seventh Framework Programme (FP7/2007-2013) with the research 
project \textit{STiMulUs}, 
ERC Grant agreement no. 278267. 
M.D. has further received funding from the European Union's Horizon 2020 
Research and 
Innovation Programme under the project \textit{ExaHyPE}, grant agreement 
number  671698 (call
FETHPC-1-2014). 
E.R. acknowledges a partial support by the Program N15 of the Presidium of 
RAS, project 121 and the Russian foundation for Basic Research (grant number 16-29-15131). 
I.P. greatly acknowledges the support by ANR-11-LABX-0040-CIMI within the 
program ANR-11-IDEX-0002-02.
The authors are grateful to the Leibniz Rechenzentrum (LRZ) for awarding 
access to the SuperMUC supercomputer based in Munich, Germany; they also
acknowledge the support of the HLRS computing center for providing access 
to the Hazel Hen supercomputer based in Stuttgart, Germany. M.D. has 
received further funding from the Italian Ministry of Education, 
University and Research (MIUR) in the frame of the Departments of Excellence 
Initiative 2018--2022 attributed to DICAM of the University of Trento  
and has been supported by the University of Trento in the frame of 
the Strategic Initiative \textit{Modeling and Simulation}.


\normalem	




\bibliographystyle{model2-names.bst}\biboptions{authoryear}





\bibliography{./library}

\end{document}